\documentclass[11pt]{revtex4}
\usepackage{mathtools}
\usepackage{hyperref}
\usepackage{pdflscape}
\usepackage{graphics}
\usepackage{subfigure}
\usepackage{epsfig}
\usepackage{amsmath,amsfonts}
\usepackage{amssymb}
\usepackage{color}
\usepackage{ulem}
\usepackage{amsmath}
\usepackage{CJKutf8}
\allowdisplaybreaks

\begin{document}
	\begin{CJK}{UTF8}{gbsn}
\title{{Gravitational-wave effects in the most general vector-tensor theory}}
%{A Unified Framework for Analyzing Gravitational Wave Effects in Modified Gravity Theories Based on the Isaacson Picture} %Analyzing gravitational wave effects in general modified gravity: an example based on the most general vector-tensor theory}

\author{Yu-Qi Dong$^{a,b}$}
\email{dongyq2023@lzu.edu.cn}

\author{Xiao-Bin Lai$^{a,b}$}
\email{laixb2024@lzu.edu.cn}

\author{Yu-Qiang Liu$^{c}$}
\email{liuyq18@lzu.edu.cn}

\author{Yu-Xiao Liu$^{a,b}$}
\email{liuyx@lzu.edu.cn (corresponding author)}

\affiliation{$^{a}$ Lanzhou Center for Theoretical Physics, Key Laboratory of Theoretical Physics  of Gansu Province, Key Laboratory of Quantum Theory and Applications of MoE, Gansu Provincial Research Center for Basic Disciplines of Quantum Physics, Lanzhou University, Lanzhou 730000, China\\
	$^{b}$ Institute of Theoretical Physics \& Research Center of Gravitation, Lanzhou University, Lanzhou 730000, China\\
	$^{c}$ Department of Basic Courses, Lanzhou Institute of Technology, Lanzhou 730050, China}

\begin{abstract}
\textbf{Abstract:} 
{In this paper, we establish a model-independent framework based on the Isaacson picture to analyze the gravitational-wave effects in the most general vector-tensor theory that yields second-order field equations. Within this framework, we derive two basic sets of equations for the Isaacson picture. These equations enable the analysis of gravitational wave polarization modes, the dispersion relation of each mode, the effective energy-momentum tensor of gravitational waves, and the memory effects. These features are closely tied to observable phenomena and have attracted considerable attention. They are expected to be detected by the next generation of gravitational wave observatories designed to test potential modifications to general relativity. Using this framework, we present the explicit expression for the effective energy-momentum tensor of gravitational waves in the most general second-order vector-tensor theory and perform a complete analysis of their polarization modes.}

\end{abstract}
	
\maketitle

\section{Introduction}
\label{sec: intro} 

Although general relativity remains the most successful theory of gravity to date, it still faces both theoretical \cite{Weinberg,N. Arkani-Hamed,R. Penrose,A. Shomer} and observational \cite{V. C. Rubin,A. G. Riess} challenges. This has motivated researchers to propose various modified gravity theories. Naturally, determining how to test these theories through experimental and observational means has become a key scientific question.

Significant differences between various theories of gravity and general relativity usually emerge in regions of strong gravitational fields. However, due to the weak nature of gravitational interactions, scientists have long lacked effective direct detection methods for observing effects in strong gravitational regions. The direct detection of gravitational waves marks a significant advancement in addressing this challenge \cite{Abbott1,Abbott2,Abbott3,Abbott4,Abbott5}. Many astrophysical sources located in strong gravitational regions emit gravitational waves, which may carry information about the strong gravitational fields in these regions. It is evident that gravitational wave detection offers a direct method for observing the effects of strong gravitational regions. Additionally, gravitational waves themselves are a direct manifestation of gravitational effects. The basic effects of gravitational waves, such as polarization, wave speed, effective energy-momentum tensor, and nonlinear memory effect, typically vary across different modified gravity theories.  As a result, gravitational wave detection has emerged as a powerful tool for testing and constraining theories of gravity. At present, numerous gravitational wave detectors are either in operation or under development \cite{J. Aasi,F. Acernese,T. Akutsu,M. Punturo,D. Reitze,P. Amaro-Seoane,Z. Luo,J. W. Mei,A. Torres-Orjuela,Z. Chen,Z. Chen2,G. Agazie,Yu-Mei Wu,Ziwei Wang,Zhu Yi,H. Xu,F. Jenet,B. W. Stappers,R. N. Manchester,B. C. Joshi,J. L. Li,L. J. Shao2}. As more gravitational wave detectors come online, we anticipate detecting numerous high-precision gravitational wave events in the near future. These events will significantly constrain the range of possible modified gravity theories.

Gravitational wave detection has become an important tool for testing theories of gravity. Meanwhile, many candidate modified gravity theories have been proposed. Therefore, from a theoretical perspective, it is natural to develop a model-independent method. This approach allows for the study of gravitational wave effects not one theory at a time, but simultaneously across all theories that satisfy certain fundamental assumptions. In this regard, the Isaacson picture \cite{Isaacson1,Isaacson2,MTW,Michele Maggiore,Lavinia Heisenberg1} effectively provides a way to establish such a model-independent analytical framework. There have already been numerous studies on the gravitational wave effects in modified gravity theories. For instance, Refs. \cite{Eardley,f(R,f(R2,Horndeski0,TeVeS,Horava,STVG,fT,dCS and EdGB,Y.Dong,Y.Dong2,S. Bahamonde,J.Lu,L.Shao,S. Nojiri,Y.Dong3,TH. Hyun,Kristen Schumacher,M.Alves,Y.Liu,Y.Dong4,Shaoqi Hou,Xiao-Bin Lai,Eva Sagi} discussed the polarization modes and wave speeds of gravitational waves, while Refs. \cite{Christopher P. L. Berry,Leo C. Stein,A. Saffer,Joao C. Lobato} addressed the effective energy-momentum tensor of gravitational waves. Additionally, the memory effect on gravitational waves is discussed in Refs. \cite{Demetrios Christodoulou,Lavinia Heisenberg1,Lavinia Heisenberg2,Arpan Hait}. In particular, our previous work \cite{Y.Dong4} provides examples of using the Isaacson picture to analyze the gravitational wave polarization modes in a model-independent way, within the most general pure metric theory and the most general scalar-tensor theory.

This work aims to establish a model-independent framework for analyzing gravitational wave effects in the most general vector-tensor theories, which yield second-order field equations. We place particular emphasis on the analysis of gravitational wave polarization modes. The motivation for considering vector-tensor theories is as follows. Lovelock's theorem \cite{David Lovelock1,David Lovelock2} states that in four-dimensional spacetime with Riemannian geometry, if gravity is to be described solely by the metric, then the only theory that can derive a second-order field equation is general relativity. Therefore, if we retain assumptions about the dimensionality and geometry of spacetime and still require the field equations to be second-order, we can modify general relativity only by introducing additional fields. In this type of modified gravity theory, the most common examples are scalar-tensor theory, which includes an additional scalar field, and vector-tensor theory, which includes an additional vector field. The most general scalar-tensor theory yielding second-order field equations has been formulated, and is known as the Horndeski theory \cite{Horndeski}. The most general action for vector-tensor theories that yields second-order field equations remains unknown. Notable examples of such theories include the generalized Proca theory \cite{L. Heisenberg3}, the bumblebee model \cite{V. A. Kostelecky}, and the Einstein-aether theory  \cite{Wei-Jian Geng}. These representative vector-tensor theories are mutually disjoint in their construction and correspond to three distinct choices of parameters in the most general action. This motivates our investigation of the most general class of vector-tensor theories in this work.
The organization of this paper is as follows: %In Sec. \ref{sec: 2}, we review the Isaacson picture and rephrase the perturbation action method for general relativity and modified gravity theories. 
In Sec. \ref{sec: 3}, {we derive the Isaacson picture in vector-tensor theory and show that knowing only the second-order term in the action with respect to perturbations in the Minkowski background is sufficient to construct a framework for the model-independent analysis of gravitational-wave effects.}
%we explain why, under an asymptotic Minkowski spacetime, it is only necessary to know the second-order perturbation action to use the perturbation action method to obtain the two sets of basic equations of Isaacson picture far from the source. 
In Sec. \ref{sec: 4}, we construct the second-order perturbation action for the most general second-order vector-tensor theory. In Sec. \ref{sec: 5}, {we derive the fundamental set of equations describing the gravitational-wave effects in the most general second-order vector-tensor theory and present the expression for the effective energy-momentum tensor of gravitational waves.} %we use the most general second-order vector-tensor theory as an example to demonstrate the perturbation action method, particularly deriving the expression of its energy-momentum tensor in an asymptotic Minkowski spacetime, far from the source.
In Sec. \ref{sec: 6}, we analyze the polarization modes of gravitational waves in the most general second-order vector-tensor theory. Finally, Sec. \ref{sec: 7} is the conclusion.

We use $c=G=1$ and adopt the metric signature $(-,+,+,+)$. The indices $(\mu,\nu,\lambda,\rho)$ range over four-dimensional spacetime indices ($0,1,2,3$), while the indices $(i,j,k,l)$ range over three-dimensional spatial indices ($1,2,3$), corresponding to the ($+x,+y,+z$) directions, respectively.

\section{Isaacson picture in an asymptotic Minkowski spacetime}
\label{sec: 3}
{As a prerequisite for analyzing gravitational-wave effects, we first derive the Isaacson picture in vector-tensor theory in this section. Specifically, we demonstrate how to derive the Isaacson framework for regions far from the source in asymptotically Minkowski spacetime. This derivation relies solely on the second-order expansion of the action for high-frequency perturbations about a Minkowski background. It should be noted that although our derivation is limited to vector-tensor theories, it can be directly extended to any modified gravity theory without introducing essential differences. In principle, the Isaacson picture in general relativity is essentially different from that in modified gravity. Nevertheless, in Appendix \ref{app: H}, we demonstrate that the case in modified gravity can be directly extended from that in general relativity.}

%In this section, we point out that knowing only the second-order term $S^{(2)}$ of the action of a theory with respect to the high-frequency perturbations allows us to derive the two sets of basic equations of the Isaacson pictures far from the source in an asymptotic Minkowski spacetime. Furthermore, in fact, knowing the perturbed $S^{(2)}$ in the Minkowski background is sufficient.

{We consider a vector-tensor theory with a vector field $\mathcal{A}^{\mu}$}:
%Without loss of generality, we use the vector-tensor theory with an additional vector field $\mathcal{A}^{\mu}$ as an example:
\begin{eqnarray}
	\label{vector-tensor theory action}
	S=\int d^{4}x \sqrt{-g} \mathcal{L}\left[g_{\mu\nu},\mathcal{A}^{\mu}\right].
\end{eqnarray}
%{It should be noted that the use of vector-tensor theory as an example here is merely for the convenience of notation. All arguments in this section can be directly rewritten to apply to cases of modified gravity theories with any additional fields. Therefore, the conclusions in this section are quite general and applicable to the vast majority of modifications to gravity theory.}
Varying the action (\ref{vector-tensor theory action}) with respect to $g_{\mu\nu}$ and $\mathcal{A}^{\mu}$, respectively, we find that 
\begin{eqnarray}
	\label{varing the vector-tensor theory action}
	\delta S
	&=&\int d^{4} x \sqrt{-g} 
	\left(
	-\mathcal{M}^{\mu\nu}
	\delta g_{\mu\nu}
	+\mathcal{N_{\mu}} \delta \mathcal{A}^{\mu}
	\right).
\end{eqnarray}
Therefore, the field equations of this theory are 
\begin{eqnarray}
	\label{vector-tensor theory field equations metric}
	\mathcal{M}_{\mu\nu} \left[g_{\mu\nu},\mathcal{A}^{\mu}\right]&=&0,
	\\
	\label{vector-tensor theory field equations vector field}
	\mathcal{N}_{\mu}\left[g_{\mu\nu},\mathcal{A}^{\mu}\right]&=&0.
\end{eqnarray}

By decomposing the fields into low-frequency and high-frequency parts
\begin{eqnarray}
	\label{perturbation vector-tensor theory}
	g_{\mu\nu}=\bar{g}_{\mu\nu}+h_{\mu\nu},\quad \mathcal{A}^{\mu}=\bar{A}^{\mu}+B^{\mu},
\end{eqnarray}	
where
\begin{eqnarray}	
	\label{relationship perturbation vector-tensor theory}
	\bar{g}_{\mu\nu} \sim \bar{A}^{\mu} \sim 1,\quad
	h_{\mu\nu} \sim B^{\mu} \sim \alpha, \quad \alpha \ll 1,
\end{eqnarray}
we can obtain the high-frequency equations of the Isaacson picture
\begin{eqnarray}
	\label{vector-tensor field equations metric high feq}
	\mathcal{M}^{(1)}_{\mu\nu}\left[\bar{g}_{\mu\nu},\bar{A}^{\mu},h_{\mu\nu},B^{\mu}\right]&=&0,
	\\
	\label{vector-tensor field equations vector field high feq}
	\mathcal{N}^{(1)}_{\mu}\left[\bar{g}_{\mu\nu},\bar{A}^{\mu},h_{\mu\nu},B^{\mu}\right]&=&0,
\end{eqnarray}
and the low-frequency equations
\begin{eqnarray}
	\label{vector-tensor field equations metric low feq}
	\mathcal{M}^{(0)}_{\mu\nu}\left[\bar{g}_{\mu\nu},\bar{A}^{\mu}\right]
	+\left\langle\mathcal{M}^{(2)}_{\mu\nu}\left[\bar{g}_{\mu\nu},\bar{A}^{\mu},h_{\mu\nu},B^{\mu}\right]\right\rangle&=&0,
	\\
	\label{vector-tensor field equations vector field low feq}
	\mathcal{N}^{(0)}_{\mu}\left[\bar{g}_{\mu\nu},\bar{A}^{\mu}\right]
	+\left\langle\mathcal{N}^{(2)}_{\mu}\left[\bar{g}_{\mu\nu},\bar{A}^{\mu},h_{\mu\nu},B^{\mu}\right]\right\rangle&=&0.
\end{eqnarray}
{Here, the symbol $\langle ... \rangle$ is an averaging operator and the symbol $(i)$ in the superscript position denotes the $i$-th order term in the expansion of the perturbation $h_{\mu\nu}$ ($i=0, 1, 2, 3, ...$). We will continue to use this notation to label the perturbation terms in the following text.}

{Taking into account that gravitational wave detectors are located in an asymptotically Minkowski spacetime far from the source, we can further simplify the above two sets of basic equations in this situation.}
%Now, let us consider the case far from the source in an asymptotic Minkowski spacetime to further rewrite the forms of the two sets of basic equations.
In an asymptotic Minkowski spacetime, as we move infinitely far from the source, the background fields should approach the Minkowski spacetime solution, i.e.,
\begin{eqnarray}
	\label{vector-tensor theory background solution in Minkowski spacrtime}
	\bar{g}_{\mu\nu} \rightarrow \eta_{\mu\nu},\quad 
	\bar{A}^{\mu}\rightarrow{A}^{\mu}=\left(A,0,0,0\right).
\end{eqnarray}	
Here, since the Minkowski spacetime is homogeneous and isotropic, we assume that the background vector field $A^{\mu}$ only has a temporal component $A$, and $A$ is a constant. Solution (\ref{vector-tensor theory background solution in Minkowski spacrtime}) should satisfy Eqs. (\ref{vector-tensor theory field equations metric}) and (\ref{vector-tensor theory field equations vector field}).

Therefore, when far from the source,  the background fields can be further decomposed into
\begin{eqnarray}
	\label{perturbation background vector-tensor theory}
	\bar{g}_{\mu\nu}=\eta_{\mu\nu}+\mathfrak{d}\bar{g}_{\mu\nu},\quad \bar{A}^{\mu}=A^{\mu}+\mathfrak{d} \bar{A}^{\mu},
\end{eqnarray}	
where
\begin{eqnarray}	
	\label{relationship perturbation background vector-tensor theory}
	\eta_{\mu\nu} \sim A^{\mu} \sim 1,\quad
	\mathfrak{d}\bar{g}_{\mu\nu} \sim \mathfrak{d} \bar{A}^{\mu} \sim \beta, \quad \beta \ll 1.
\end{eqnarray}
For convenience in discussion, we have assumed that both $\mathfrak{d}\bar{g}_{\mu\nu}$ and $\mathfrak{d} \bar{A}^{\mu}$ are of the order of $\beta$. Removing this assumption will not affect the subsequent results.

From Eq. (\ref{perturbation background vector-tensor theory}), we can further expand the equations for the perturbations $\mathfrak{d}\bar{g}_{\mu\nu}$ and $\mathfrak{d} \bar{A}^{\mu}$. For example, for $\mathcal{M}^{(0)}_{\mu\nu}$, we have
\begin{eqnarray}	
	\label{expand M0}
	\mathcal{M}^{(0)}_{\mu\nu}\left[\bar{g}_{\mu\nu},\bar{A}^{\mu}\right]
	=
	\mathcal{M}^{(0,0)}_{\mu\nu}\left[\eta_{\mu\nu},A^{\mu}\right]
	+\mathcal{M}^{(0,1)}_{\mu\nu}\left[\eta_{\mu\nu},A^{\mu},\mathfrak{d}\bar{g}_{\mu\nu},\mathfrak{d} \bar{A}^{\mu}\right]
	+\sum_{i = 2}^{\infty} \mathcal{M}^{(0,i)}_{\mu\nu}.
\end{eqnarray}
Here, the symbol $(0,i)$ in the upper right corner of the letter $\mathcal{M}$ denotes the $i$-th order term in the expansion of the perturbations $\mathfrak{d}\bar{g}_{\mu\nu}$ and $\mathfrak{d} \bar{A}^{\mu}$. The method of expressing the other terms is similar. It can be seen that
\begin{eqnarray}	
	\label{M00=M0}
	\mathcal{M}^{(0,0)}_{\mu\nu}\left[\eta_{\mu\nu},A^{\mu}\right]
	=
	\mathcal{M}^{(0)}_{\mu\nu}\left[\eta_{\mu\nu},A^{\mu}\right]
	=
	\mathcal{M}_{\mu\nu}\left[\eta_{\mu\nu},A^{\mu}\right]
	=
	0,
\end{eqnarray}
and
\begin{eqnarray}	
	\label{M01=M1}
	\mathcal{M}^{(0,1)}_{\mu\nu}\left[\eta_{\mu\nu},A^{\mu},\mathfrak{d}\bar{g}_{\mu\nu},\mathfrak{d} \bar{A}^{\mu}\right]
	=
	\mathcal{M}^{(1)}_{\mu\nu}\left[\eta_{\mu\nu},A^{\mu},\mathfrak{d}\bar{g}_{\mu\nu},\mathfrak{d} \bar{A}^{\mu}\right].
\end{eqnarray}

In Eq. (\ref{expand M0}), each $\mathcal{M}^{(0,i)}_{\mu\nu}$ with $i \textgreater 0$ can be ignored compared to $\mathcal{M}^{(0,0)}_{\mu\nu}$. Thus, it can be seen that when we take the leading-order of the high-frequency equations, we have
\begin{eqnarray}
	\label{vector-tensor field equations metric high feq leading-order}
	\mathcal{M}^{(1)}_{\mu\nu}\left[\eta_{\mu\nu},A^{\mu},h_{\mu\nu},B^{\mu}\right]&=&0,
	\\
	\label{vector-tensor field equations vector field high feq leading-order}
	\mathcal{N}^{(1)}_{\mu}\left[\eta_{\mu\nu},A^{\mu},h_{\mu\nu},B^{\mu}\right]&=&0.
\end{eqnarray}
{This set of equations can be used to derive the polarization modes of gravitational waves and the dispersion relation of each mode.} For the leading-order of the low-frequency equations, we have
\begin{eqnarray}
	\label{vector-tensor field equations metric low feq leading-order}
	\mathcal{M}^{(1)}_{\mu\nu}\left[\eta_{\mu\nu},A^{\mu},\mathfrak{d}\bar{g}_{\mu\nu},\mathfrak{d}\bar{A}^{\mu}\right]
	+\left\langle\mathcal{M}^{(2)}_{\mu\nu}\left[\eta_{\mu\nu},A^{\mu},h_{\mu\nu},B^{\mu}\right]\right\rangle&=&0,
	\\
	\label{vector-tensor field equations vector field low feq leading-order}
	\mathcal{N}^{(1)}_{\mu}\left[\eta_{\mu\nu},A^{\mu},\mathfrak{d}\bar{g}_{\mu\nu},\mathfrak{d}\bar{A}^{\mu}\right]
	+\left\langle\mathcal{N}^{(2)}_{\mu}\left[\eta_{\mu\nu},A^{\mu},h_{\mu\nu},B^{\mu}\right]\right\rangle&=&0.
\end{eqnarray}
Here, the leading-order of the effective energy-momentum tensor of gravitational waves is
\begin{eqnarray}
	\label{leadingorder effective energy-momentum tensor of gravitational waves in MGT}
	t_{\mu\nu}
	= 
	-2 
	\left\langle 
	\mathcal{M}^{(2)}_{\mu\nu}\left[\eta_{\mu\nu},A^{\mu},h_{\mu\nu},B^{\mu}\right] 
	\right\rangle.
\end{eqnarray}
{This set of equations can be used to derive the effective energy-momentum tensor of gravitational waves and to analyze their memory effects. When the field equations are second-order, the following assumption is also required to hold:}
\begin{eqnarray}
	\label{assumption 0}
    \beta \sim \left(\frac{f_{H}}{f_{L}}\alpha\right)^{2},
\end{eqnarray}
{where $f_H$ and $f_L$ are the characteristic frequencies of the high-frequency gravitational waves and the low-frequency background, respectively.}

{The perturbation action method constitutes an important approach within the Isaacson picture. As detailed in Appendix \ref{app: G}, this method reveals that all quantities related to the above equations can be derived using the second-order perturbation action $S^{(2)}_{flat}\left[\eta_{\mu\nu},A^{\mu},h_{\mu\nu},B^{\mu}\right] $ in the Minkowski background:}
\begin{eqnarray}
	\label{M1=deltaS flat/delta h 0}
	\mathcal{M}^{(1)}_{\mu\nu}\left[\eta_{\mu\nu},A^{\mu},h_{\mu\nu},B^{\mu}\right]
	\!&=&\!
	-\eta_{\mu\lambda}\eta_{\nu\rho}
	\frac{\delta S_{flat}^{(2)}}{\delta h_{\lambda\rho}}
	=-\frac{\delta S_{flat}^{(2)}}{\delta h^{\mu\nu}},
	\\
	\label{N1=deltaS flat/delta B 0}
	\mathcal{N}^{(1)}_{\mu}\left[\eta_{\mu\nu},A^{\mu},h_{\mu\nu},B^{\mu}\right]
	\!&=&\!
	\frac{\delta S_{flat}^{(2)}}{\delta B^{\mu}},
	\\
	\label{M2=deltaS flat/delta g 0}
	\left\langle\mathcal{M}^{(2)}_{\mu\nu}\left[\eta_{\mu\nu},A^{\mu},h_{\mu\nu},B^{\mu}\right]\right\rangle
	\!&=&\!
	-\eta_{\mu\lambda}\eta_{\nu\rho}
	\left\langle\frac{\delta S_{flat}^{(2)}}{\delta \eta_{\lambda\rho}}\right\rangle
	=\left\langle\frac{\delta S_{flat}^{(2)}}{\delta \eta^{\mu\nu}}\right\rangle,
	\\
	\label{N2=deltaS flat/delta A 0}
	\left\langle\mathcal{N}^{(2)}_{\mu}\left[\eta_{\mu\nu},A^{\mu},h_{\mu\nu},B^{\mu}\right]\right\rangle
	\!&=&\!
	\left\langle\frac{\delta S_{flat}^{(2)}}{\delta A^{\mu}}\right\rangle.
\end{eqnarray}
Especially,
\begin{eqnarray}
	\label{t_munu S2flat 0}
	t_{\mu\nu}
	= 
	-2\left\langle\frac{\delta S_{flat}^{(2)}}{\delta \eta^{\mu\nu}}\right\rangle.
\end{eqnarray}
Here, varying the action with respect to $\eta_{\mu\nu}$ and $A^{\mu}$ means formally considering $\eta_{\mu\nu}$ and $A^{\mu}$ as variables during the variation, and then substituting their actual values after obtaining the field equations. {For the proof of this statement and a more detailed discussion, see Appendix \ref{app: F}.}

\section{Second-order action of the most general vector-tensor theory}
\label{sec: 4}
In the previous section, we explained that as long as the second-order perturbation action in the Minkowski background is known, various gravitational-wave effects can be analyzed. Now, in order to study the gravitational-wave effects of the most general modified gravity theories that satisfy certain common assumptions, it is not necessary to find the most general action that satisfies these assumptions, but only to construct the most general second-order perturbation action. Compared to the former, the latter is often much easier. This most general second-order perturbation action will contain numerous theoretical parameters. The experimental detection of gravitational waves can provide the range of values for these parameters without considering a specific theory that satisfies the assumptions. The theoretical work only requires determining the relationship between a specific theory and the parameters in the most general second-order perturbation action. This can avoid the duplication of theoretical and experimental work and provides the possibility of using gravitational wave detection to test gravity theories independently of specific models. {Therefore, in this section, we construct the most general second-order perturbation action for the theory, which will be used to analyze gravitational-wave effects in subsequent sections.} %In this section, we use the most general second-order vector-tensor theory as an example to demonstrate how to construct the most general second-order perturbation action and analyze their gravitational wave effects in subsequent sections. 
Throughout this and following sections, indices are raised and lowered using $\eta^{\mu\nu}$ and $\eta_{\mu\nu}$.
 
We consider vector-tensor theory with an additional vector field and continue to use the symbols from the previous section. For the theory under consideration, we make the following assumptions:
\begin{itemize}
	\item [(1)] Spacetime is represented by a four-dimensional (pseudo) Riemannian manifold.
	\item [(2)] The theory satisfies the principle of least action.
	\item [(3)] The theory is generally covariant.
	\item [(4)] The field equations are second-order.
	\item [(5)] The action of a free particle is $\int ds=\int \sqrt{|g_{\mu\nu}dx^\mu dx^\nu|}$.
\end{itemize}
For assumption (1), a four-dimensional spacetime aligns most closely with our life experience. If spacetime is higher-dimensional, additional explanation would be required to address why we cannot observe these extra dimensions. For simplicity, we continue to use the concept of Riemannian geometry as employed in general relativity. Assumption (2) is necessary for constructing the second-order perturbation action. Assumption (3) ensures the equivalence of all reference frames. Higher than second-order field equations often lead to the Ostrogradski instability \cite{M.Ostrogradsky,R.P.Woodard,H.Motohashi,A.Ganz}. For simplicity, we apply assumption (4). Finally, assumption (5) requires that free particles have minimal coupling with the metric. This implies that we do not need to redefine concepts such as the polarizations of gravitational waves; instead, we can still use the standard definition.

{Following a method similar to that in Ref. \cite{Y.Dong4}, the most general second-order perturbative action in vector-tensor theory satisfying the above assumptions takes the form:}
\begin{eqnarray}
	\label{the most general second-order perturbation action in vector-tensor theory}
	S^{(2)}_{flat}=S^{(2)}_{0}+S^{(2)}_{1}+S^{(2)}_{2}=\int d^4 x \sqrt{-\eta} \left(\mathcal{L}_{0}+\mathcal{L}_{1}+\mathcal{L}_{2}\right),
\end{eqnarray}
where
\begin{eqnarray}
	\label{L0 gauge}
	\mathcal{L}_{0}
	\!&=&\!
	A_{(0)} A^{\mu}A^{\nu}A^{\lambda}A^{\rho}h_{\mu\nu}h_{\lambda\rho}
	+4 A_{(0)} A^{\mu}A^{\nu}A^{\lambda}h_{\mu\nu}B_{\lambda}
	+4 A_{(0)} A_{\mu}A_{\nu}B^{\mu}B^{\nu},
\end{eqnarray}
\begin{eqnarray}
	\label{L1 gauge}
	\mathcal{L}_{1}
	\!&=&\!
	A_{(1)} \left(E^{\mu\lambda\sigma\gamma}A^{\nu}A^{\rho}A_{\sigma}\partial_{\gamma}h_{\mu\nu}\right)h_{\lambda\rho}
	+B_{(1)}\left(\left(A \cdot \partial\right)A_{\mu}A_{\nu}h\right)h^{\mu\nu}
	\nonumber \\
	\!&-&\!
	2B_{(1)}\left(A^{\mu}A^{\nu}\partial^{\lambda}h_{\mu\nu}\right)B_{\lambda}
	+2A_{(1)}\left(E^{\mu\lambda\sigma\gamma}A_{\sigma}\partial_{\gamma}A^{\nu}h_{\mu\nu}\right)B_{\lambda}
	\nonumber \\
	\!&+&\!
	2B_{(1)}\left(\left(A \cdot \partial\right)A^{\mu}h\right)B_{\mu}
	\nonumber \\
	\!&-&\!
	4B_{(1)}\left(A_{\mu}\partial_{\nu}B^{\mu}\right)B^{\nu}
	-A_{(1)}\left(E^{\mu\nu\lambda\rho}\partial_{\lambda}A_{\rho}B_{\mu}\right)B_{\nu},
\end{eqnarray}
\begin{eqnarray}
	\label{L2 gauge}
	\mathcal{L}_{2}
	\!&=&\!
	A_{(2)} \left(\Box A^{\mu}A^{\nu}A^{\lambda}A^{\rho}h_{\mu\nu}\right)h_{\lambda\rho}
	+B_{(2)} \left(\left(A \cdot \partial\right)^{2}A^{\mu}A^{\nu}A^{\lambda}A^{\rho}h_{\mu\nu}\right)h_{\lambda\rho}
	\nonumber \\
	\!&+&\!
	C_{(2)} \left(\left(A \cdot \partial\right)A^{\mu}A^{\nu}A^{\lambda}\partial^{\rho}h_{\mu\nu}\right)h_{\lambda\rho}
	+D_{(2)} \left(A^{\mu}A^{\nu}\partial^{\lambda}\partial^{\rho}h_{\mu\nu}\right)h_{\lambda\rho}
	\nonumber \\
	\!&+&\!
	E_{(2)} \left(A^{\mu}A^{\lambda}\partial^{\nu}\partial^{\rho}h_{\mu\nu}\right)h_{\lambda\rho}
	-E_{(2)} \left(\Box A^{\lambda}A_{\rho}h_{\mu\lambda}\right)h^{\mu\rho}
	\nonumber \\
	\!&+&\!
	F_{(2)} \left(\left(A \cdot \partial\right)^{2} A^{\lambda}A_{\rho}h_{\mu\lambda}\right)h^{\mu\rho}
	+G_{(2)} \left(\left(A \cdot \partial\right)A^{\lambda}\partial_{\rho}h_{\mu\lambda}\right)h^{\mu\rho}
	\nonumber \\
	\!&-&\!
	2H_{(2)} \left(\partial^{\lambda}\partial_{\rho}h_{\mu\lambda}\right)h^{\mu\rho}
	+H_{(2)} \left(\Box h_{\mu\nu}\right)h^{\mu\nu}
	+2H_{(2)} \left(\partial_{\mu}\partial_{\nu}h\right)h^{\mu\nu}
	-H_{(2)} \left(\Box h\right)h
	\nonumber \\
	\!&+&\!
	I_{(2)} \left(\left(A \cdot \partial\right)^{2}h_{\mu\nu}\right)h^{\mu\nu}
	-D_{(2)} \left(\Box A_{\mu}A_{\nu}h\right)h^{\mu\nu}
	+J_{(2)} \left(\left(A \cdot \partial\right)^{2}A_{\mu}A_{\nu}h\right)h^{\mu\nu}
	\nonumber \\
	\!&-&\!
	G_{(2)} \left(\left(A \cdot \partial\right)A_{\mu}\partial_{\nu}h\right)h^{\mu\nu}
	+K_{(2)} \left(\left(A \cdot \partial\right)^{2}h\right)h
	\nonumber \\
	\!&+&\!
	\left(4A_{(2)}+C_{(2)}\right)\left(\Box A^{\mu}A^{\nu}A^{\lambda}h_{\mu\nu}\right)B_{\lambda}
	+4B_{(2)}\left(\left(A \cdot \partial\right)^{2}A^{\mu}A^{\nu}A^{\lambda}h_{\mu\nu}\right)B_{\lambda}
	\nonumber \\
	\!&+&\!
	\left(C_{(2)}+2J_{(2)}\right)\left(\left(A \cdot \partial\right)A^{\mu}A^{\nu}\partial^{\lambda}h_{\mu\nu}\right)B_{\lambda}
	+2\left(C_{(2)}+F_{(2)}\right)\left(\left(A \cdot \partial\right)A^{\mu}\partial^{\nu}A^{\lambda}h_{\mu\nu}\right)B_{\lambda}
	\nonumber \\
	\!&+&\!
	\left(2E_{(2)}-G_{(2)}\right)\left(A^{\mu}\partial^{\nu}\partial^{\lambda}h_{\mu\nu}\right)B_{\lambda}
	+\left(2D_{(2)}+G_{(2)}\right)\left(\partial^{\mu}\partial^{\nu}A^{\lambda}h_{\mu\nu}\right)B_{\lambda}
	\nonumber \\
	\!&+&\!
	\left(-2E_{(2)}+G_{(2)}\right)\left(\Box A^{\mu}h_{\mu\lambda}\right)B^{\lambda}
	+2F_{(2)}\left(\left(A \cdot \partial\right)^{2}A^{\mu}h_{\mu\lambda}\right)B^{\lambda}
	\nonumber \\
	\!&+&\!
	\left(G_{(2)}+4I_{(2)}\right)\left(\left(A \cdot \partial\right)\partial^{\mu}h_{\mu\lambda}\right)B^{\lambda}
	+\left(-2D_{(2)}-G_{(2)}\right)\left(\Box A^{\mu} h\right)B_{\mu}
	\nonumber \\
	\!&+&\!
	2J_{(2)}\left(\left(A \cdot \partial\right)^{2}A^{\mu}h\right)B_{\mu}
	+\left(-G_{(2)}+4K_{(2)}\right)\left(\left(A \cdot \partial\right)\partial^{\mu} h\right)B_{\mu}
	\nonumber \\
	\!&+&\!
	\left(4A_{(2)}+2C_{(2)}+F_{(2)}\right)\left(\Box A_{\mu}A_{\nu}B^{\mu}\right)B^{\nu}
	+4B_{(2)}\left(\left(A \cdot \partial\right)^{2}A_{\mu}A_{\nu}B^{\mu}\right)B^{\nu}
	\nonumber \\
	\!&+&\!
	\left(2C_{(2)}+2F_{(2)}+4J_{(2)}\right)\left(\left(A \cdot \partial\right)A_{\mu}\partial_{\nu}B^{\mu}\right)B^{\nu}
	\nonumber \\
	\!&+&\!
	\left(E_{(2)}+2I_{(2)}-G_{(2)}+4K_{(2)}\right)\left(\partial_{\mu}\partial_{\nu}B^{\mu}\right)B^{\nu}
	\nonumber \\
	\!&+&\!
	\left(-E_{(2)}+G_{(2)}+2I_{(2)}\right)\left(\Box B_{\mu}\right)B^{\mu}
	+F_{(2)}\left(\left(A \cdot \partial\right)^{2}B_{\mu}\right)B^{\mu}.
\end{eqnarray}
{Here, $E^{\mu\nu\lambda\rho}$ ($E^{0123}=1$) is the four-dimensional Levi-Civita totally antisymmetric tensor, $(A\cdot \partial)\coloneqq A^{\mu}\partial_{\mu}$, $\Box$ is the d'Alembert operator, and $A_{(0)},\dots, K_{(2)}$ are free parameters. For the detailed construction of the above action, see Appendix \ref{app: E}.} We hope that this theory, as an extension of general relativity, includes the Einstein-Hilbert term, which necessitates
\begin{eqnarray}
	\label{H2 neq 0}
	H_{(2)} \neq 0.
\end{eqnarray}

From Eqs. ({\ref{L0 gauge}}), (\ref{L1 gauge}), and (\ref{L2 gauge}), it can be seen that when the background vector field is zero, i.e., $A=0$, there is no coupling term for $h_{\mu\nu}$ and $B^{\mu}$ in the action, and the terms related to $h_{\mu\nu}$ are the same as those in general relativity. This indicates that when $A=0$, the properties of gravitational waves in the most general second-order vector-tensor theory are exactly the same as those in general relativity. So, when we analyze gravitational waves in the following text, we only consider the case where $A \neq 0$.

\section{Isaacson picture in the most general vector-tensor theory}
\label{sec: 5}
In this section, we demonstrate the derivation of the Isaacson picture for the most general second-order vector-tensor theory. In other words, we use $S^{(2)}_{flat}$ to derive two sets of basic equations.

Firstly, in order to obtain Eqs. (\ref{vector-tensor field equations metric high feq leading-order}) and (\ref{vector-tensor field equations vector field high feq leading-order}), which characterize the propagation of gravitational waves, according to Eqs. (\ref{M1=deltaS flat/delta h 0}) and (\ref{N1=deltaS flat/delta B 0}), we need to vary the action (\ref{the most general second-order perturbation action in vector-tensor theory}) with respect to the perturbations.

Varying the action with respect to the perturbation $h^{\mu\nu}$, we obtain
\begin{eqnarray}
	\label{delta S2flat/delta h}
	\frac{\delta S^{(2)}_{flat}}{\delta h^{\mu\nu}}
	\!&=&\!
	2A_{(0)} A_{\mu}A_{\nu}A^{\lambda}A^{\rho}h_{\lambda\rho}
	+4A_{(0)} A_{\mu}A_{\nu}A_{\lambda}B^{\lambda}
	\nonumber \\
	\!&+&\!
	A_{(1)} E^{\lambda~\sigma\gamma}_{~\mu}A^{\rho}A_{\nu}A_{\sigma}\partial_{\gamma}h_{\lambda\rho}
   +A_{(1)} E^{\lambda~\sigma\gamma}_{~\nu}A^{\rho}A_{\mu}A_{\sigma}\partial_{\gamma}h_{\lambda\rho}
   \nonumber \\
   \!&+&\!
   B_{(1)}\left(A \cdot \partial\right)A_{\mu}A_{\nu}h
   -B_{(1)}\eta_{\mu\nu}\left(A \cdot \partial\right)A_{\lambda}A_{\rho}h^{\lambda\rho}
   \nonumber \\
   \!&+&\!
   2B_{(1)}A_{\mu}A_{\nu}\partial_{\lambda}B^{\lambda}
   -2B_{(1)}\eta_{\mu\nu}\left(A \cdot \partial\right)A^{\lambda}B_{\lambda}
   \nonumber \\
   \!&+&\!
   A_{(1)}E^{\lambda~\sigma\gamma}_{~\mu}A_{\sigma}A_{\nu}\partial_{\gamma}B_{\lambda}
   +A_{(1)}E^{\lambda~\sigma\gamma}_{~\nu}A_{\sigma}A_{\mu}\partial_{\gamma}B_{\lambda}
   \nonumber \\
   \!&+&\!
   2A_{(2)}A_{\mu}A_{\nu}A^{\lambda}A^{\rho}\Box h_{\lambda\rho}
   +2B_{(2)}\left(A \cdot \partial\right)^{2}A_{\mu}A_{\nu}A^{\lambda}A^{\rho}h_{\lambda\rho}
   \nonumber \\
   \!&+&\!
   \frac{1}{2}C_{(2)}\left(A \cdot \partial\right)A_{\mu}\partial_{\nu}A^{\lambda}A^{\rho}h_{\lambda\rho}
   +\frac{1}{2}C_{(2)}\left(A \cdot \partial\right)A_{\nu}\partial_{\mu}A^{\lambda}A^{\rho}h_{\lambda\rho}
   \nonumber \\
   \!&+&\!
   C_{(2)}\left(A \cdot \partial\right)A_{\mu}A_{\nu}A^{\lambda}\partial^{\rho}h_{\lambda\rho}
   +D_{(2)}\partial_{\mu}\partial_{\nu}A^{\lambda}A^{\rho}h_{\lambda\rho}
   +D_{(2)}A_{\mu}A_{\nu}\partial_{\lambda}\partial_{\rho}h^{\lambda\rho}
   \nonumber \\
   \!&+&\!
   E_{(2)}A_{\mu}\partial_{\nu}A^{\lambda}\partial^{\rho}h_{\lambda\rho}
   +E_{(2)}A_{\nu}\partial_{\mu}A^{\lambda}\partial^{\rho}h_{\lambda\rho}
   -E_{(2)}A^{\lambda}A_{\nu}\Box h_{\mu\lambda}
   -E_{(2)}A^{\lambda}A_{\mu}\Box h_{\nu\lambda}
   \nonumber \\
   \!&+&\!
   F_{(2)}\left(A \cdot \partial\right)^{2}A^{\lambda}A_{\nu}h_{\mu\lambda}
   +F_{(2)}\left(A \cdot \partial\right)^{2}A^{\lambda}A_{\mu}h_{\nu\lambda}
   \nonumber \\
   \!&+&\!
   \frac{1}{2}G_{(2)}\left(A \cdot \partial\right)A^{\lambda}\partial_{\nu}h_{\mu\lambda}
   +\frac{1}{2}G_{(2)}\left(A \cdot \partial\right)A^{\lambda}\partial_{\mu}h_{\nu\lambda}
   \nonumber \\
   \!&+&\!
   \frac{1}{2}G_{(2)}\left(A \cdot \partial\right)A_{\nu}\partial^{\lambda}h_{\mu\lambda}
   +\frac{1}{2}G_{(2)}\left(A \cdot \partial\right)A_{\mu}\partial^{\lambda}h_{\nu\lambda}
   \nonumber \\
   \!&-&\!
   2H_{(2)}\partial^{\lambda}\partial_{\nu}h_{\mu\lambda}
   -2H_{(2)}\partial^{\lambda}\partial_{\mu}h_{\nu\lambda}
   +2H_{(2)}\Box h_{\mu\nu}
   \nonumber \\
   \!&+&\!
   2H_{(2)}\partial_{\mu}\partial_{\nu}h
   +2H_{(2)}\eta_{\mu\nu}\partial_{\lambda}\partial_{\rho}h^{\lambda\rho}
   -2H_{(2)}\eta_{\mu\nu}\Box h
   \nonumber \\
   \!&+&\!
   2I_{(2)}\left(A \cdot \partial\right)^{2}h_{\mu\nu}
   -D_{(2)}A_{\mu}A_{\nu}\Box h
   -D_{(2)}\eta_{\mu\nu}A_{\lambda}A_{\rho}\Box h^{\lambda\rho}
   \nonumber \\
   \!&+&\!
   J_{(2)}\left(A \cdot \partial\right)^{2}A_{\mu}A_{\nu}h
   +J_{(2)}\eta_{\mu\nu}\left(A \cdot \partial\right)^{2}A_{\lambda}A_{\rho}h^{\lambda\rho}
   -\frac{1}{2}G_{(2)}\left(A \cdot \partial\right)A_{\mu}\partial_{\nu}h
   \nonumber \\
   \!&-&\!
   \frac{1}{2}G_{(2)}\left(A \cdot \partial\right)A_{\nu}\partial_{\mu}h
   -G_{(2)}\eta_{\mu\nu}\left(A \cdot \partial\right)A_{\lambda}\partial_{\rho}h^{\lambda\rho}
   +2K_{(2)}\eta_{\mu\nu}\left(A \cdot \partial\right)^{2}h
   \nonumber \\
   \!&+&\!
   \left(4A_{(2)}+C_{(2)}\right)A_{\mu}A_{\nu}A^{\lambda}\Box B_{\lambda}
   +4B_{(2)}\left(A \cdot \partial\right)^{2}A_{\mu}A_{\nu}A^{\lambda}B_{\lambda}
   \nonumber \\
   \!&+&\!
   \left(C_{(2)}+2J_{(2)}\right)\left(A \cdot \partial\right)A_{\mu}A_{\nu}\partial_{\lambda}B^{\lambda}
   +\left(C_{(2)}+F_{(2)}\right)\left(A \cdot \partial\right)A_{\mu}\partial_{\nu}A^{\lambda}B_{\lambda}
   \nonumber \\
   \!&+&\!
   \left(C_{(2)}+F_{(2)}\right)\left(A \cdot \partial\right)A_{\nu}\partial_{\mu}A^{\lambda}B_{\lambda}
   +\frac{1}{2}\left(2E_{(2)}-G_{(2)}\right)A_{\mu}\partial_{\nu}\partial_{\lambda}B^{\lambda}
   \nonumber \\
   \!&+&\!
   \frac{1}{2}\left(2E_{(2)}-G_{(2)}\right)A_{\nu}\partial_{\mu}\partial_{\lambda}B^{\lambda}
   +\left(2D_{(2)}+G_{(2)}\right)\partial_{\mu}\partial_{\nu}A^{\lambda}B_{\lambda}
   \nonumber \\
   \!&+&\!
   \frac{1}{2}\left(-2E_{(2)}+G_{(2)}\right)A_{\mu}\Box B_{\nu}
   +\frac{1}{2}\left(-2E_{(2)}+G_{(2)}\right)A_{\nu}\Box B_{\mu}
   \nonumber \\
   \!&+&\!
   F_{(2)}\left(A \cdot \partial\right)^{2}A_{\mu}B_{\nu}
   +F_{(2)}\left(A \cdot \partial\right)^{2}A_{\nu}B_{\mu}
   \nonumber \\
   \!&+&\!
   \frac{1}{2}\left(G_{(2)}+4I_{(2)}\right)\left(A \cdot \partial\right)\partial_{\mu}B_{\nu}
   +\frac{1}{2}\left(G_{(2)}+4I_{(2)}\right)\left(A \cdot \partial\right)\partial_{\nu}B_{\mu}
   \nonumber \\
   \!&+&\!
   \left(-2D_{(2)}-G_{(2)}\right)\eta_{\mu\nu}A^{\lambda}\Box B_{\lambda}
   +2J_{(2)}\eta_{\mu\nu}\left(A \cdot \partial\right)^{2}A^{\lambda}B_{\lambda}
   \nonumber \\
   \!&+&\!
   \left(-G_{(2)}+4K_{(2)}\right)\eta_{\mu\nu}\left(A \cdot \partial\right)\partial_{\lambda}B^{\lambda}
   \nonumber \\
   \!&=&\!
   -\mathcal{M}^{(1)}_{\mu\nu}\left[\eta_{\mu\nu},A^{\mu},h_{\mu\nu},B^{\mu}\right]
   =0.
\end{eqnarray}
And by varying the action with respect to the perturbation $B^{\mu}$, we have
\begin{eqnarray}
	\label{delta S2flat/delta B}
	\frac{\delta S^{(2)}_{flat}}{\delta B^{\mu}}
	\!&=&\!
	4A_{(0)}A_{\mu}A^{\lambda}A^{\rho}h_{\lambda\rho}
	+8A_{(0)}A_{\mu}A_{\lambda}B^{\lambda}
	\nonumber \\
	\!&-&\!
	2B_{(1)}A^{\lambda}A^{\rho}\partial_{\mu}h_{\lambda\rho}
	+2B_{(1)}\left(A \cdot \partial\right)A_{\mu}h
	+2A_{(1)}E^{\lambda~\sigma\gamma}_{~\mu}A_{\sigma}A^{\nu}\partial_{\gamma}h_{\lambda\nu}
	\nonumber \\
	\!&-&\!
	4B_{(1)}A_{\lambda}\partial_{\mu}B^{\lambda}
	+4B_{(1)}A_{\mu}\partial_{\lambda}B^{\lambda}
	-2A_{(1)}E^{\nu~\lambda\rho}_{~\mu}\partial_{\lambda}A_{\rho}B_{\nu}
	\nonumber \\
	\!&+&\!
	\left(4A_{(2)}+C_{(2)}\right)A_{\mu}A^{\lambda}A^{\rho}\Box h_{\lambda\rho}
	+4B_{(2)}\left(A \cdot \partial\right)^{2}A_{\mu}A^{\lambda}A^{\rho}h_{\lambda\rho}
	\nonumber \\
	\!&+&\!
	\left(C_{(2)}+2J_{(2)}\right)\left(A \cdot \partial\right)A^{\lambda}A^{\rho}\partial_{\mu}h_{\lambda\rho}
	+2\left(C_{(2)}+F_{(2)}\right)\left(A \cdot \partial\right)A_{\mu}A^{\lambda}\partial^{\rho}h_{\lambda\rho}
	\nonumber \\
	\!&+&\!
	\left(2E_{(2)}-G_{(2)}\right)A^{\lambda}\partial_{\mu}\partial^{\rho}h_{\lambda\rho}
	+\left(2D_{(2)}+G_{(2)}\right)A_{\mu}\partial^{\lambda}\partial^{\rho}h_{\lambda\rho}
	\nonumber \\
	\!&+&\!
	\left(-2E_{(2)}+G_{(2)}\right)A^{\lambda}\Box h_{\mu\lambda}
	+2F_{(2)}\left(A \cdot \partial\right)^{2}A^{\lambda}h_{\mu\lambda}
	\nonumber \\
	\!&+&\!
	\left(G_{(2)}+4I_{(2)}\right)\left(A \cdot \partial\right)\partial^{\lambda}h_{\mu\lambda}
	+\left(-2D_{(2)}-G_{(2)}\right)A_{\mu}\Box h
	\nonumber \\
	\!&+&\!
	2J_{(2)}\left(A \cdot \partial\right)^{2}A_{\mu}h
	+\left(-G_{(2)}+4K_{(2)}\right)\left(A \cdot \partial\right)\partial_{\mu}h
	\nonumber \\
	\!&+&\!
	2\left(4A_{(2)}+2C_{(2)}+F_{(2)}\right)A_{\mu}A_{\lambda}\Box B^{\lambda}
	+8B_{(2)}\left(A \cdot \partial\right)^{2}A_{\mu}A_{\lambda}B^{\lambda}
	\nonumber \\
	\!&+&\!
	\left(2C_{(2)}+2F_{(2)}+4J_{(2)}\right)\left(A \cdot \partial\right)A_{\lambda}\partial_{\mu}B^{\lambda}
	\nonumber \\
	\!&+&\!
	\left(2C_{(2)}+2F_{(2)}+4J_{(2)}\right)\left(A \cdot \partial\right)A_{\mu}\partial_{\lambda}B^{\lambda}
	\nonumber \\
	\!&+&\!
	2\left(E_{(2)}+2I_{(2)}-G_{(2)}+4K_{(2)}\right)\partial_{\mu}\partial_{\lambda}B^{\lambda}
	\nonumber \\
	\!&+&\!
	2\left(-E_{(2)}+G_{(2)}+2I_{(2)}\right)\Box B_{\mu}
	+2F_{(2)}\left(A \cdot \partial\right)^{2}B_{\mu}
	\nonumber \\
	\!&=&\!
	\mathcal{N}^{(1)}_{\mu}\left[\eta_{\mu\nu},A^{\mu},h_{\mu\nu},B^{\mu}\right]
	=0.
\end{eqnarray}
These two equations are crucial for analyzing the polarization modes of gravitational waves in the next section.

Now, consider the derivation of Eqs. (\ref{vector-tensor field equations metric low feq leading-order}) and (\ref{vector-tensor field equations vector field low feq leading-order}). $\mathcal{M}^{(1)}_{\mu\nu}\left[\eta_{\mu\nu},A^{\mu},\mathfrak{d}\bar{g}_{\mu\nu},\mathfrak{d}\bar{A}^{\mu}\right]$ and $\mathcal{N}^{(1)}_{\mu}\left[\eta_{\mu\nu},A^{\mu},\mathfrak{d}\bar{g}_{\mu\nu},\mathfrak{d}\bar{A}^{\mu}\right]$ in the equations can be directly obtained from the variable substitution in Eqs. (\ref{delta S2flat/delta B}) and (\ref{delta S2flat/delta h}). According to Eqs. (\ref{M2=deltaS flat/delta g 0}) and (\ref{N2=deltaS flat/delta A 0}), the averaged terms in Eqs. (\ref{vector-tensor field equations metric low feq leading-order}) and (\ref{vector-tensor field equations vector field low feq leading-order}) need to be obtained by varying the action with respect to the background fields. Therefore, it is necessary to clearly state the form of $S^{(2)}_{flat}$ explicitly containing $\eta_{\mu\nu}$. At this point, it should be noted that in curved spacetime, the definition of the four-dimensional totally antisymmetric tensor is \cite{landau}
\begin{eqnarray}
	\label{four-dimensional totally antisymmetric tensor in curved spacetime}
    \mathcal{E}^{\mu\nu\lambda\rho}
    \coloneqq 
    \frac{1}{\sqrt{-g}}E^{\mu\nu\lambda\rho}.
\end{eqnarray}
Therefore, when writing $S^{(2)}_{flat}$ that explicitly includes $\eta_{\mu\nu}$, in addition to inserting $\sqrt{-\eta}$ before $d^{4} x$, all occurrences of $E^{\mu\nu\lambda\rho}$ should be replaced with $E^{\mu\nu\lambda\rho}/\sqrt{-\eta}$.

Varying the action with respect to the background fields is straightforward but tedious. Here, we provide only the expression for the effective energy-momentum tensor of gravitational waves as an example and omit the variation of the background vector field. Using Eq. (\ref{t_munu S2flat 0}), the effective energy-momentum tensor for gravitational waves is expressed as
\begin{eqnarray}
	\label{t_munu S2flat in the most general vector-tensor theory}
	t_{\mu\nu}
	\!&=&\! 
	\bigg\langle
   4A_{(0)}A^{\lambda}A^{\rho}A_{\mu}h_{\lambda\rho}B_{\nu}
   +4A_{(0)}A^{\lambda}A^{\rho}A_{\nu}h_{\lambda\rho}B_{\mu}
   +8A_{(0)}A_{\nu}A^{\lambda}B_{\mu}B_{\lambda}
   +8A_{(0)}A_{\mu}A^{\lambda}B_{\nu}B_{\lambda}
   \nonumber \\
   \!&+&\!
   A_{(1)}\left(E^{\sigma\lambda~\gamma}_{~~~\mu}A^{\omega}A^{\rho}A_{\nu}\partial_{\gamma}h_{\sigma\omega}\right)h_{\lambda\rho}
   +A_{(1)}\left(E^{\sigma\lambda~\gamma}_{~~~\nu}A^{\omega}A^{\rho}A_{\mu}\partial_{\gamma}h_{\sigma\omega}\right)h_{\lambda\rho}
   \nonumber \\
   \!&-&\!
   2B_{(1)}\left(\left(A \cdot \partial\right)A^{\lambda}A^{\rho}h_{\mu\nu}\right)h_{\lambda\rho}
   \nonumber \\
   \!&+&\!
   2A_{(1)}\left(E^{\sigma\lambda~\gamma}_{~~~\mu}A_{\nu}\partial_{\gamma}A^{\rho}h_{\sigma\rho}\right)B_{\lambda}
   +2A_{(1)}\left(E^{\sigma\lambda~\gamma}_{~~~\nu}A_{\mu}\partial_{\gamma}A^{\rho}h_{\sigma\rho}\right)B_{\lambda}
   \nonumber \\
   \!&+&\!
   2A_{(1)}\left(E^{\lambda~\sigma\gamma}_{~\mu}A_{\sigma}\partial_{\gamma}A^{\rho}h_{\lambda\rho}\right)B_{\nu}
   +2A_{(1)}\left(E^{\lambda~\sigma\gamma}_{~\nu}A_{\sigma}\partial_{\gamma}A^{\rho}h_{\lambda\rho}\right)B_{\mu}
   \nonumber \\
   \!&-&\!
   4B_{(1)}\left(\left(A \cdot \partial\right)A_{\lambda}h_{\mu\nu}\right)B^{\lambda}
   +2B_{(1)}\left(\left(A \cdot \partial\right)A_{\mu}h\right)B_{\nu}
   +2B_{(1)}\left(\left(A \cdot \partial\right)A_{\nu}h\right)B_{\mu}
   \nonumber \\
   \!&-&\!
   4B_{(1)}\left(A_{\nu}\partial_{\lambda}B_{\mu}\right)B^{\lambda}
   -4B_{(1)}\left(A_{\mu}\partial_{\lambda}B_{\nu}\right)B^{\lambda}
   \nonumber \\
   \!&-&\!
   A_{(1)}\left(E^{\rho\gamma\lambda}_{~~~~\mu}\partial_{\lambda}A_{\nu}B_{\rho}\right)B_{\gamma}
   -A_{(1)}\left(E^{\rho\gamma\lambda}_{~~~~\nu}\partial_{\lambda}A_{\mu}B_{\rho}\right)B_{\gamma}
   \nonumber \\
   \!&-&\!
   A_{(1)}\left(E_{\mu}^{~\gamma\lambda\rho}\partial_{\lambda}A_{\rho}B_{\nu}\right)B_{\gamma}
   -A_{(1)}\left(E_{\nu}^{~\gamma\lambda\rho}\partial_{\lambda}A_{\rho}B_{\mu}\right)B_{\gamma}
   \nonumber \\
   \!&-&\!
   A_{(1)}\left(E^{\gamma~\lambda\rho}_{~\nu}\partial_{\lambda}A_{\rho}B_{\gamma}\right)B_{\mu}
   -A_{(1)}\left(E^{\gamma~\lambda\rho}_{~\mu}\partial_{\lambda}A_{\rho}B_{\gamma}\right)B_{\nu}
   \nonumber \\
   \!&-&\!
   2A_{(2)}\left(\partial_{\mu}\partial_{\nu}A^{\lambda}A^{\rho}A^{\sigma}A^{\gamma}h_{\sigma\gamma}\right)h_{\lambda\rho}
   \nonumber \\
   \!&-&\!
   C_{(2)}\left(\left(A \cdot \partial\right)A^{\rho}A^{\gamma}A^{\lambda}\partial_{\mu}h_{\rho\gamma}\right)h_{\lambda\nu}
   -C_{(2)}\left(\left(A \cdot \partial\right)A^{\rho}A^{\gamma}A^{\lambda}\partial_{\nu}h_{\rho\gamma}\right)h_{\lambda\mu}
   \nonumber \\
   \!&-&\!
   2D_{(2)}\left(A^{\lambda}A^{\rho}\partial_{\mu}\partial_{\gamma}h_{\lambda\rho}\right)h_{\nu}^{~\gamma}
   -2D_{(2)}\left(A^{\lambda}A^{\rho}\partial_{\nu}\partial_{\gamma}h_{\lambda\rho}\right)h_{\mu}^{~\gamma}
   \nonumber \\
   \!&+&\!
   2E_{(2)}\left(\partial_{\mu}\partial_{\nu}A^{\lambda}A^{\rho}h^{\gamma}_{~\lambda}\right)h_{\gamma\rho}
   -E_{(2)}\left(A^{\sigma}A^{\lambda}A_{\mu}\partial_{\gamma}h_{\sigma\nu}\right)h_{\lambda}^{~\gamma}
   -E_{(2)}\left(A^{\sigma}A^{\lambda}A_{\nu}\partial_{\gamma}h_{\sigma\mu}\right)h_{\lambda}^{~\gamma}
   \nonumber \\
   \!&-&\!
   E_{(2)}\left(A^{\sigma}A^{\lambda}\partial_{\gamma}\partial_{\mu}h_{\sigma}^{~\gamma}\right)h_{\lambda\nu}
   -E_{(2)}\left(A^{\sigma}A^{\lambda}\partial_{\gamma}\partial_{\nu}h_{\sigma}^{~\gamma}\right)h_{\lambda\mu}
   +2E_{(2)}\left(\Box A^{\lambda}A^{\rho}h_{\mu\lambda}\right)h_{\nu\rho}
   \nonumber \\
   \!&-&\!
   2F_{(2)}\left(\left(A \cdot \partial\right)^{2}A^{\lambda}A^{\rho}h_{\mu\lambda}\right)h_{\nu\rho}
   \nonumber \\
   \!&-&\!
   G_{(2)}\left(\left(A \cdot \partial\right)A^{\lambda}\partial^{\rho}h_{\mu\lambda}\right)h_{\nu\rho}
   -G_{(2)}\left(\left(A \cdot \partial\right)A^{\lambda}\partial^{\rho}h_{\nu\lambda}\right)h_{\mu\rho}
   \nonumber \\
   \!&-&\!
   G_{(2)}\left(\left(A \cdot \partial\right)A^{\lambda}\partial_{\mu}h^{\rho}_{~\lambda}\right)h_{\rho\nu}
   -G_{(2)}\left(\left(A \cdot \partial\right)A^{\lambda}\partial_{\nu}h^{\rho}_{~\lambda}\right)h_{\rho\mu}
   \nonumber \\
   \!&+&\!
   2H_{(2)}\left(\partial_{\mu}\partial_{\rho}h^{\lambda}_{~\nu}\right)h_{\lambda}^{~\rho}
   +2H_{(2)}\left(\partial_{\nu}\partial_{\rho}h^{\lambda}_{~\mu}\right)h_{\lambda}^{~\rho}
   \nonumber \\
   \!&+&\!
   2H_{(2)}\left(\partial^{\lambda}\partial^{\rho}h_{\mu\lambda}\right)h_{\nu\rho}
   +2H_{(2)}\left(\partial^{\lambda}\partial^{\rho}h_{\nu\lambda}\right)h_{\mu\rho}
   \nonumber \\
   \!&+&\!
   2H_{(2)}\left(\partial^{\lambda}\partial_{\mu}h^{\rho}_{~\lambda}\right)h_{\rho\nu}
   +2H_{(2)}\left(\partial^{\lambda}\partial_{\nu}h^{\rho}_{~\lambda}\right)h_{\rho\mu}
   -2H_{(2)}\left(\partial_{\mu}\partial_{\nu}h^{\lambda\rho}\right)h_{\lambda\rho}
   \nonumber \\
   \!&+&\!
   2H_{(2)}\left(\partial_{\mu}\partial_{\nu}h\right)h
   -4H_{(2)}\left(\Box h_{\mu}^{~\lambda}\right)h_{\nu\lambda}
   -4H_{(2)}\left(\partial_{\lambda}\partial_{\rho}h_{\mu\nu}\right)h^{\lambda\rho}
   \nonumber \\
   \!&-&\!
   4H_{(2)}\left(\partial_{\mu}\partial^{\lambda}h\right)h_{\nu\lambda}
   -4H_{(2)}\left(\partial_{\nu}\partial^{\lambda}h\right)h_{\mu\lambda}
   +4H_{(2)}\left(\Box h\right)h_{\mu\nu}
   \nonumber \\
   \!&-&\!
   2I_{(2)}\left(\left(A \cdot \partial\right)^{2}h_{\mu}^{~\lambda}\right)h_{\nu\lambda}
   -2I_{(2)}\left(\left(A \cdot \partial\right)^{2}h_{\nu}^{~\lambda}\right)h_{\mu\lambda}
   \nonumber \\
   \!&+&\!
   2D_{(2)}\left(\Box A^{\lambda}A^{\rho}h_{\mu\nu}\right)h_{\lambda\rho}
   +2D_{(2)}\left(\partial_{\mu}\partial_{\nu}A^{\lambda}A^{\rho}h\right)h_{\lambda\rho}
   \nonumber \\
   \!&-&\!
   2J_{(2)}\left(\left(A \cdot \partial\right)^{2}A^{\lambda}A^{\rho}h_{\mu\nu}\right)h_{\lambda\rho}
   +2G_{(2)}\left(\left(A \cdot \partial\right)A^{\lambda}A^{\rho}h_{\mu\nu}\right)h_{\lambda\rho}
   \nonumber \\
   \!&+&\!
   G_{(2)}\left(\left(A \cdot \partial\right)A^{\lambda}\partial_{\mu}h\right)h_{\lambda\nu}
   +G_{(2)}\left(\left(A \cdot \partial\right)A^{\lambda}\partial_{\nu}h\right)h_{\lambda\mu}
   -4K_{(2)}\left(\left(A \cdot \partial\right)^{2}h\right)h_{\mu\nu}
   \nonumber \\
   \!&+&\!
   \left(4A_{(2)}+C_{(2)}\right)\left(\Box A^{\lambda}A^{\rho}A_{\mu}h_{\lambda\rho}\right)B_{\nu}
   +\left(4A_{(2)}+C_{(2)}\right)\left(\Box A^{\lambda}A^{\rho}A_{\nu}h_{\lambda\rho}\right)B_{\mu}
   \nonumber \\
   \!&-&\!
   2\left(4A_{(2)}+C_{(2)}\right)\left(\partial_{\mu}\partial_{\nu}A^{\lambda}A^{\rho}A_{\gamma}h_{\lambda\rho}\right)B^{\gamma}
   \nonumber \\
   \!&+&\!
   4B_{(2)}\left(\left(A \cdot \partial\right)^{2}A^{\lambda}A^{\rho}A_{\mu}h_{\lambda\rho}\right)B_{\nu}
   +4B_{(2)}\left(\left(A \cdot \partial\right)^{2}A^{\lambda}A^{\rho}A_{\nu}h_{\lambda\rho}\right)B_{\mu}
   \nonumber \\
   \!&-&\!
   2\left(C_{(2)}+F_{(2)}\right)\left(\left(A \cdot \partial\right)A^{\lambda}\partial_{\mu}A_{\rho}h_{\lambda\nu}\right)B^{\rho}
   -2\left(C_{(2)}+F_{(2)}\right)\left(\left(A \cdot \partial\right)A^{\lambda}\partial_{\nu}A_{\rho}h_{\lambda\mu}\right)B^{\rho}
   \nonumber \\
   \!&+&\!
   2\left(C_{(2)}+F_{(2)}\right)\left(\left(A \cdot \partial\right)A^{\lambda}\partial^{\rho}A_{\mu}h_{\lambda\rho}\right)B_{\nu}
   +2\left(C_{(2)}+F_{(2)}\right)\left(\left(A \cdot \partial\right)A^{\lambda}\partial^{\rho}A_{\nu}h_{\lambda\rho}\right)B_{\mu}
   \nonumber \\
   \!&-&\!
   \left(2E_{(2)}-G_{(2)}\right)\left(A^{\lambda}\partial_{\mu}\partial_{\rho}h_{\lambda\nu}\right)B^{\rho}
   -\left(2E_{(2)}-G_{(2)}\right)\left(A^{\lambda}\partial_{\nu}\partial_{\rho}h_{\lambda\mu}\right)B^{\rho}
   \nonumber \\
   \!&-&\!
   2\left(-2E_{(2)}+G_{(2)}\right)\left(\partial_{\mu}\partial_{\nu}A^{\rho}h_{\rho\lambda}\right)B^{\lambda}
   \nonumber \\
   \!&-&\!
   2\left(2D_{(2)}+G_{(2)}\right)\left(\partial_{\nu}\partial_{\lambda}A_{\rho}h_{\mu}^{~\lambda}\right)B^{\rho}
   -2\left(2D_{(2)}+G_{(2)}\right)\left(\partial_{\mu}\partial_{\lambda}A_{\rho}h_{\nu}^{~\lambda}\right)B^{\rho}
   \nonumber \\
   \!&+&\!
   \left(2D_{(2)}+G_{(2)}\right)\left(\partial_{\lambda}\partial_{\rho}A_{\mu}h^{\lambda\rho}\right)B_{\nu}
   +\left(2D_{(2)}+G_{(2)}\right)\left(\partial_{\lambda}\partial_{\rho}A_{\nu}h^{\lambda\rho}\right)B_{\mu}
   \nonumber \\
   \!&-&\!
   \left(G_{(2)}+4I_{(2)}\right)\left(\left(A \cdot \partial\right)\partial_{\mu}h_{\nu\lambda}\right)B^{\lambda}
   -\left(G_{(2)}+4I_{(2)}\right)\left(\left(A \cdot \partial\right)\partial_{\nu}h_{\mu\lambda}\right)B^{\lambda}
   \nonumber \\
   \!&+&\!
   2\left(2D_{(2)}+G_{(2)}\right)\left(\partial_{\mu}\partial_{\nu}A_{\lambda}h\right)B^{\lambda}
   +2\left(2D_{(2)}+G_{(2)}\right)\left(\Box A_{\lambda}h_{\mu\nu}\right)B^{\lambda}
   \nonumber \\
   \!&+&\!
   \left(-2D_{(2)}-G_{(2)}\right)\left(\Box A_{\mu}h\right)B_{\nu}
   +\left(-2D_{(2)}-G_{(2)}\right)\left(\Box A_{\nu}h\right)B_{\mu}
   \nonumber \\
   \!&-&\!
   4J_{(2)}\left(\left(A \cdot \partial\right)^{2}A_{\lambda}h_{\mu\nu}\right)B^{\lambda}
   +2J_{(2)}\left(\left(A \cdot \partial\right)^{2}A_{\mu}h\right)B_{\nu}
   +2J_{(2)}\left(\left(A \cdot \partial\right)^{2}A_{\nu}h\right)B_{\mu}
   \nonumber \\
   \!&-&\!
   2\left(-G_{(2)}+4K_{(2)}\right)\left(\left(A \cdot \partial\right)\partial_{\lambda}h_{\mu\nu}\right)B^{\lambda}
   +2\left(4A_{(2)}+2C_{(2)}+F_{(2)}\right)\left(\Box A_{\mu}A^{\lambda}B_{\nu}\right)B_{\lambda}
   \nonumber \\
   \!&+&\!
   2\left(4A_{(2)}+2C_{(2)}+F_{(2)}\right)\left(\Box A_{\nu}A^{\lambda}B_{\mu}\right)B_{\lambda}
   \nonumber \\
   \!&-&\!
   2\left(4A_{(2)}+2C_{(2)}+F_{(2)}\right)\left(\partial_{\mu}\partial_{\nu}A^{\lambda}A^{\rho}B_{\lambda}\right)B_{\rho}
   \nonumber \\
   \!&+&\!
   8B_{(2)}\left(\left(A \cdot \partial\right)^{2}A_{\mu}A^{\lambda}B_{\nu}\right)B_{\lambda}
   +8B_{(2)}\left(\left(A \cdot \partial\right)^{2}A_{\nu}A^{\lambda}B_{\mu}\right)B_{\lambda}
   \nonumber \\
   \!&+&\!
   \left(2C_{(2)}+2F_{(2)}+4J_{(2)}\right)\left(\left(A \cdot \partial\right)A_{\mu}\partial_{\lambda}B_{\nu}\right)B^{\lambda}
   \nonumber \\
   \!&+&\!
   \left(2C_{(2)}+2F_{(2)}+4J_{(2)}\right)\left(\left(A \cdot \partial\right)A_{\nu}\partial_{\lambda}B_{\mu}\right)B^{\lambda}
   \nonumber \\
   \!&+&\!
   2\left(-E_{(2)}+G_{(2)}+2I_{(2)}\right)\left(\Box B_{\mu}\right)B_{\nu}
   -2\left(-E_{(2)}+G_{(2)}+2I_{(2)}\right)\left(\partial_{\mu}\partial_{\nu}B_{\lambda}\right)B^{\lambda}
   \nonumber \\
   \!&+&\!
   2F_{(2)}\left(\left(A \cdot \partial\right)^{2}B_{\mu}\right)B_{\nu}
   -2\left(
   {\mathcal{L}}'_{0}+{\mathcal{L}}'_{1}+{\mathcal{L}}'_{2}
   \right)
   \nonumber \\
   \!&+&\!
   \eta_{\mu\nu}
   \left(
   \tilde{\mathcal{L}}_{0}+\tilde{\mathcal{L}}_{1}+\tilde{\mathcal{L}}_{2}
   \right)
   \bigg\rangle.
\end{eqnarray}
Here, $\tilde{\mathcal{L}}_{0},~\tilde{\mathcal{L}}_{1}$, and $\tilde{\mathcal{L}}_{2}$ in the last line of Eq. (\ref{t_munu S2flat in the most general vector-tensor theory}) are the sums of the remaining terms in ${\mathcal{L}}_{0},~{\mathcal{L}}_{1}$, and ${\mathcal{L}}_{2}$ after removing all terms containing $E^{\mu\nu\lambda\rho}$, respectively. Since we now consider the background fields as formalized variables rather than known constant fields, the parameters $A_{(0)}$, $A_{(1)}$, $B_{(1)}$ etc., in action (\ref{the most general second-order perturbation action in vector-tensor theory}) are actually functions of $\eta_{\mu\nu}A^{\mu}A^{\nu}$. Therefore, when formalizing the variation with respect to $\eta^{\mu\nu}$, it is necessary to account for the derivative terms of parameters such as $\delta A_{(0)}/\delta \eta^{\mu\nu}$. All such terms in $\mathcal{L}_{0}$, $\mathcal{L}_{1}$, and $\mathcal{L}_{2}$ are represented by $\mathcal{L}'_{0}$, $\mathcal{L}'_{1}$ and $\mathcal{L}'_{2}$ in Eq. (\ref{t_munu S2flat in the most general vector-tensor theory}), respectively. {It can be seen that such a term is proportional to $A_{\mu}A_{\nu}$.}

{The effective energy-momentum tensor of gravitational waves is, in principle, gauge-invariant. In the Isaacson picture, once the perturbations to the background are decomposed into low-frequency and high-frequency components, the physical effects of gravitational waves are governed by Eqs. (\ref{vector-tensor field equations metric high feq leading-order})–(\ref{vector-tensor field equations vector field low feq leading-order}). Under the gauge transformation, after decomposing $\xi^{\mu}$ into its low-frequency component $\xi_{L}^{\mu}$ and high-frequency component $\xi_{H}^{\mu}$, we can see that, neglecting higher-order small quantities, the corresponding perturbations transform as follows:}
\begin{eqnarray}
	\label{gauge transformation low and high}
	h_{\mu\nu} \!&\rightarrow&\! h_{\mu\nu}-\partial_{\mu}\xi_{H\nu}-\partial_{\nu}\xi_{H\mu}, ~~
	B^{\mu} \!\rightarrow\!
	B^{\mu}+A^{\nu}\partial_{\nu}\xi_{H}^{\mu},\\
	\mathfrak{d} g_{\mu\nu} \!&\rightarrow&\! \mathfrak{d} g_{\mu\nu}-\partial_{\mu}\xi_{L\nu}-\partial_{\nu}\xi_{L\mu}, ~~
	\mathfrak{d} \bar{A}^{\mu} \!\rightarrow\!
	\mathfrak{d} \bar{A}^{\mu}+A^{\nu}\partial_{\nu}\xi_{L}^{\mu}.
\end{eqnarray} 
{Therefore, in Eq. (\ref{vector-tensor field equations metric low feq leading-order}), $\mathcal{M}^{(1)}_{\mu\nu}\left[\eta_{\mu\nu},A^{\mu},\mathfrak{d}\bar{g}_{\mu\nu},\mathfrak{d}\bar{A}^{\mu}\right]$ is gauge invariant, which further indicates that $t_{\mu\nu}$ should be gauge invariant. A more formally rigorous proof of the gauge invariance of $t_{\mu\nu}$ can be derived from the general covariance of the field equations. General covariance ensures that the second-order perturbed field equations are gauge invariant up to second order. To demonstrate this, we consider the gauge transformation of the second-order perturbed field equations. By decomposing both the background perturbations and $\xi^{\mu}$ into high-frequency and low-frequency parts,  extracting the low-frequency part of the expression, and invoking Eq. (\ref{assumption 0}), it can be shown that the variation of $t_{\mu\nu}$ under the gauge transformation is of higher order and thus negligible.}%It should be noted that here we use all the terms in $\mathcal{M}^{(1)}_{\mu\nu}\left[\eta_{\mu\nu},A^{\mu},\mathfrak{d}\bar{g}_{\mu\nu},\mathfrak{d}\bar{A}^{\mu}\right]$, which requires an implicit assumption that all terms in $\mathcal{M}^{(1)}_{\mu\nu}\left[\eta_{\mu\nu},A^{\mu},\mathfrak{d}\bar{g}_{\mu\nu},\mathfrak{d}\bar{A}^{\mu}\right]$ are of the same order of magnitude. This ensures that, in the high-frequency equations (\ref{delta S2flat/delta h}) and (\ref{delta S2flat/delta B}), the terms with fewer derivative operators are small compared to those with more derivative operators. This assumption is reasonably justified, as it implies that the mass of the graviton should be sufficiently small. A rigorous proof of the gauge invariance of $t_{\mu\nu}$ can, in principle, be derived from the general covariance of the field equations, which implies that the second-order perturbation field equations possess gauge invariance up to second order. By writing down the expression for the gauge transformation of the second-order perturbation field equations, decomposing both the background perturbations and $\xi^{\mu}$ into high-frequency and low-frequency parts, taking the low-frequency part of the expression, and using Eq. (\ref{assumption 0}) and the implicit assumption we just mentioned to neglect all small quantities, the gauge invariance of $t_{\mu\nu}$ can be proven. The same assumption also allows us to neglect all non-second-derivative terms in Eq. (\ref{t_munu S2flat in the most general vector-tensor theory}).}

\section{Polarization modes of gravitational waves in the most general vector-tensor theory}
\label{sec: 6}
In this section, we use Eqs. (\ref{delta S2flat/delta h}) and (\ref{delta S2flat/delta B}) to analyze the polarization modes and the speed of gravitational waves in the most general vector-tensor theory.

Due to the equivalence principle, there is no fundamental difference between the motion of a single free particle and the motion of a free-falling body. Therefore, it is not possible to detect the presence of gravitational waves using a single test particle. To detect gravitational waves, it is necessary to measure the change of the relative position between two test particles. 

For the above considerations, the polarization modes of gravitational waves are defined by the different relative motion modes between two test particles. Assumption (5) in Sec. \ref{sec: 4} requires minimal coupling between free particles and the metric, allowing the relative motion of two free test particles in asymptotic Minkowski spacetime far from the source to be described by the geodesic deviation equation \cite{MTW}:
\begin{eqnarray}
	\label{equation of geodesic deviation}
	\frac{d^{2}\eta_{i}}{dt^{2}}=-R^{(1)}_{i0j0}\eta^{j}.
\end{eqnarray}
Here, $\eta_{i}$ represents the relative displacement of the two test particles. % and $R^{(1)}_{i0j0}$ denotes the linear order of the $i0j0$ component of the Riemannian tensor. 
From Eq. (\ref{equation of geodesic deviation}), we observe that knowing $R^{(1)}_{i0j0}$ allows us to determine the relative motion of the two test particles. Hence, the polarization modes of gravitational waves can be completely defined by the linear order of the $i0j0$ component of the Riemannian tensor $R^{(1)}_{i0j0}$ \cite{Eardley}.

\begin{figure*}[htbp]
	\makebox[\textwidth][c]{\includegraphics[width=1.2\textwidth]{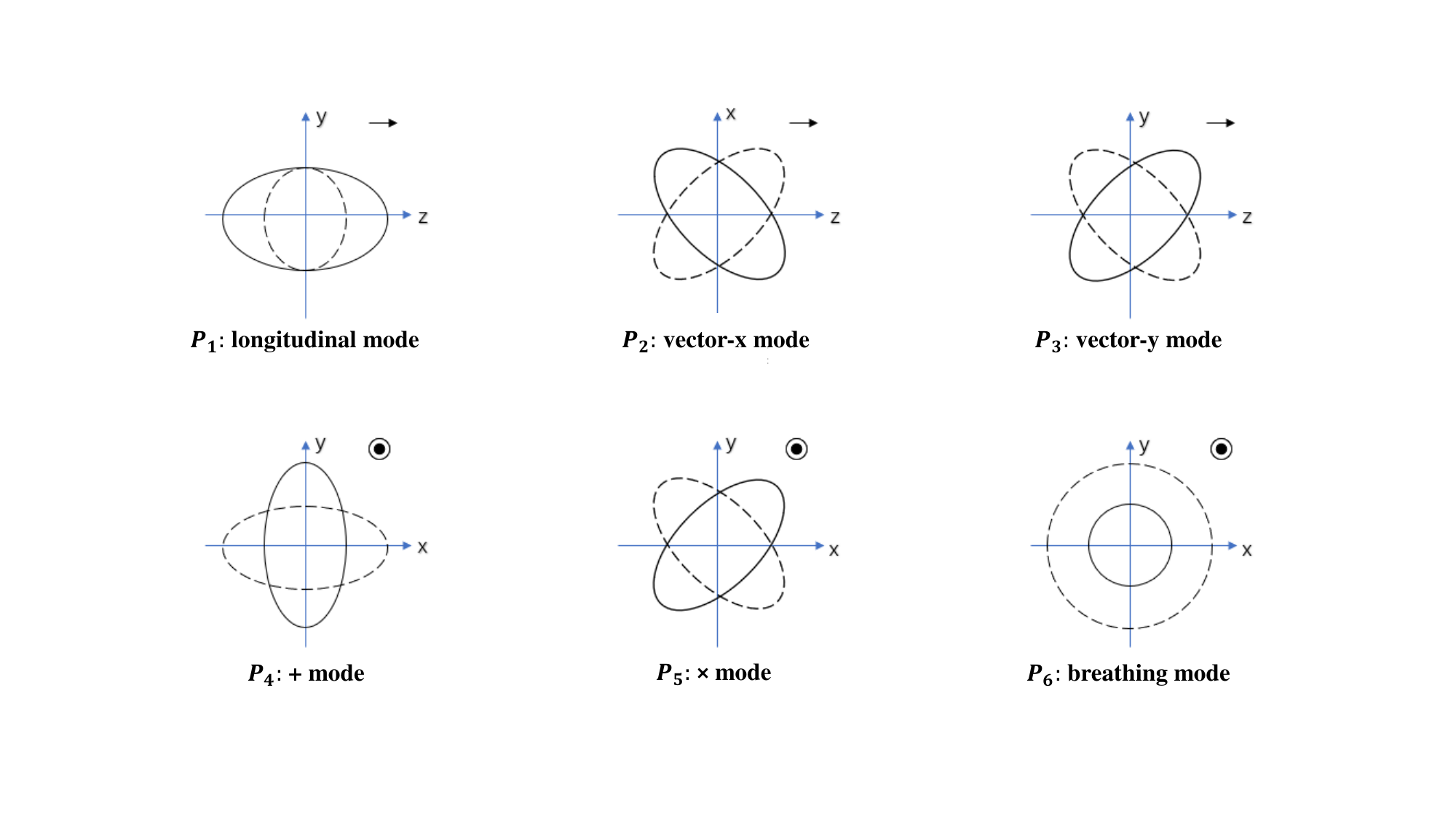}}
	\caption{Six polarization modes of gravitational waves \cite{Eardley}. The gravitational wave propagates along the $+z$ direction. The test particles move periodically only within the two-dimensional plane shown in the figure. The solid and dashed lines correspond to states with a phase difference of $\pi$.}%The solid and dotted lines are the cases of a circle of test particles when the phases of the waves are $0$ and $\pi$, respectively. There is no relative motion between test particles in the direction not shown by the third axis.}
	\label{fig: 1}
\end{figure*}

Specifically, we can consider monochromatic plane gravitational waves propagating along the $+z$ direction in a Minkowski background without loss of generality. (Since the propagation equations (\ref{delta S2flat/delta h}) and (\ref{delta S2flat/delta B}) of gravitational waves are linear, it is generally possible to express gravitational waves as a superposition of monochromatic plane wave solutions via the Fourier transform.) In such a situation, $R^{(1)}_{i0j0}$ takes the form of a monochromatic plane wave:
\begin{eqnarray}
	\label{Ri0j0=AEeikx}
	R^{(1)}_{i0j0}=\mathbb{A} E_{ij} e^{ikx}.
\end{eqnarray}
Here, $k^{\mu}$ is a four-wavevector, $\mathbb{A}$ represents the intensity of the wave, and $E_{ij}$ contains all polarization information and satisfies 
\begin{eqnarray}
	\label{EE=1}
	E_{ij}E^{ij}=1.
\end{eqnarray}
Due to the fact that $E_{ij}$ is a symmetric $4\times4$ matrix, it has at most six independent components in four-dimensional spacetime. Therefore, gravitational waves in four-dimensional spacetime can only have up to six independent polarization modes: $P_1, ..., P_6$. Their definition is as follows \cite{Eardley}: 
\begin{eqnarray}
	\label{P1-P6}
	{R}^{(1)}_{i0j0}=\begin{pmatrix}
		P_{4}+P_{6} & P_{5} & P_{2}\\
		P_{5}       & -P_{4}+P_{6}  & P_{3}\\
		P_{2}       &  P_{3}   &   P_{1}
	\end{pmatrix}.
\end{eqnarray}
The polarization mode of any gravitational wave can be represented as a linear combination of these six modes. We illustrate these six polarization modes of gravitational waves in Fig. \ref{fig: 1}.

The gauge invariant method \cite{James. Bardeen,R. Jackiw,Eanna E Flanagan} can help us analyze Eqs. (\ref{delta S2flat/delta h}) and (\ref{delta S2flat/delta B}) more easily. We have detailed in our previous paper \cite{Y.Dong3} how to use this method to analyze the polarization modes of gravitational waves. Due to the theory being generally covariant, the left-hand sides of Eqs. (\ref{delta S2flat/delta h}) and (\ref{delta S2flat/delta B}) are gauge invariant. Therefore, we can aim to combine the perturbations into some gauge-invariant variables and rephrase the field equations using these gauge-invariant variables. Then, the analysis of the polarization modes of gravitational waves boils down to solving for these gauge-invariant variables. This method can eliminate redundant gauge degrees of freedom.

To identify possible gauge invariant variables, we first uniquely decompose the perturbations as follows:
\begin{eqnarray}
	\label{decompose perturbations}
	B^{0}&=&B^{0},\nonumber \\
	B^{i}&=&\partial^{i}\omega+\mu^{i},\nonumber \\
	h_{00}&=&h_{00}, \\
	h_{0i}&=&\partial_{i}\gamma+\beta_{i},\nonumber\\
	h_{ij}&=&h^{TT}_{ij}+\partial_{i}\epsilon_{j}+\partial_{j}\epsilon_{i}
	+\frac{1}{3}\delta_{ij}H+(\partial_{i}\partial_{j}-\frac{1}{3}\delta_{ij}\Delta)\zeta. \nonumber
\end{eqnarray}
Here,
\begin{eqnarray}
	&\partial_{i}\mu^{i}=\partial_{i}\beta^{i}=\partial_{i}\epsilon^{i}=0,  \\
	&\delta^{ij}h^{TT}_{ij}=\partial^{i}h^{TT}_{ij}=0.
\end{eqnarray}
This decomposition uniquely separates a spatial vector into a spatial scalar component and a transverse spatial vector component. Similarly, it uniquely decomposes a spatial tensor into two spatial scalar components, a transverse spatial vector component, and a transverse traceless spatial tensor component.

Then, we recombine these quantities to obtain the following gauge invariant transverse traceless spatial tensor, transverse spatial vectors, and spatial scalars:
\begin{eqnarray}
	\label{gauge invariant}
	h^{TT}_{ij}\!&=&\!h^{TT}_{ij},
	\nonumber \\
	\Xi_{i}\!&=&\!\beta_{i}-\partial_{0}\epsilon_{i}, 
	\nonumber \\
	\Sigma_{i}\!&=&\!\mu_{i}+A\partial_{0}\epsilon_{i},
    \nonumber \\
	\phi \,\!&=&\! -\frac{1}{2}h_{00}+\partial_{0}\gamma-\frac{1}{2}\partial_{0}\partial_{0}\zeta,  
	\\
	\Theta \!&=&\! \frac{1}{3}\left(H-\Delta\zeta\right),  
	\nonumber \\
	\Omega\!&=&\! B^{0}-A\partial_{0}\gamma+\frac{1}{2}A\partial_{0}\partial_{0}\zeta,  
	\nonumber \\
	\Psi \!&=&\! \omega+\frac{1}{2}A\partial_{0}\zeta. \nonumber
\end{eqnarray}
Equations (\ref{delta S2flat/delta h}) and (\ref{delta S2flat/delta B}) can be rephrased using these variables.

Since $R^{(1)}_{i0j0}$ is gauge invariant, it can also be represented by gauge invariants:
\begin{eqnarray}
	\label{Ri0j0 gauge invariant}
	R^{(1)}_{i0j0}=-\frac{1}{2}\partial_{0}\partial_{0}h^{TT}_{ij}
	+\frac{1}{2}\partial_{0}\partial_{i}\Xi_{j}
	+\frac{1}{2}\partial_{0}\partial_{j}\Xi_{i}
	+\partial_{i}\partial_{j}\phi
	-\frac{1}{2}\delta_{ij}\partial_{0}\partial_{0}\Theta.
\end{eqnarray}
Therefore, using Eq. (\ref{P1-P6}), the six polarization modes of gravitational waves satisfy the following relationship with gauge invariants:
\begin{eqnarray}
	\label{P1-P6 gauge invariant}
		P_{1}\!&=&\!\partial_{3}\partial_{3}\phi-\frac{1}{2}\partial_{0}\partial_{0}\Theta, \quad
		P_{2}=\frac{1}{2}\partial_{0}\partial_{3}\Xi_{1}, \nonumber \\
		P_{3}\!&=&\!\frac{1}{2}\partial_{0}\partial_{3}\Xi_{2},  \quad \quad\quad\quad\,\,
		P_{4}=-\frac{1}{2}\partial_{0}\partial_{0}h^{TT}_{11}, \\
		P_{5}\!&=&\!-\frac{1}{2}\partial_{0}\partial_{0}h^{TT}_{12}, \quad\quad~~~
		P_{6}=-\frac{1}{2}\partial_{0}\partial_{0}\Theta. \nonumber 
\end{eqnarray}
It can be seen that not all gauge invariants in Eq. (\ref{gauge invariant}) contribute to the polarization modes of gravitational waves. The $+$ mode $P_{4}$ and the $\times$ mode $P_{5}$ are only related to tensor $h^{TT}_{ij}$; therefore, they are called tensor modes. Similarly, the vector-$x$ mode $P_{2}$ and the vector-$y$ mode $P_{3}$ are referred to as vector modes, and the longitudinal mode $P_{1}$ and the breathing mode $P_{6}$ are referred to as scalar modes.

The same principle used for the unique decomposition in Eq. (\ref{decompose perturbations}) can be applied to decompose the left-hand side of Eqs. (\ref{delta S2flat/delta h}) and (\ref{delta S2flat/delta B}) into tensor, vector, and scalar parts. Further decompose these equations into tensor, vector, and scalar equations. Due to $SO(3)$ symmetry of the Minkowski background, the decomposed tensor equation depends only on gauge invariant tensor \cite{Weinberg 2}. Similarly, vector and scalar equations depend solely on gauge invariant vectors and scalars, respectively. This achieves the decoupling of the equations, enabling the solution of the equations class by class.

Now, we can analyze tensor, vector, and scalar equations to determine the properties of tensor, vector, and scalar modes of gravitational waves and their corresponding wave speeds, respectively. In the following text, we consider the case of gravitational waves propagating along the $+z$ direction without loss of generality.

\subsection{Tensor modes}
\label{tensor modes}
We first analyze the tensor modes of gravitational waves. The tensor equation describing the tensor modes is given by the transverse traceless part of the $ij$ component in Eq. (\ref{delta S2flat/delta h}):
\begin{eqnarray}
 	\label{tensor mode equation}
 	\left[
 	-\left(H_{(2)}-I_{(2)}A^{2}\right)\partial_{0}^{2}+H_{(2)}\Delta
 	\right]
 	h^{TT}_{ij}
 	=0.
\end{eqnarray}
We study its monochromatic plane wave solution
\begin{eqnarray}
	\label{hijtt=hijtt eikx}
	h^{TT}_{ij}=\mathring{h}^{TT}_{ij} e^{ikx},
\end{eqnarray}
where $\mathring{h}^{TT}_{ij}$ is a constant tensor. 
In this case, the condition for the existence of non-zero solutions requires
\begin{eqnarray}
	\label{condition for the existence of non-zero solutions tensor mode equation}
	\left(H_{(2)}-I_{(2)}A^{2}\right) k_{0}^{2}-H_{(2)}k_{3}^{2}
	=0.
\end{eqnarray}
Further from Eq. (\ref{P1-P6 gauge invariant}), this indicates that the most general second-order vector-tensor theory allows for the existence of two tensor modes, the $+$ mode and the $\times$ mode, and their wave speeds $v_{T}$ satisfy
\begin{eqnarray}
	\label{vT=}
	v_{T}^{2}=\frac{k_{0}^{2}}{k_{3}^2}=\frac{H_{(2)}}{H_{(2)}-I_{(2)}A^{2}}.
\end{eqnarray}

GW170817 and GRB170817A require the speed of tensor modes to satisfy \cite{B.P.Abbott000,B.P.Abbott111}
\begin{eqnarray} 	
	\label{GW170817}
	-3\times10^{-15} \leq v_{T}-1 \leq 7\times10^{-16}.
\end{eqnarray}
Therefore, we have
\begin{eqnarray}
	\label{I2A^2/H2 less 10-15}
	\left | \frac{I_{(2)}A^{2}}{H_{(2)}}\right |
	\!\! ~\lesssim \!\! ~ 10^{-15}.
\end{eqnarray}
Specifically, when tensor gravitational waves propagate at the speed of light, we have $I_{(2)}=0$.

\subsection{Vector modes}
\label{vector modes}
Using Eqs. (\ref{delta S2flat/delta h}) and (\ref{delta S2flat/delta B}), we can derive two independent vector equations that describe the vector mode gravitational waves. They are
\begin{eqnarray}
	\label{vector mode equations 1}
	\left(G_{(2)}A^{2}+4H_{(2)}\right)\Xi_{i}
	+\left(G_{(2)}+4I_{(2)}\right)A\Sigma_{i}
	=0,
\end{eqnarray}
\begin{eqnarray}
	\label{vector mode equations 2}
	\!&&\! 2A_{(1)}E^{0ijk}A^{2}\partial_{k}\Xi_{j}
	+
	2A_{(1)}E^{0ijk}A\partial_{k}\Sigma_{j}
	\nonumber \\
	\!&+&\!
	\left(2E_{(2)}-2G_{(2)}-4I_{(2)}+2F_{(2)}A^{2}\right)A\partial_{0}^{2}\Xi_{i}
	+\left(G_{(2)}-2E_{(2)}\right)A\Delta \Xi_{i}
	\nonumber \\
	\!&-&\!
	2\left(-E_{(2)}+G_{(2)}+2I_{(2)}-F_{(2)}A^{2}\right)\partial_{0}^{2}\Sigma_{i}
	+2\left(-E_{(2)}+G_{(2)}+2I_{(2)}\right)\Delta \Sigma_{i}
    =0.
\end{eqnarray}
From the above equations, it can be seen that unlike pure metric theory or scalar-tensor theory \cite{Y.Dong4}, when the tensor mode gravitational wave propagates at the speed of light, i.e., 
$I_{(2)}=0$, $\Xi_{i}$ is generally not zero. Therefore, from Eq. (\ref{P1-P6 gauge invariant}), in vector-tensor theory, vector mode gravitational waves may still exist even when tensor mode gravitational waves propagate at the speed of light. However, in the special case where $G_{(2)}=0$, from Eqs. (\ref{vector mode equations 1}) and (\ref{H2 neq 0}), it can be deduced that $I_{(2)}=0$ implies $\Xi_{i}=0$. In this case, if the tensor mode propagates at the speed of light, there is no vector mode in the theory. Furthermore, it should be noted that when $A_{(1)} \neq 0$, there is a term in Eq. (\ref{vector mode equations 2}) that includes $E^{\mu\nu\lambda\rho}$. In such a situation, parity symmetry is broken, leading to different properties for left-handed and right-handed gravitational waves.

To solve Eqs. (\ref{vector mode equations 1}) and (\ref{vector mode equations 2}), and analyze vector mode gravitational waves, it is necessary to classify and discuss the parameter space. In Appendix \ref{app: B}, we systematically classify the parameter space and analyze in detail the properties of vector modes of gravitational waves in the theory for each possible choice of parameters. In principle, this part should be included in the main text to present a complete account of the theoretical analysis; however, due to its length, we have placed it in the appendix.
%To avoid excessive length, the detailed classification of vector modes is provided in Appendix \ref{app: B}.

It should be noted that in the analysis of vector mode gravitational waves, whether $A_{(1)}$ is zero— that is, whether there is a term containing $E^{\mu\nu\lambda\rho}$— determines whether the properties of left-handed and right-handed vector modes are the same. When $A_{(1)} \neq 0$, vector mode gravitational waves exhibit superluminal phenomena. Therefore, these cases can only be made reasonable by adding additional mechanisms to prevent exceeding the speed of light. This might imply that there should be no term containing $E^{\mu\nu\lambda\rho}$ in the second-order perturbation action.

\subsection{Scalar modes}
\label{scalar modes}
Now, let us analyze scalar mode gravitational waves. Using Eqs. (\ref{delta S2flat/delta h}) and (\ref{delta S2flat/delta B}), we can derive four independent scalar equations that describe the scalar mode gravitational waves:
\begin{eqnarray}
	\label{scalar mode equation 1}
	4B_{(1)}A^{2}\phi
	+4B_{(1)}A\Omega
	+\Lambda_{1}\partial_{0}\phi
	+\Lambda_{2}\partial_{0}\Theta
	+\Lambda_{3}\partial_{0}\Omega
	+\Lambda_{4}\partial_{0}^{2}\Psi
	+\Lambda_{5}\Delta\Psi
	=0,
	\\
	-4A^{3}A_{(0)}\left(\Omega+A\phi\right)
	+3B_{(1)}A^{3}\partial_{0}\Theta
	+2B_{(1)}A^{2}\Delta\Psi
	+K_{1}\partial_{0}^{2}\phi
	+K_{2}\Delta\phi
	\nonumber \\
	\label{scalar mode equation 2}
	+K_{3}\partial_{0}^{2}\Theta
	+K_{4}\Delta\Theta
	+K_{5}\partial_{0}^{2}\Omega
	+K_{6}\Delta\Omega
	+K_{7}\partial_{0}\Delta\Psi
	=0,
	\\
	\label{scalar mode equation 3}
	M_{1}\phi
	+M_{2}\Theta
	+M_{3}\Omega
	+M_{4}\partial_{0}\Psi
	=0,
	\\
	\label{scalar mode equation 4}
	N_{1}\phi
	+2H_{(2)}\Theta
	+N_{3}\Omega
	+N_{4}\partial_{0}\Psi
	=0.
\end{eqnarray}
Here,
\begin{eqnarray}
	\label{scalar mode parameters}
	\Lambda_{1}\!&=&\!
	-2C_{(2)}A^{3}
	-4J_{(2)}A^{3}
	+4E_{(2)}A
	-4G_{(2)}A
	+8K_{(2)}A,
	\nonumber \\
	\Lambda_{2}\!&=&\!
	-2G_{(2)}A
	+4I_{(2)}A
	+12K_{(2)}A,
	\nonumber \\
	\Lambda_{3}\!&=&\!
	2E_{(2)}
	+4I_{(2)}
	-2G_{(2)}
	+8K_{(2)}
	-2C_{(2)}A^{2}
	-2F_{(2)}A^{2}
	-4J_{(2)}A^{2},
	\nonumber \\
	\Lambda_{4}\!&=&\!
	2E_{(2)}
	-2G_{(2)}
	-4I_{(2)}
	+2F_{(2)}A^{2},
	\nonumber \\
	\Lambda_{5}\!&=&\!
	8I_{(2)}
	+8K_{(2)},
	\nonumber \\
	K_{1}\!&=&\!
	4A_{(2)}A^{4}
	-4B_{(2)}A^{6}
	+4C_{(2)}A^{4}
	+4J_{(2)}A^{4}
	+4F_{(2)}A^{4}
	-4K_{(2)}A^{2}
	-4I_{(2)}A^{2},
	\nonumber \\
	K_{2}\!&=&\!
	-4A_{(2)}A^{4}
	-4E_{(2)}A^{2}
	-4D_{(2)}A^{2},
	\nonumber \\
	K_{3}\!&=&\!
	3D_{(2)}A^{2}
	+3J_{(2)}A^{4}
	+3G_{(2)}A^{2}
	-6K_{(2)}A^{2},
	\nonumber \\
	K_{4}\!&=&\!
	-2D_{(2)}A^{2}+4H_{(2)},
	\nonumber \\
	K_{5}\!&=&\!
	4A_{(2)}A^{3}
	+4C_{(2)}A^{3}
	-4B_{(2)}A^{5}
	+4J_{(2)}A^{3}
	+4F_{(2)}A^{3}
	-4K_{(2)}A
	-4I_{(2)}A,
	\nonumber \\
	K_{6}\!&=&\!
	-4A_{(2)}A^{3}
	-C_{(2)}A^{3}
	-2E_{(2)}A
	-2D_{(2)}A,
	\nonumber \\
	K_{7}\!&=&\!
	C_{(2)}A^{3}
	+2J_{(2)}A^{3}
	-2E_{(2)}A
	+2G_{(2)}A
	-4K_{(2)}A,
	\nonumber \\
	M_{1}\!&=&\!
	C_{(2)}A^{4}
	-2D_{(2)}A^{2}
	-2E_{(2)}A^{2},
	\nonumber \\
	M_{2}\!&=&\!
	G_{(2)}A^{2}
	+4H_{(2)},
	\nonumber \\
	M_{3}\!&=&\!
	C_{(2)}A^{3}
	+F_{(2)}A^{3}
	-E_{(2)}A
	-G_{(2)}A
	-2D_{(2)}A
	-2I_{(2)}A,
	\nonumber \\
	M_{4}\!&=&\!
	-E_{(2)}A
	+G_{(2)}A
	-F_{(2)}A^{3}
	+2I_{(2)}A,
	\nonumber \\
	N_{1}\!&=&\!
	-2D_{(2)}A^{2}
	+4H_{(2)},
	\nonumber \\
	N_{3}\!&=&\!
	-2D_{(2)}A
	-G_{(2)}A,
	\nonumber \\
	N_{4}\!&=&\!
	G_{(2)}A
	+4I_{(2)}A.\nonumber
\end{eqnarray}

Due to Eq. (\ref{H2 neq 0}), Eq. (\ref{scalar mode equation 4}) can be rewritten as 
\begin{eqnarray}
   \label{scalar mode Theta}
   \Theta
   =
   -\frac{N_{1}\phi+N_{3}\Omega+N_{4}\partial_{0}\Psi}{2H_{(2)}}.
\end{eqnarray}
By substituting Eq. (\ref{scalar mode Theta}) into Eqs. (\ref{scalar mode equation 1}), (\ref{scalar mode equation 2}), and (\ref{scalar mode equation 3}), we obtain the following equations:
\begin{eqnarray}
	\label{scalar mode equation 1 new}
	4B_{(1)}A^{2}\phi
	+4B_{(1)}A\Omega
	+\Lambda_{1}'\partial_{0}\phi
	+\Lambda_{3}'\partial_{0}\Omega
	+\Lambda_{4}'\partial_{0}^{2}\Psi
	+\Lambda_{5}\Delta\Psi
	=0,
	\\
	-4A^{3}A_{(0)}\left(\Omega+A\phi\right)
	-\frac{3B_{(1)}A^{3}}{2H_{(2)}}
	\left(
	N_{1}\partial_{0}\phi
	+N_{3}\partial_{0}\Omega
	+N_{4}\partial_{0}^{2}\Psi
	\right)
	+2B_{(1)}A^{2}\Delta\Psi
	\nonumber \\
	\label{scalar mode equation 2 new}
	+K_{1}'\partial_{0}^{2}\phi
	+K_{2}'\Delta\phi
	+K_{5}'\partial_{0}^{2}\Omega
	+K_{6}'\Delta\Omega
	+K_{7}'\partial_{0}\Delta\Psi
	-\frac{K_{3}N_{4}}{2H_{(2)}}\partial_{0}^{3}\Psi
	=0,
	\\
	\label{scalar mode equation 3 new}
	M_{1}'\phi
	+M_{3}'\Omega
	+M_{4}'\partial_{0}\Psi
	=0.
\end{eqnarray}
Here,
\begin{eqnarray}
   \label{scalar mode parameters 2}
   \Lambda_{1}'\!&=&\!
   \Lambda_{1}
   -\frac{N_{1}\Lambda_{2}}{2H_{(2)}},\quad
   \Lambda_{3}'=
   \Lambda_{3}
   -\frac{N_{3}\Lambda_{2}}{2H_{(2)}},
   \nonumber \\
   \Lambda_{4}'\!&=&\!
   \Lambda_{4}
   -\frac{N_{4}\Lambda_{2}}{2H_{(2)}},\quad
   K_{1}'=
   K_{1}
   -\frac{N_{1}K_{3}}{2H_{(2)}},
   \nonumber \\
   K_{2}'\!&=&\!
   K_{2}
   -\frac{N_{1}K_{4}}{2H_{(2)}},\quad
   K_{5}'=
   K_{5}
   -\frac{N_{3}K_{3}}{2H_{(2)}},
   \nonumber \\
   K_{6}'\!&=&\!
   K_{6}
   -\frac{N_{3}K_{4}}{2H_{(2)}},\quad
   K_{7}'=
   K_{7}
   -\frac{N_{4}K_{4}}{2H_{(2)}},
   \nonumber \\
   M_{1}'\!&=&\!
   M_{1}
   -\frac{N_{1}M_{2}}{2H_{(2)}},\quad
   M_{3}'=
   M_{3}
   -\frac{N_{3}M_{2}}{2H_{(2)}},
   \nonumber \\
   M_{4}'\!&=&\!
   M_{4}
   -\frac{N_{4}M_{2}}{2H_{(2)}}.
\end{eqnarray}
Equations (\ref{scalar mode equation 1 new})-(\ref{scalar mode equation 3 new}) and relationship (\ref{scalar mode Theta}) provide a complete description of scalar mode gravitational waves. Similar to the case of the vector modes, a detailed and comprehensive analysis of the scalar modes has been placed in Appendix \ref{app: C} to avoid making the main text overly lengthy.
%To avoid excessive length, the detailed classification of scalar modes is provided in Appendix \ref{app: C}.

The results indicate that for a specified dispersion relation, the characteristics of scalar mode gravitational waves fall into one of three categories: (1) the absence of scalar mode gravitational waves; (2) the presence of two independent polarization modes for scalar gravitational waves: the breathing mode and the longitudinal mode; (3) scalar gravitational waves exhibiting only one polarization mode, which is a combination of two modes: a pure longitudinal mode (dictated by $\phi$) and a mixed mode comprising both breathing and longitudinal modes, with equal amplitudes for each (determined by $\Theta$). In the third case, the two mixed modes typically show a phase difference.
﻿

In Appendix \ref{app: D}, we use the gravitational wave polarization modes of generalized Proca theory as an example to illustrate the validity of the analysis in this section.

\section{Conclusion}
\label{sec: 7}

{In this paper, we have established a model-independent framework for analyzing the gravitational-wave effects in the most general second-order vector-tensor theory through the Isaacson picture.}
%We used the most general second-order vector-tensor theory as an example to demonstrate how to use our developed method to construct a model-independent theoretical framework for studying gravitational wave effects.
After constructing the most general second-order perturbation action of the second-order vector-tensor theory in the Minkowski background (assuming spatial isotropy of the background, thus requiring the vector field background to have only a non-zero temporal component), we proceeded to derive the two sets of basic equations in the Isaacson picture and the effective energy-momentum tensor of gravitational waves. Subsequently, we focused on analyzing the polarization modes of gravitational waves and the corresponding dispersion relations in the most general second-order vector-tensor theory.

Compared to the pure metric theory (which describes gravity solely through the metric) and the scalar-tensor theory, the analysis of polarization modes of gravitational waves in the vector-tensor theory is quite complex. The polarization modes of gravitational waves in the vector-tensor theory generally do not satisfy the general properties found in the pure metric theory and the scalar-tensor theory as Ref. \cite{Y.Dong4}. Additionally, another important difference is that the second-order perturbation action of the second-order vector-tensor theory allows for the appearance of the four-dimensional Levi-Civita totally antisymmetric tensor $E^{\mu\nu\lambda\rho}$, which does not appear in the pure metric theory and the scalar-tensor theory. The terms containing $E^{\mu\nu\lambda\rho}$ in the action only affect vector mode gravitational waves. However, such terms can cause vector modes to exceed the speed of light in certain spectral ranges of the wave vector. Therefore, without introducing additional mechanisms to suppress superluminal phenomena, these terms would lead to unreasonable physical implications. For this consideration, perhaps we should remove all terms containing $E^{\mu\nu\lambda\rho}$ from the second-order perturbation action. For tensor mode gravitational waves, nonvanishing background vector fields often cause the wave speed of the tensor modes to deviate from the speed of light. Therefore, we can constrain the parameter space of the theory using gravitational wave events such as GW170817. For scalar modes, the cases become very complex. However, generally speaking, for a given dispersion relation, the properties of scalar mode gravitational waves satisfy one of the following three cases: (1) no scalar mode gravitational waves; (2) scalar gravitational waves have two independent polarization modes: the breathing mode and the longitudinal mode; (3) scalar mode gravitational waves have only one polarization mode, which is a mixture of two modes: a pure longitudinal mode (determined by $\phi$), and a mixed mode of breathing mode and longitudinal mode, with equal amplitude for both (determined by $\Theta$). In the last case, the two mixed modes generally exhibit a phase difference.

{It should be noted that although in this paper we have only analyzed the gravitational-wave effects in the most general second-order vector-tensor theory, the Isaacson picture can be directly used to establish a model-independent framework for analyzing gravitational-wave effects in a broader range of modified gravity theories, without requiring any substantial further work. In fact, this method is highly general. For the vast majority of modified gravity theories, including those with Lagrange multiplier theories or metric-affine theories that modify Riemannian geometry, the action can formally be viewed as a functional of the metric and a series of additional fields. Therefore, these theories are all included within the scope of the Isaacson picture.} Furthermore, as long as the energy-momentum tensor of the matter field $T_{\mu\nu}$ is defined according to 
\begin{eqnarray}
	\label{varing the E-H action2}
	\delta S_{m}
	&=&\frac{1}{2}\int d^{4}x \sqrt{-g} T^{\mu\nu} \delta g_{\mu\nu},
\end{eqnarray}
all discussions in this paper regarding the effective energy-momentum tensor of gravitational waves retain their physical significance.
%It should be pointed out that our examples of vector-tensor theory do not include theories such as Einstein-aether theory, where Lagrange multipliers are introduced to ensure the vector field has unit length. However, since Lagrange multipliers are formally equivalent to scalar fields, the theory with Lagrange multipliers is not fundamentally different. It can still be directly used to analyze the gravitational wave effects of such theories within the method we have developed. In fact, the method we developed is highly general. For the vast majority of modified gravity theories, including those with Lagrange multipliers theories or metric-affine theories that modify Riemannian geometry, the action can be formally treated as a functional of the metric and a series of additional fields. As a result, these theories are encompassed within the scope of the method we discussed. In principle, they can still be analyzed in a generalized manner, similar to the vector-tensor theory example presented in this paper. Furthermore, as long as the energy-momentum tensor of the matter field $T_{\mu\nu}$ is defined according to Eq. (\ref{varing the E-H action2}), all discussions in this paper regarding the effective energy-momentum tensor of gravitational waves retain their physical significance.

There are still many areas worth studying. Firstly, our study only provides a detailed analysis of the polarization modes of gravitational waves within the framework of the most general second-order vector-tensor theory. The memory effect of gravitational waves also constitutes an important area of research. Investigating this aspect entails solving the low-frequency equation within the Isaacson picture, which will be a subject of future research. Simplifying the effective energy-momentum tensor (\ref{t_munu S2flat in the most general vector-tensor theory}) of gravitational waves derived in this paper using the on-shell condition is also an important yet complex problem. Secondly, in addition to the most general second-order vector-tensor theory, the most general pure metric theory, and the most general scalar-tensor theory analyzed in our previous paper \cite{Y.Dong4}, there are other important classes of modified gravity theories, such as metric-affine theory, that require the development of model-independent frameworks for analyzing gravitational-wave effects. In principle, this can be achieved using the method presented in this paper, although further detailed work is needed. Combining such model-independent theoretical frameworks with specific experiments is also a crucial research topic.

Furthermore, it is known that starting from the second-order perturbation action of general relativity, we can systematically reconstruct the complete Einstein-Hilbert action. A natural question arises: can theories such as vector-tensor theory similarly originate from second-order perturbation actions and derive their complete actions? This remains an area requiring further investigation. Additionally, exploring which parameter selections are viable in physics for the action is also an important research topic. An important example in this regard is finding the conditions for the theory to be ghost-free. This problem is expected to be addressed in the next phase of work using the program introduced in Ref. {\cite{Will Barker}}. Under the dual constraints of theory and experiment, we believe that a viable theory of gravity can eventually be discovered in the future.

\section*{Acknowledgments}
This work is supported in part by the National Key Research and Development Program of China (Grant No. 2020YFC2201503), the National Natural Science Foundation of China (Grants No. 123B2074, No. 12475056,  and No. 12247101), Gansu Province's Top Leading Talent Support Plan, the Fundamental Research Funds for the Central Universities (Grant No. lzujbky-2024-jdzx06), the Natural Science Foundation of Gansu Province (No. 22JR5RA389), and the `111 Center' under Grant No. B20063.

\appendix
\section{Isaacson picture in modified gravity theory}
\label{app: H}

We consider a modified gravity theory that satisfies the following form：
\begin{eqnarray}
	\label{MGT action}
	S=\int d^{4}x \sqrt{-g} \mathcal{L}\left[g_{\mu\nu},\Phi^{A}\right]
	+S_{m}\left[g_{\mu\nu},\Psi_{m}\right],
\end{eqnarray}
where $A=1, 2, ..., N$. This theory has $N$ additional fields, labeled with the superscript $A$. As long as each component of a vector field or a tensor field is treated as an additional field, it can be seen that the action (\ref{MGT action}) can describe vector-tensor theory and tensor-tensor theory.

Varying the action (\ref{MGT action}) with respect to $g_{\mu\nu}$ and $\Phi^{A}$, respectively, we have
\begin{eqnarray}
	\label{varing the MGT action}
	\delta S
	&=&\int d^{4} x \sqrt{-g} 
	\left[
	\left(-\mathcal{M}^{\mu\nu}+\frac{1}{2}T^{\mu\nu}\right)
	\delta g_{\mu\nu}
	+\mathcal{N_{A}} \delta \Phi^{A}
	\right].
\end{eqnarray}
Therefore, the field equations of this modified gravity theory are 
\begin{eqnarray}
	\label{MGT field equations metric}
	\mathcal{M}_{\mu\nu}&=&\frac{1}{2}T_{\mu\nu},
	\\
	\label{MGT field equations phiA}
	\mathcal{N}_{A}&=&0.
\end{eqnarray}
Among them, the index in Eq. (\ref{MGT field equations metric}) have been lowered using the metric $g_{\mu\nu}$. 

Similar to general relativity, we decompose the metric $g_{\mu\nu}$ and additional fields $\Phi^{A}$ into low-frequency background and high-frequency perturbation parts:
\begin{eqnarray}
	\label{perturbation MGT}
	g_{\mu\nu}=\bar{g}_{\mu\nu}+h_{\mu\nu},\quad \Phi^{A}=\bar{\Phi}^{A}+\varphi^{A}.
\end{eqnarray}
Just as in general relativity, the condition $\bar{g}_{\mu\nu} \sim 1, 
h_{\mu\nu} \sim \alpha, \alpha \ll 1,$ can always be applied to the metric field without loss of generality. For additional fields, we can always redefine them such that $\bar{\Phi}^{A}$ is set to the order of $1$. Additionally, we assume that the orders of magnitude of $\varphi^{A}$ and $h_{\mu\nu}$ are the same, i.e. 
\begin{eqnarray}
	\label{Phi-1,varPhi<<1}
	\bar{\Phi}^{A} \sim 1,\quad
	\varphi^{A} \sim h_{\mu\nu} \sim \alpha.
\end{eqnarray}
In addition, we also assume that both the low-frequency components and high-frequency of $\Phi^{A}$ have the same typical frequencies as the corresponding components of $g_{\mu\nu}$.

Using Eq. (\ref{MGT field equations metric}), we expand $\mathcal{M}_{\mu\nu}$ and $\mathcal{N}_{A}$ for the small perturbations:
\begin{eqnarray}
	\label{expand M}
	\mathcal{M}_{\mu\nu}
	\!&=&\!\mathcal{M}^{(0)}_{\mu\nu}\left[\bar{g}_{\mu\nu},\bar{\Phi}^{A}\right]
	+\mathcal{M}^{(1)}_{\mu\nu}
	\left[\bar{g}_{\mu\nu},\bar{\Phi}^{A},h_{\mu\nu},\varphi^{A}\right]
	+\mathcal{M}^{(2)}_{\mu\nu}
	\left[\bar{g}_{\mu\nu},\bar{\Phi}^{A},h_{\mu\nu},\varphi^{A}\right]
	+...~,
	\\
	\label{expand N}
	\mathcal{N}_A
	\!&=&\!\mathcal{N}_{A}^{(0)}\left[\bar{g}_{\mu\nu},\bar{\Phi}^{A}\right]
	+\mathcal{N}_{A}^{(1)}
	\left[\bar{g}_{\mu\nu},\bar{\Phi}^{A},h_{\mu\nu},\varphi^{A}\right]
	+\mathcal{N}_{A}^{(2)}
	\left[\bar{g}_{\mu\nu},\bar{\Phi}^{A},h_{\mu\nu},\varphi^{A}\right]
	+...~.
\end{eqnarray}
Here, we only write the expansion up to the second-order term.

To obtain the two sets of basic equations in the Isaacson picture, we first use Eqs. (\ref{expand M}) and (\ref{expand N}) to expand the field equations (\ref{MGT field equations metric}) and (\ref{MGT field equations phiA}). Next, we perform the averaging operation $\langle ... \rangle$ on the expanded equations to separate them into two sets: the high-frequency and low-frequency parts. Finally, we retain only the leading-order terms in these equations. In the following, we use $f_{H}$ and $f_{L}$ to denote the characteristic frequencies of the high-frequency and low-frequency parts, respectively.

It should be pointed out that when retaining the leading-order term, the third-order and higher-order perturbation terms in the expansion cannot be simply ignored, as in general relativity. In general relativity, each term in the Einstein tensor $G_{\mu\nu}$ contains only two derivative operators. This leads to the following order of magnitude relationship:
\begin{eqnarray}
	\label{order of magnitude relationships of Gi}
	G^{(i)}_{\mu\nu} \sim f_{H}^{2} \alpha^{i}, \qquad i \neq 0.
\end{eqnarray}
Since $\alpha \ll 1$, it can be seen that any $G^{(i)}_{\mu\nu}$ corresponding to $i \geq 3$ is always much smaller compared to $G^{(2)}_{\mu\nu}$ and can be ignored at the leading order. However, in modified gravity theories, even if the field equations are required to be second-order, it does not guarantee that each term in the equations has at most two derivative operators (an example is Horndeski theory \cite{Tsutomu Kobayashi}). Therefore, we cannot simply ignore the third-order and higher-order perturbation terms in the field equations within modified gravity theories.

For any modified gravity theory that can derive second-order field equations, the structure of each term in the field equations can be formally written as
\begin{eqnarray}
	\label{formally trem of field Eq}
	\left(\partial\partial X \right)^{n_{1}} \left(\partial X \right)^{n_{2}} X^{n_{3}}.
\end{eqnarray}
Here, the character $X$ formally refers to the dynamic fields, i.e., the metric $g_{\mu\nu}$ and additional fields $\Psi^{A}$, and $n_{1}$, $n_{2}$ and $n_{3}$ are natural numbers. The meaning of the above equation is that, in this term of the field equation, $n_1$ fields are taken as second-order derivatives, $n_2$ fields are taken as first-order derivatives, and $n_3$ fields are not taken as derivatives. Since the field equations are second-order, there will be no components of the form $\partial^k X$ where $k \geq 3$.

We take the case of $n_1=3$, $n_2=1$, $n_3=1$ as an example to illustrate the different points of magnitude analysis in modified gravity theories compared to general relativity. For this scenario, the formal expression of this term is:
\begin{eqnarray}
	\label{formally trem of field Eq n1=3 n2=1 n3=1}
	\left(\partial\partial X \right)^{3} \left(\partial X \right) X.
\end{eqnarray}
After expanding this term into perturbations, it can be seen that the magnitude of the second-order perturbation term is $f_{H}^{4}f_{L}^{3}\alpha^2$, while the magnitude of the third-order perturbation term is
\begin{eqnarray}
	\label{formally trem of field Eq n1=3 n2=1 n3=1 magnitude of the third-order perturbation term}
	\left(f_{H}^{4}f_{L}^{3}\alpha^2\right) \left(\frac{f_{H}^{2}}{f_{L}^2}\alpha\right).
\end{eqnarray}
It can be seen that only when 
\begin{eqnarray}
	\label{fH2/fL2a<<1}
	\frac{f_{H}^{2}}{f_{L}^2}\alpha \ll 1,
\end{eqnarray}
the third-order perturbation term can be considered small. However, while $\alpha$ is small, $f_{H}/f_{L}$ is large, so it cannot be assumed that this condition is always satisfied. 

Although we only provide a special example here, it is not difficult to see that the condition of at most second-order partial derivatives of the dynamic field appearing in Eq. (\ref{formally trem of field Eq}) and the relationship $f_{H}/f_{L} \gg 1$ ensure the following proposition: for any values of $\left(n_1, n_2, n_3\right)$, as long as the condition (\ref{fH2/fL2a<<1}) is satisfied, the higher-order perturbation term of Eq. (\ref{formally trem of field Eq}) is much smaller compared to the second-order perturbation term and can be ignored at leading-order. Therefore, as long as the condition (\ref{fH2/fL2a<<1}) holds, we can ignore the perturbation terms higher than second-order in the field equations.

In fact, for the gravitational wave events we observe, the condition (\ref{fH2/fL2a<<1}) is always satisfied. We use the example mentioned in Ref. \cite{Lavinia Heisenberg1} to illustrate this point. Reference \cite{Lavinia Heisenberg1} points out that for gravitational waves generated by a binary merger with a total mass of $10^2 M_{\odot}$, we have $\alpha\sim 10^{-22}, f_{H} \sim 10^2$ Hz, $f_{L}\sim 10$ Hz. For gravitational waves generated by a binary merger with a total mass of $10^5 M_{\odot}$, the result is $\alpha\sim 10^{-19}, f_{H} \sim 10^{-1}$ Hz, $f_{L}\sim 10^{-2}$ Hz. These two types of gravitational wave events can be detected by ground-based and space gravitational wave detectors, and they satisfy $f_{H}^2/f_{L}^2 \alpha \sim 10^{-20}$ and $f_{H}^2/f_{L}^2 \alpha \sim 10^{-17}$, respectively. All of them satisfy the condition (\ref{fH2/fL2a<<1}).

%Regarding the condition (\ref{fH2/fL2a<<1}), there are two additional specific points in need of clarification. The first point is that 
In a modified gravity theory, we consider cases where the field equations are not necessarily second-order but could be of $N$-th order. Thus, the sufficient condition for ignoring perturbation terms higher than second-order in the field equations is changed from (\ref{fH2/fL2a<<1}) to
\begin{eqnarray}
	\label{fHN/fLNa<<1}
	\left(\frac{f_{H}}{f_{L}}\right)^{N}\alpha \ll 1.
\end{eqnarray}
As can be seen from the example in the previous paragraph, this condition can be satisfied for theories with $N \textless 19$. Therefore, in high-order derivative theories where $N \textless 19$, we can still ignore higher-order perturbation terms. This includes the vast majority of common modified gravity theories.

%The second point needing clarification is that the condition (\ref{fH2/fL2a<<1}) appears to contradict Eq. (\ref{order of magnitude relationships of G0 and G2}). According to Eq. (\ref{order of magnitude relationships of G0 and G2}), the condition that requires the order of magnitude of $G^{(0)}_{\mu\nu}$ and $G^{(2)}_{\mu\nu}$ to be comparable is 
%\begin{eqnarray}
%	\label{fH2/fL2a2=1}
%\frac{f_{H}^{2}}{f_{L}^{2}}\alpha^2 \sim 1.
%\end{eqnarray}
%However, the condition (\ref{fH2/fL2a<<1}) results in 
%\begin{eqnarray}
%	\label{fH2/fL2a2 ll 1}
%	\frac{f_{H}^{2}}{f_{L}^{2}}\alpha^{2} \ll 1.
%\end{eqnarray}
%The key to resolving this contradiction is to note that the estimation of $G^{(0)}_{\mu\nu}$ in Eq. (\ref{order of magnitude relationships of G0,G1,G2}) is quite rough. For the gravitational wave events we observe, although $\bar{g}_{\mu\nu} \sim 1$, its amplitude generally does not vary by 1. The background metric typically exhibits only a slight deviation from the Minkowski metric, i.e.,
%\begin{eqnarray}
%	\label{gbar=eta+deltah}
%	\bar{g}_{\mu\nu}=\eta_{\mu\nu}+\mathfrak{d}\bar{g}_{\mu\nu},\quad  
%    \mathfrak{d}\bar{g}_{\mu\nu} \sim \beta \ll 1.
%\end{eqnarray}
%This ensures that the order of magnitude of $G^{(0)}_{\mu\nu}$ satisfies
%\begin{eqnarray}
%    \label{order G0 new}
%	G^{(0)}_{\mu\nu} \sim f_{L}^2 \beta \ll f_{L}^2.
%\end{eqnarray}
%The relationship (\ref{order G0 new}) modifies Eq. (\ref{order of magnitude relationships of G0 and G2}), thereby changing the condition (\ref{fH2/fL2a2=1}) to the condition (\ref{fH2/fL2a2 ll 1}) and resolving the contradiction.

Now, we can derive two sets of basic equations in the Isaacson picture. For high-frequency equations, we have
\begin{eqnarray}
	\label{MGT field equations metric high feq}
	\mathcal{M}^{(1)}_{\mu\nu}&=&\frac{1}{2}T^{H,(0)}_{\mu\nu},
	\\
	\label{MGT field equations phiA high feq}
	\mathcal{N}^{(1)}_{A}&=&0.
\end{eqnarray}
And for low-frequency equations, the result is 
\begin{eqnarray}
	\label{MGT field equations metric low feq}
	\mathcal{M}^{(0)}_{\mu\nu}+\left\langle\mathcal{M}^{(2)}_{\mu\nu}\right\rangle&=&\frac{1}{2}T^{L,(0)}_{\mu\nu},
	\\
	\label{MGT field equations phiA low feq}
	\mathcal{N}^{(0)}_{A}+\left\langle\mathcal{N}^{(2)}_{A}\right\rangle&=&0.
\end{eqnarray}
Similarly, we can define the effective energy-momentum tensor of gravitational waves in modified gravity theories as
\begin{eqnarray}
	\label{effective energy-momentum tensor of gravitational waves in MGT}
	t_{\mu\nu}
	\coloneqq 
	-2 \left\langle \mathcal{M}^{(2)}_{\mu\nu} \right\rangle.
\end{eqnarray}

\section{Perturbation action method}
\label{app: G}

\subsection{Perturbation action method and perturbed field equations}
\label{Perturbation action method and perturbation field equation method}
In this subsection, we use a simple example to introduce the perturbation action method and explain its relationship with the perturbed field equations.

Consider the following action:
\begin{eqnarray}
	\label{Lphi action}
	S\left[\phi\right]=\int d^{4}x \mathcal{L}\left[\phi\right].
\end{eqnarray}
By varying the action (\ref{Lphi action}) with respect to $\phi$, we have
\begin{eqnarray}
	\label{varying Lphi action}
	\delta S=\int d^{4}x \mathcal{F}\left[\phi\right] \delta \phi.
\end{eqnarray}
Therefore, the field equation is
\begin{eqnarray}
	\label{field eq of Lphi action}
	\mathcal{F}\left[\phi\right]=0.
\end{eqnarray}

After dividing $\phi$ into the background part and the perturbation part, 
\begin{eqnarray}
	\label{phi=phi0+varphi}
	\phi=\phi_{0}+\varphi,
\end{eqnarray}
we can expand the field equation for the perturbation as follows:
\begin{eqnarray}
	\label{expand field eq of Lphi action}
	\mathcal{F}\left[\phi_{0}+\varphi\right]
	=
	\mathcal{F}^{(0)}\left[\phi_{0}\right]
	+\mathcal{F}^{(1)}\left[\phi_{0},\varphi\right]
	+\mathcal{F}^{(2)}\left[\phi_{0},\varphi\right]
	+\sum_{i=3}^{\infty}\mathcal{F}^{(i)}\left[\phi_{0},\varphi\right].
\end{eqnarray}
Similarly, we can also expand the action as
\begin{eqnarray}
	\label{expand Lphi action}
	S\left[\phi_{0}+\varphi\right]
	=
	S^{(0)}\left[\phi_{0}\right]
	+S^{(1)}\left[\phi_{0},\varphi\right]
	+S^{(2)}\left[\phi_{0},\varphi\right]
	+\sum_{i=3}^{\infty}S^{(i)}\left[\phi_{0},\varphi\right],
\end{eqnarray}
where 
\begin{eqnarray}
	\label{Lphi action Si and Li}
	S^{(i)}\left[\phi_{0},\varphi\right]
	=
	\int d^4 x \mathcal{L}^{(i)}\left[\phi_{0},\varphi\right].
\end{eqnarray}

To illustrate the relationship between (\ref{expand field eq of Lphi action}) and (\ref{expand Lphi action}), we need to consider varying the action (\ref{expand Lphi action}) with respect to 
$\phi_{0}$ and $\varphi$, respectively:
\begin{eqnarray}
	\label{varying Lphi action with phi0}
	\delta S \left[\phi_{0}+\varphi\right]&=&\int d^{4}x \mathcal{F}\left[\phi_{0}+\varphi\right] \delta \phi_{0}, \nonumber \\
	\delta S \left[\phi_{0}+\varphi\right]&=&\int d^{4}x \mathcal{F}\left[\phi_{0}+\varphi\right] \delta \varphi.
\end{eqnarray}
The field equations obtained from both are $\mathcal{F}\left[\phi_{0}+\varphi\right]=0$. This can be easily observed using the chain rule of composite function differentiation or directly from the position symmetry of $\phi_{0}$ and $\varphi$ in the action (\ref{expand Lphi action}). Therefore, we have
\begin{eqnarray}
	\label{varying Lphi action with phi0 equivalently}
	\mathcal{F}\left[\phi_{0}+\varphi\right]
	=
	\frac{\delta S}{\delta \phi_{0}}=\frac{\delta S}{\delta \varphi}.
\end{eqnarray}

Using Eq. (\ref{varying Lphi action with phi0 equivalently}), we can see that $\mathcal{F}^{(i)}$ in the field equation can only be derived from varying $S^{(i+1)}$ with respect to $\varphi$, or from varying $S^{(i)}$ with respect to $\phi_{0}$, i.e.,
\begin{eqnarray}
	\label{Fi=dSi+1/varphi=dSi/phi0}
	\mathcal{F}^{(i)}
	=
	\frac{\delta S^{(i+1)}}{\delta \varphi}=\frac{\delta S^{(i)}}{\delta \phi_0},
	\qquad i \in \mathbb{N}.
\end{eqnarray}
The above equation provides the relationship between the perturbation action method and the perturbed field equations. Especially, to determine the field equation up to the second-order perturbation term, one only needs to know $S^{(1)}$ and $S^{(2)}$.

\subsection{Perturbation action method in general relativity}
\label{Perturbation action method in GR}

In this subsection, we consider how to obtain the two sets of basic equations of the Isaacson picture in general relativity under vacuum using the perturbation action method. To solve this problem, we only need to obtain the relationship between $G^{(0)}_{\mu\nu}$, $G^{(1)}_{\mu\nu}$,  $G^{(2)}_{\mu\nu}$, and the variation of the perturbed action ($G_{\mu\nu}$ is the Einstein tensor).

We start with the action of general relativity in vacuum:
\begin{eqnarray}
	\label{E-H action in vacuum}
	S=\frac{1}{16\pi}\int d^{4}x \sqrt{-g} R.
\end{eqnarray}
Varying this action with respect to the metric $g_{\mu\nu}$, we obtain
%The most important thing to note is that varying this action with respect to the metric $g_{\mu\nu}$ does not directly lead to Einstein field equation, but rather to
\begin{eqnarray}
	\label{varying E-H action in vacuum}
	\frac{\delta S}{\delta g_{\mu\nu}}=-\frac{1}{16\pi} \sqrt{-g} G^{\mu\nu}=0.
\end{eqnarray}
It can be seen that the quantity obtained by directly varying the action differs from the Einstein tensor by a factor proportional to $\sqrt{-g}$. Due to the presence of this factor, the relationship between the variation of the perturbed action and the perturbed Einstein tensor is not simply order-by-order correspondence, but rather a more complex one.

Specifically, when using $g_{\mu\nu}=\bar{g}_{\mu\nu}+h_{\mu\nu}$ to perturb the action (\ref{E-H action in vacuum}), we have
\begin{eqnarray}
	\label{relationship between G and perturbation action 0 order}
	\frac{\delta S^{(0)}}{\delta \bar{g}_{\mu\nu}}
	\!&=&\!\frac{\delta S^{(1)}}{\delta h_{\mu\nu}}
	=-\frac{1}{16\pi} \sqrt{-g}^{(0)} G^{(0)\mu\nu},
	\\
	\label{relationship between G and perturbation action 1 order}
	\frac{\delta S^{(1)}}{\delta \bar{g}_{\mu\nu}}
	\!&=&\!\frac{\delta S^{(2)}}{\delta h_{\mu\nu}}
	=-\frac{1}{16\pi} 
	\left(
	\sqrt{-g}^{(0)} G^{(1)\mu\nu}
	+\sqrt{-g}^{(1)} G^{(0)\mu\nu}
	\right),
	\\
	\label{relationship between G and perturbation action 2 order}
	\frac{\delta S^{(2)}}{\delta \bar{g}_{\mu\nu}}
	\!&=&\!\frac{\delta S^{(3)}}{\delta h_{\mu\nu}}
	=-\frac{1}{16\pi} 
	\left(
	\sqrt{-g}^{(0)} G^{(2)\mu\nu}
	+\sqrt{-g}^{(1)} G^{(1)\mu\nu}
	+\sqrt{-g}^{(2)} G^{(0)\mu\nu}
	\right).
\end{eqnarray}
Or equivalently, 
\begin{eqnarray}
	\label{relationship between G and perturbation action 0 order write G}
	G^{(0)\mu\nu}
	\!&=&\!
	- \frac{16\pi}{\sqrt{-g}^{(0)}} \frac{\delta S^{(0)}}{\delta \bar{g}_{\mu\nu}},
	\\
	\label{relationship between G and perturbation action 1 order write G}
	G^{(1)\mu\nu}
	\!&=&\!
	- \frac{16\pi}{\sqrt{-g}^{(0)}} 
	\left(
	\frac{\delta S^{(2)}}{\delta h_{\mu\nu}}
	-\frac{\sqrt{-g}^{(1)}}{\sqrt{-g}^{(0)}} \frac{\delta S^{(0)}}{\delta \bar{g}_{\mu\nu}}
	\right),
	\\
	\label{relationship between G and perturbation action 2 order write G}
	G^{(2)\mu\nu}
	\!&=&\!
	- \frac{16\pi}{\sqrt{-g}^{(0)}} 
	\left[
	\frac{\delta S^{(2)}}{\delta \bar{g}_{\mu\nu}}
	-\frac{\sqrt{-g}^{(1)}}{\sqrt{-g}^{(0)}} 
	\left(
	\frac{\delta S^{(2)}}{\delta h_{\mu\nu}}
	-\frac{\sqrt{-g}^{(1)}}{\sqrt{-g}^{(0)}} \frac{\delta S^{(0)}}{\delta \bar{g}_{\mu\nu}}
	\right)
	-\frac{\sqrt{-g}^{(2)}}{\sqrt{-g}^{(0)}} \frac{\delta S^{(0)}}{\delta \bar{g}_{\mu\nu}}  
	\right].
\end{eqnarray}

For more commonly used lower indices, due to $G_{\mu\nu}=g_{\mu\lambda}g_{\nu\rho}G^{\lambda\rho}$, we have
\begin{eqnarray}
	\label{G _munu with 0 order}
	G^{(0)}_{\mu\nu}
	\!&=&\!
	\bar{g}_{\mu\lambda}\bar{g}_{\nu\rho}G^{(0)\lambda\rho},
	\\
	\label{G _munu with 1 order}
	G^{(1)}_{\mu\nu}
	\!&=&\!
	\bar{g}_{\mu\lambda}\bar{g}_{\nu\rho}G^{(1)\lambda\rho}
	+h_{\mu\lambda}\bar{g}_{\nu\rho}G^{(0)\lambda\rho}
	+\bar{g}_{\mu\lambda}h_{\nu\rho}G^{(0)\lambda\rho},
	\\
	\label{G _munu with 2 order}
	G^{(2)}_{\mu\nu}
	\!&=&\!
	\bar{g}_{\mu\lambda}\bar{g}_{\nu\rho}G^{(2)\lambda\rho}
	+h_{\mu\lambda}\bar{g}_{\nu\rho}G^{(1)\lambda\rho}
	+\bar{g}_{\mu\lambda}h_{\nu\rho}G^{(1)\lambda\rho}
	+h_{\mu\lambda}h_{\nu\rho}G^{(0)\lambda\rho}.
\end{eqnarray}
It should be pointed out that we cannot directly use $\bar{g}_{\mu\nu}$ to lower indices, which can not correctly change $G^{(i)\mu\nu}$ to $G^{(i)}_{\mu\nu}$, where $i\textgreater 0$. From Eqs. (\ref{relationship between G and perturbation action 0 order write G})-(\ref{G _munu with 2 order}), we can use the variation of the perturbed action to represent the two sets of basic equations in the Isaacson picture.

In the Isaacson picture, the effective energy-momentum tensor of gravitational waves in general relativity is defined as
\begin{eqnarray}
	\label{effective energy-momentum tensor of gravitational waves in GR deltaS/delta g}
	t_{\mu\nu}
	\coloneqq 
	-\frac{1}{8\pi} \left\langle G^{(2)}_{\mu\nu} \right\rangle
	=-2 \left\langle\left(\frac{1}{\sqrt{-g}}\frac{\delta S}{\delta g^{\mu\nu}}\right)^{(2)}\right\rangle.
\end{eqnarray}
It can be seen that when using the perturbation action method, we should use Eqs. (\ref{G _munu with 2 order}) and (\ref{relationship between G and perturbation action 0 order write G})-(\ref{relationship between G and perturbation action 2 order write G}) to calculate the effective energy-momentum tensor (\ref{effective energy-momentum tensor of gravitational waves in GR deltaS/delta g}) of gravitational waves. 

In some papers, such as \cite{Leo C. Stein} and \cite{Lavinia Heisenberg1}, the effective energy-momentum tensor of gravitational waves in the perturbation action method is defined as
\begin{eqnarray}
	\label{effective energy-momentum tensor of gravitational waves in some paper}
	\tilde{t}_{\mu\nu}
	\coloneqq 
	-2 \left\langle\frac{1}{\sqrt{-\bar{g}}}\frac{\delta S^{(2)}}
	{\delta \bar{g}^{\mu\nu}}\right\rangle,
\end{eqnarray}
where $\bar{g}^{\mu\nu}$ is the inverse of the background metric $\bar{g}_{\mu\nu}$ and $\bar{g}$ is the determinant of the $\bar{g}_{\mu\nu}$. It should be pointed out that $\sqrt{-g}^{(0)}=\sqrt{-\bar{g}}$. The definition (\ref{effective energy-momentum tensor of gravitational waves in some paper}) seems reasonable. To obtain the effective energy-momentum tensor of gravitational waves in quadratic form (or understood as the leading-order), a natural approach is to analogize the idea that varying a matter field action with respect to the metric $g^{\mu\nu}$ will yield the energy-momentum tensor of that matter field. Therefore, if the background metric $\bar{g}_{\mu\nu}$ is regarded as a `metric' in a certain sense, and the perturbation $h_{\mu\nu}$ is regarded as a matter field, the method of defining the effective energy-momentum of a quadratic form of gravitational wave is, of course, by varying the second-order perturbation action $S^{(2)}$ with respect to the background metric, as described in the definition (\ref{effective energy-momentum tensor of gravitational waves in some paper}). When considering the quantum case, i.e., the graviton, the Heisenberg uncertainty principle prevents us from defining a local gravitational wave energy-momentum tensor. This is why an averaging operator $\langle ... \rangle$ is required in the definition (\ref{effective energy-momentum tensor of gravitational waves in some paper}) \cite{Michele Maggiore}.

However, the two definitions (\ref{effective energy-momentum tensor of gravitational waves in GR deltaS/delta g}) and (\ref{effective energy-momentum tensor of gravitational waves in some paper}) are generally not equivalent:
\begin{eqnarray}
	\label{t neq tilde t}
	t_{\mu\nu}
	\neq
	\tilde{t}_{\mu\nu}.
\end{eqnarray}
When substituting Eqs. (\ref{relationship between G and perturbation action 0 order write G})-(\ref{relationship between G and perturbation action 2 order write G}) into Eq. (\ref{G _munu with 2 order}) and expanding the parentheses, we can see that $t_{\mu\nu}$ has 9 terms. And one of them is
\begin{eqnarray}
	\label{one of 9 terms}
	2
	\left\langle
	\frac{1}{\sqrt{-g}^{(0)}} 
	\bar{g}_{\mu\lambda}\bar{g}_{\nu\rho}
	\frac{\delta S^{(2)}}{\delta \bar{g}_{\lambda\rho}}
	\right\rangle
	=
	-2
	\left\langle
	\frac{1}{\sqrt{-\bar{g}}} 
	\frac{\delta S^{(2)}}{\delta \bar{g}^{\mu\nu}}
	\right\rangle
	=
	\tilde{t}_{\mu\nu}.
\end{eqnarray}
The remaining 8 terms are only related to $S^{(0)}$ and $S^{(1)}$, thus generally not equal to $0$. This proves the relationship (\ref{t neq tilde t}). It should also be pointed out that we have used the condition $\delta(\bar{g}^{\mu\lambda}\bar{g}_{\lambda\nu})=\delta\bar{g}^{\mu\lambda}\bar{g}_{\lambda\nu}+\bar{g}^{\mu\lambda}\delta\bar{g}_{\lambda\nu}=0$ in the derivation of the first equal sign in Eq. (\ref{one of 9 terms}).

In this paper, we do not use Eq. (\ref{effective energy-momentum tensor of gravitational waves in some paper}) but still use Eq. (\ref{effective energy-momentum tensor of gravitational waves in GR deltaS/delta g}) to define the effective energy-momentum tensor of gravitational waves in general relativity using the perturbation action method, to ensure consistency with the results obtained from the Isaacson's derivation.

Although generally $t_{\mu\nu} \neq \tilde{t}_{\mu\nu}$, it can be proven that, when considering asymptotic Minkowski spacetime far from the source and on-shell gravitational waves, we have $t_{\mu\nu}=\tilde{t}_{\mu\nu}$. When considering asymptotic Minkowski spacetime, the background metric $\bar{g}_{\mu\nu}$ far from the source satisfies $\bar{g}_{\mu\nu}=\eta_{\mu\nu}+\mathfrak{d}\bar{g}_{\mu\nu}$. The contribution of $\mathfrak{d}\bar{g}_{\mu\nu}$ to the effective energy-momentum tensor of gravitational waves is negligible and can be ignored. Therefore, it is always possible to set $\bar{g}_{\mu\nu}=\eta_{\mu\nu}$ in the calculation.
%the background metric $\bar{g}_{\mu\nu}$ far from the source can be approximated by the Minkowski metric $\eta_{\mu\nu}$. 
From Eq. (\ref{G _munu with 0 order}), this leads to $G^{(0)}_{\mu\nu}=G^{(0)\mu\nu}=0$. The condition that gravitational waves are on-shell leads to $G^{(1)}_{\mu\nu}=0$. From Eq. (\ref{G _munu with 1 order}), this further leads to $G^{(1)\mu\nu}=0$. According to Eqs. (\ref{relationship between G and perturbation action 0 order})-(\ref{relationship between G and perturbation action 2 order}), these conditions result in the remaining 8 terms in the expansion of $t_{\mu\nu}$ being zero, thereby resulting in $t_{\mu\nu}=\tilde{t}_{\mu\nu}$.

In Appendix \ref{app: A}, we verify the correctness of the derivation presented in this subsection by calculating $t_{\mu\nu}$ and $\tilde{t}_{\mu\nu}$ within the framework of general relativity.

\subsection{Perturbation action method in modified gravity theory}
\label{Perturbation action method in MGT}
In modified gravity theory, the situation is entirely analogous to that of general relativity. As long as it is noted that the variation of the action still has an $\sqrt{-g}$ factor, i.e., %does not directly lead to what we usually call the field equation, i.e.,
\begin{eqnarray}
	\label{varying MGT action in vacuum}
	\frac{\delta S}{\delta g_{\mu\nu}}=-\sqrt{-g} \mathcal{M}^{\mu\nu}=0,\qquad
	\frac{\delta S}{\delta \Phi^{A}}=\sqrt{-g} \mathcal{N}_{A}=0,
\end{eqnarray}
the two sets of basic equations in the Isaacson picture of modified gravity theory can be derived using the perturbation action method in a completely parallel manner, following the steps outlined in the previous subsection. Especially, for the effective energy-momentum tensor of gravitational waves,
\begin{eqnarray}
	\label{effective energy-momentum tensor of gravitational waves in MGT deltaS/delta g}
	t_{\mu\nu}
	\coloneqq 
	-2 \langle \mathcal{M}^{(2)}_{\mu\nu} \rangle
	=-2 \left\langle\left(\frac{1}{\sqrt{-g}}\frac{\delta S}{\delta g^{\mu\nu}}\right)^{(2)}\right\rangle.
\end{eqnarray}
All the conclusions from the previous subsection are equally applicable to the case of modified gravity theory. Therefore, we need not elaborate further.

\section{A proof that the Isaacson picture can be derived from $S^{(2)}_{flat}$}
\label{app: F}

Using the perturbation action method, since $\mathcal{M}^{(0)}_{\mu\nu}\left[\eta_{\mu\nu},A^{\mu}\right]=\mathcal{N}^{(0)}_{\mu}\left[\eta_{\mu\nu},A^{\mu}\right]=0$, we have
\begin{eqnarray}
	\label{M1=deltaS/delta h}
	\mathcal{M}^{(1)}_{\mu\nu}\left[\eta_{\mu\nu},A^{\mu},h_{\mu\nu},B^{\mu}\right]
	\!&=&\!
	-\eta_{\mu\lambda}\eta_{\nu\rho}
	\left(
	\frac{1}{\sqrt{-\bar{g}}}
	\frac{\delta S^{(2)}}{\delta h_{\lambda\rho}}
	\right)
	\Bigg|_{\bar{g}_{\mu\nu}=\eta_{\mu\nu},\bar{A}^{\mu}=A^{\mu}},
	\\
	\label{N1=deltaS/delta B}
	\mathcal{N}^{(1)}_{\mu}\left[\eta_{\mu\nu},A^{\mu},h_{\mu\nu},B^{\mu}\right]
	\!&=&\!
	\left(
	\frac{1}{\sqrt{-\bar{g}}}
	\frac{\delta S^{(2)}}{\delta B^{\mu}}
	\right)
	\Bigg|_{\bar{g}_{\mu\nu}=\eta_{\mu\nu},\bar{A}^{\mu}=A^{\mu}}.
\end{eqnarray}
Here, $S^{(2)}$ is the second-order term of the action with respect to the high-frequency perturbations $h_{\mu\nu}$ and $B^{\mu}$. Furthermore, if we require gravitational waves to be on-shell, i.e., Eqs. (\ref{vector-tensor field equations metric high feq leading-order}) and (\ref{vector-tensor field equations vector field high feq leading-order}) hold, then we also have 
\begin{eqnarray}
	\label{M2=deltaS/delta g}
	\mathcal{M}^{(2)}_{\mu\nu}\left[\eta_{\mu\nu},A^{\mu},h_{\mu\nu},B^{\mu}\right]
	\!&=&\!
	-\eta_{\mu\lambda}\eta_{\nu\rho}
	\left(
	\frac{1}{\sqrt{-\bar{g}}}
	\frac{\delta S^{(2)}}{\delta \bar{g}_{\lambda\rho}}
	\right)
	\Bigg|_{\bar{g}_{\mu\nu}=\eta_{\mu\nu},\bar{A}^{\mu}=A^{\mu}},
	\\
	\label{N2=deltaS/delta A}
	\mathcal{N}^{(2)}_{\mu}\left[\eta_{\mu\nu},A^{\mu},h_{\mu\nu},B^{\mu}\right]
	\!&=&\!
	\left(
	\frac{1}{\sqrt{-\bar{g}}}
	\frac{\delta S^{(2)}}{\delta \bar{A}^{\mu}}
	\right)
	\Bigg|_{\bar{g}_{\mu\nu}=\eta_{\mu\nu},\bar{A}^{\mu}=A^{\mu}}.
\end{eqnarray}
It can be seen that once the second-order perturbation term $S^{(2)}\left[\bar{g}_{\mu\nu},\bar{A}^{\mu},h_{\mu\nu},B^{\mu}\right]$ is known, Eqs. (\ref{vector-tensor field equations metric high feq leading-order})-(\ref{vector-tensor field equations vector field low feq leading-order}) can be derived.

The structure of $S^{(2)}\left[\bar{g}_{\mu\nu},\bar{A}^{\mu},h_{\mu\nu},B^{\mu}\right]$ can always be represented as follows:
\begin{eqnarray}
	\label{structure of S2}
	S^{(2)}\left[\bar{g}_{\mu\nu},\bar{A}^{\mu},h_{\mu\nu},B^{\mu}\right]
	=
	\int d^4 x \sqrt{-\bar{g}}
	\left[\mathcal{L}_{A}^{(2)}
	+\left(\partial \bar{g}\right)\mathcal{L}_{B}^{(2)}
	+\left(\partial \bar{A}\right)\mathcal{L}_{C}^{(2)}
	\right].
\end{eqnarray}
Among these, the term related to $\mathcal{L}_{A}^{(2)}$ represents the set of terms in $S^{(2)}$ that do not contain the partial derivative of the background fields. The terms related to $\mathcal{L}_{B}^{(2)}$ and $\mathcal{L}_{C}^{(2)}$ represent those that contain the partial derivative of $\bar{g}_{\mu\nu}$ and $\bar{A}^\mu$, respectively. They can naturally be written in the form of the partial derivative of the background field multiplied by another quantity, as described in Eq. (\ref{structure of S2}). (The representation here is only a rough indication. Accurate representation requires providing specific indices and the number of derivative operators of $\partial \bar{g}$ and $\partial \bar{A}$.) In $S^{(2)}$, the assignment of terms involving both the partial derivative of $\bar{g}_{\mu\nu}$ and $\bar{A}^\mu$ is not unique. Such terms can be freely allocated to either $\mathcal{L}_{B}^{(2)}$ or $\mathcal{L}_{C}^{(2)}$ terms, and this arbitrariness does not affect the reasoning in this article.

When we take the Minkowski spacetime solution (\ref{vector-tensor theory background solution in Minkowski spacrtime}) for the background fields, we observe that $S^{(2)}$ becomes
\begin{eqnarray}
	\label{S2 flat}
	S^{(2)}_{flat}\left[\eta_{\mu\nu},A^{\mu},h_{\mu\nu},B^{\mu}\right]
	\coloneqq S^{(2)}\big|_{\bar{g}_{\mu\nu}=\eta_{\mu\nu},\bar{A}^{\mu}=A^{\mu}}
	=
	\int d^4 x \sqrt{-\eta}
	\mathcal{L}_{A,flat}^{(2)},
\end{eqnarray}
where $\eta$ is the determinant of $\eta_{\mu\nu}$ and
\begin{eqnarray}
	\label{LAflat is}
	\mathcal{L}_{A,flat}^{(2)}\left[\eta_{\mu\nu},A^{\mu},h_{\mu\nu},B^{\mu}\right]
	\coloneqq
	\mathcal{L}_{A}^{(2)}\left[\bar{g}_{\mu\nu},\bar{A}^{\mu},h_{\mu\nu},B^{\mu}\right]\big|_{\bar{g}_{\mu\nu}=\eta_{\mu\nu},\bar{A}^{\mu}=A^{\mu}}.
\end{eqnarray}
In the following text, we will prove that
\begin{eqnarray}
	\label{M1=deltaS flat/delta h}
	\mathcal{M}^{(1)}_{\mu\nu}\left[\eta_{\mu\nu},A^{\mu},h_{\mu\nu},B^{\mu}\right]
	\!&=&\!
	-\eta_{\mu\lambda}\eta_{\nu\rho}
	\frac{\delta S_{flat}^{(2)}}{\delta h_{\lambda\rho}}
	=-\frac{\delta S_{flat}^{(2)}}{\delta h^{\mu\nu}},
	\\
	\label{N1=deltaS flat/delta B}
	\mathcal{N}^{(1)}_{\mu}\left[\eta_{\mu\nu},A^{\mu},h_{\mu\nu},B^{\mu}\right]
	\!&=&\!
	\frac{\delta S_{flat}^{(2)}}{\delta B^{\mu}},
	\\
	\label{M2=deltaS flat/delta g}
	\left\langle\mathcal{M}^{(2)}_{\mu\nu}\left[\eta_{\mu\nu},A^{\mu},h_{\mu\nu},B^{\mu}\right]\right\rangle
	\!&=&\!
	-\eta_{\mu\lambda}\eta_{\nu\rho}
	\left\langle\frac{\delta S_{flat}^{(2)}}{\delta \eta_{\lambda\rho}}\right\rangle
	=\left\langle\frac{\delta S_{flat}^{(2)}}{\delta \eta^{\mu\nu}}\right\rangle,
	\\
	\label{N2=deltaS flat/delta A}
	\left\langle\mathcal{N}^{(2)}_{\mu}\left[\eta_{\mu\nu},A^{\mu},h_{\mu\nu},B^{\mu}\right]\right\rangle
	\!&=&\!
	\left\langle\frac{\delta S_{flat}^{(2)}}{\delta A^{\mu}}\right\rangle.
\end{eqnarray}
Here, varying the action with respect to $\eta_{\mu\nu}$ and $A^{\mu}$ means formally considering $\eta_{\mu\nu}$ and $A^{\mu}$ as variables during the variation, and then substituting their actual values after obtaining the field equations. This is also the reason why we keep $\sqrt{-\eta}$ in Eq. (\ref{S2 flat}) instead of directly taking 1. If these relationships hold, it means that only by knowing $S^{(2)}_{flat}$ can we derive the Isaacson picture in asymptotic Minkowski spacetime far from the source.

Since Eqs. (\ref{M1=deltaS flat/delta h}) and (\ref{N1=deltaS flat/delta B}) do not involve the variation of the action with respect to the background fields, it is easy to prove that Eqs. (\ref{M1=deltaS flat/delta h}) and (\ref{N1=deltaS flat/delta B}) are true from Eqs. (\ref{M1=deltaS/delta h}) and (\ref{N1=deltaS/delta B}). To prove that Eqs. (\ref{M2=deltaS flat/delta g}) and (\ref{N2=deltaS flat/delta A}) are true, we note that only the terms related to $\mathcal{L}_{A}^{(2)}$ in Eq. (\ref{structure of S2}) do not include the derivative of the background fields. Thus, we have
\begin{eqnarray}
	\label{deltaS flat/delta eta=deltaS LA/delta g}
	\frac{\delta S_{flat}^{(2)}}{\delta \eta_{\mu\nu}}
	\!&=&\!
	\left(
	\frac{1}{\sqrt{-\bar{g}}}
	\frac{\delta \int d^4 x \sqrt{-\bar{g}}\mathcal{L}_{A}^{(2)}}{\delta \bar{g}_{\mu\nu}}
	\right)
	\Bigg|_{\bar{g}_{\mu\nu}=\eta_{\mu\nu},\bar{A}^{\mu}=A^{\mu}},
	\\
	\label{deltaS flat/delta A=deltaS LA/delta A}
	\frac{\delta S_{flat}^{(2)}}{\delta A^{\mu}}
	\!&=&\!
	\left(
	\frac{1}{\sqrt{-\bar{g}}}
	\frac{\delta \int d^4 x \sqrt{-\bar{g}}\mathcal{L}_{A}^{(2)}}{\delta \bar{A}^{\mu}}
	\right)
	\Bigg|_{\bar{g}_{\mu\nu}=\eta_{\mu\nu},\bar{A}^{\mu}=A^{\mu}}.
\end{eqnarray}
It can be seen from Eqs. (\ref{M2=deltaS/delta g}) and (\ref{N2=deltaS/delta A}) that to prove Eqs. (\ref{M2=deltaS flat/delta g}) and (\ref{N2=deltaS flat/delta A}), it is sufficient to prove that the following relationships are true:
\begin{eqnarray}
	\label{deltaS LB+LC/delta g avg=0}
	\left\langle
	\frac{1}{\sqrt{-\bar{g}}}
	\frac{\delta \int d^4 x \sqrt{-\bar{g}}\left[
		\left(\partial \bar{g}\right)\mathcal{L}_{B}^{(2)}
		+\left(\partial \bar{A}\right)\mathcal{L}_{C}^{(2)}
		\right]}{\delta \bar{g}_{\mu\nu}}
	\right\rangle
	\Bigg|_{\bar{g}_{\mu\nu}=\eta_{\mu\nu},\bar{A}^{\mu}=A^{\mu}}
	\!&=&\!
	0,
	\\
	\label{deltaS LB+LC/delta A avg=0}
	\left\langle
	\frac{1}{\sqrt{-\bar{g}}}
	\frac{\delta \int d^4 x \sqrt{-\bar{g}}\left[
		\left(\partial \bar{g}\right)\mathcal{L}_{B}^{(2)}
		+\left(\partial \bar{A}\right)\mathcal{L}_{C}^{(2)}
		\right]}{\delta \bar{A}^{\mu}}
	\right\rangle
	\Bigg|_{\bar{g}_{\mu\nu}=\eta_{\mu\nu},\bar{A}^{\mu}=A^{\mu}}
	\!&=&\!
	0.
\end{eqnarray}
Here, the same rough representation as in Eq. (\ref{structure of S2}) is used. This representation does not affect our proof.

Actually, it can be proven that 
\begin{eqnarray}
	\label{deltaS LB/delta g avg=0}
	\left\langle
	\frac{\delta \int d^4 x \sqrt{-\bar{g}}
		\left(\partial \bar{g}\right)\mathcal{L}_{B}^{(2)}}{\delta \bar{g}_{\mu\nu}}
	\right\rangle
	\Bigg|_{\bar{g}_{\mu\nu}=\eta_{\mu\nu},\bar{A}^{\mu}=A^{\mu}}
	\!&=&\!
	0,
	\\
	\label{deltaS LC/delta g avg=0}
	\left\langle
	\frac{\delta \int d^4 x \sqrt{-\bar{g}}
		\left(\partial \bar{A}\right)\mathcal{L}_{C}^{(2)}}{\delta \bar{g}_{\mu\nu}}
	\right\rangle
	\Bigg|_{\bar{g}_{\mu\nu}=\eta_{\mu\nu},\bar{A}^{\mu}=A^{\mu}}
	\!&=&\!
	0,
	\\
	\label{deltaS LB/delta A avg=0}
	\left\langle
	\frac{\delta \int d^4 x \sqrt{-\bar{g}}
		\left(\partial \bar{g}\right)\mathcal{L}_{B}^{(2)}
	}{\delta \bar{A}^{\mu}}
	\right\rangle
	\Bigg|_{\bar{g}_{\mu\nu}=\eta_{\mu\nu},\bar{A}^{\mu}=A^{\mu}}
	\!&=&\!
	0,
	\\
	\label{deltaS LC/delta A avg=0}
	\left\langle
	\frac{\delta \int d^4 x \sqrt{-\bar{g}}
		\left(\partial \bar{A}\right)\mathcal{L}_{C}^{(2)}
	}{\delta \bar{A}^{\mu}}
	\right\rangle
	\Bigg|_{\bar{g}_{\mu\nu}=\eta_{\mu\nu},\bar{A}^{\mu}=A^{\mu}}
	\!&=&\!
	0.
\end{eqnarray}
It is easy to see that as long as the above equations hold, Eqs. (\ref{deltaS LB+LC/delta g avg=0}) and (\ref{deltaS LB+LC/delta A avg=0}) are true. Now, we prove Eq. (\ref{deltaS LB/delta g avg=0}), and the proofs for other equations are entirely similar, so we do not elaborate further. Note that 
\begin{eqnarray}
	\label{deltaS LB}
	\delta \int d^4 x \sqrt{-\bar{g}}
	\left(\partial \bar{g}\right)\mathcal{L}_{B}^{(2)}
	\!&=&\!\int d^4 x 
	\left(\partial \bar{g}\right)\mathcal{L}_{B}^{(2)}\delta \sqrt{-\bar{g}}
	+\int d^4 x \sqrt{-\bar{g}}
	\left(\partial \bar{g}\right)\delta \mathcal{L}_{B}^{(2)}
	\nonumber \\
	\!&+&\!\int d^4 x \sqrt{-\bar{g}}
	\mathcal{L}_{B}^{(2)}\left(\partial \delta \bar{g}\right),
\end{eqnarray}
in the Minkowski background, only $\int d^4 x \sqrt{-\bar{g}}
\mathcal{L}_{B}^{(2)}\left(\partial \delta \bar{g}\right)$ is not zero. Its variation with respect to the background metric $\bar{g}_{\mu\nu}$ gives
\begin{eqnarray}
	\label{deltaS LB 3/delta g}
	\pm \left(\partial\right)\left(\sqrt{-\bar{g}}\mathcal{L}_{B}^{(2)}\right).
\end{eqnarray}
Here, the sign is related to the number of partial derivatives of the background metric in $\left(\partial \bar{g}\right)$. Therefore, to prove Eq. (\ref{deltaS LB/delta g avg=0}), it is only necessary to prove that
\begin{eqnarray}
	\label{deltaS LB 3/delta g avg=0}
	\left\langle \left(\partial\right)\left(\mathcal{L}_{B}^{(2)}\right) \right\rangle=0.
\end{eqnarray}
In the previous rough representation, we omitted the indices and the number of derivative operators in $\left(\partial\right)$. When we write back to a precise expression, Eq. (\ref{deltaS LB 3/delta g avg=0}) can always be divided into two cases. The first case is that the index of at least one partial derivative in $\left(\partial\right)$ is the same as an index in $\mathcal{L}_{B}^{(2)}$. In this case, $\left(\partial\right)\left(\mathcal{L}_{B}^{(2)}\right)$ can always be expressed in the form of a tensor divergence $\partial_{\lambda}X^{\lambda\mu\nu}$. Therefore, its average is zero. The second case is that all indices in $\left(\partial\right)$ are different from 
$\mathcal{L}_{B}^{(2)}$. In such a situation, $\left(\partial\right)\left(\mathcal{L}_{B}^{(2)}\right)$ can always be expressed as $\partial^{\mu}X^{\nu}$ or $\partial^{\nu}X^{\mu}$. As long as we define the unit vectors in the Minkowski background:
\begin{eqnarray}
	\label{et,ex,ey,ez}
	e_{t\mu}=\left(1,0,0,0\right),~		e_{x\mu}=\left(0,1,0,0\right),~	
	e_{y\mu}=\left(0,0,1,0\right),~		e_{x\mu}=\left(0,0,0,1\right),
\end{eqnarray}
there are
\begin{eqnarray}
	\label{partial X}
	\partial^{0}X^{\nu}\!&=&\!\partial^{\mu}\left(e_{t\mu}X^{\nu}\right),~
	\partial^{1}X^{\nu}=\partial^{\mu}\left(e_{x\mu}X^{\nu}\right),
	\nonumber \\
	\partial^{2}X^{\nu}\!&=&\!\partial^{\mu}\left(e_{y\mu}X^{\nu}\right),~
	\partial^{3}X^{\nu}=\partial^{\mu}\left(e_{z\mu}X^{\nu}\right).
\end{eqnarray}
$\partial^{\nu}X^{\mu}$ can also be represented in a similar form. We have the form of tensor divergence again, so its average is zero. This completes the proof.

Finally, we examine whether integration by parts of $S^{(2)}_{flat}$ will affect the calculation of $\left\langle\mathcal{M}^{(2)}_{\mu\nu}\left[\eta_{\mu\nu},A^{\mu},h_{\mu\nu},B^{\mu}\right]\right\rangle$ and $\left\langle\mathcal{N}^{(2)}_{\mu}\left[\eta_{\mu\nu},A^{\mu},h_{\mu\nu},B^{\mu}\right]\right\rangle$. We first assume that $\mathcal{L}_{A}^{(2)}$ in Eq. (\ref{structure of S2}) can be represented in the following form: (Here, the indics are also omitted.)
\begin{eqnarray}
	\label{LA=X P Y}
	\mathcal{L}_{A}^{(2)}=\mathcal{X} \partial \mathcal{Y}.
\end{eqnarray}
Therefore,
\begin{eqnarray}
	\label{S2 flat X P Y}
	S^{(2)}_{flat}
	=
	\int d^4 x \sqrt{-\eta}
	\mathcal{X}_{flat} \partial \mathcal{Y}_{flat},
\end{eqnarray}
where
\begin{eqnarray}
	\label{X, Yflat are}
	\mathcal{X}_{flat}\left[\eta_{\mu\nu},A^{\mu},h_{\mu\nu},B^{\mu}\right]
	=
	\mathcal{X}\left[\bar{g}_{\mu\nu},\bar{A}^{\mu},h_{\mu\nu},B^{\mu}\right]\big|_{\bar{g}_{\mu\nu}=\eta_{\mu\nu},\bar{A}^{\mu}=A^{\mu}},
	\nonumber \\
	\mathcal{Y}_{flat}\left[\eta_{\mu\nu},A^{\mu},h_{\mu\nu},B^{\mu}\right]
	=
	\mathcal{Y}\left[\bar{g}_{\mu\nu},\bar{A}^{\mu},h_{\mu\nu},B^{\mu}\right]\big|_{\bar{g}_{\mu\nu}=\eta_{\mu\nu},\bar{A}^{\mu}=A^{\mu}}.
\end{eqnarray}
After integration by parts of Eq. (\ref{S2 flat X P Y}), we obtain a new action $\bar{S}^{(2)}_{flat}$:
\begin{eqnarray}
	\label{bar S2 flat X P Y}
	\bar{S}^{(2)}_{flat}\left[\eta_{\mu\nu},A^{\mu},h_{\mu\nu},B^{\mu}\right]
	=
	-\int d^4 x \sqrt{-\eta}
	\mathcal{Y}_{flat} \partial \mathcal{X}_{flat}.
\end{eqnarray}
Now, we prove that the following relationships are true:
\begin{eqnarray}
	\label{deltaS flat/delta g=delta barS flat/delta g}
	\left\langle\frac{\delta S_{flat}^{(2)}}{\delta \eta_{\mu\nu}}\right\rangle
	\!&=&\!
	\left\langle\frac{\delta \bar{S}_{flat}^{(2)}}{\delta \eta_{\mu\nu}}\right\rangle,
	\\
	\label{deltaS flat/delta A=delta barS flat/delta A}
	\left\langle\frac{\delta S_{flat}^{(2)}}{\delta A^{\mu}}\right\rangle
	\!&=&\!
	\left\langle\frac{\delta \bar{S}_{flat}^{(2)}}{\delta A^{\mu}}\right\rangle.
\end{eqnarray}

To prove them, we first note that the results obtained from varying the following two actions with respect to the background fields are the same:
\begin{eqnarray}
	\label{SA and SAbar}
	S_{A}
	\!&=&\!
	\int d^4 x \sqrt{-\bar{g}}
	\mathcal{X} \partial \mathcal{Y},
	\nonumber\\
	\bar{S}_{A}
	\!&=&\!
	-\int d^4 x \mathcal{Y} \partial \left(\sqrt{-\bar{g}}
	\mathcal{X}\right).
\end{eqnarray}
As long as we note that 
\begin{eqnarray}
	\label{S2 flat=SA flat, bar S2 flat= bar SA flat}
	S^{(2)}_{flat}
	=S_A\big|_{\bar{g}_{\mu\nu}=\eta_{\mu\nu},\bar{A}^{\mu}=A^{\mu}},~
	\bar{S}^{(2)}_{flat}
	=\bar{S}_A\big|_{\bar{g}_{\mu\nu}=\eta_{\mu\nu},\bar{A}^{\mu}=A^{\mu}},
\end{eqnarray}
combined with the proof for Eqs. (\ref{M1=deltaS flat/delta h})-(\ref{N2=deltaS flat/delta A}), we can prove Eqs. (\ref{deltaS flat/delta g=delta barS flat/delta g}) and (\ref{deltaS flat/delta A=delta barS flat/delta A}). 

It should be noted that although Eqs. (\ref{deltaS flat/delta g=delta barS flat/delta g}) and (\ref{deltaS flat/delta A=delta barS flat/delta A}) are true, we generally have
\begin{eqnarray}
	\label{deltaS flat/delta g=delta barS flat/delta g not avg}
	\frac{\delta S_{flat}^{(2)}}{\delta \eta_{\mu\nu}}
	\!&\neq&\!
	\frac{\delta \bar{S}_{flat}^{(2)}}{\delta \eta_{\mu\nu}},
	\\
	\label{deltaS flat/delta A=delta barS flat/delta A not avg}
	\frac{\delta S_{flat}^{(2)}}{\delta A^{\mu}}
	\!&\neq&\!
	\frac{\delta \bar{S}_{flat}^{(2)}}{\delta A^{\mu}}.
\end{eqnarray}
The above proof allows us to derive the Isaacson picture far from the source in asymptotic Minkowski spacetime from $S^{(2)}_{flat}$ based on the difference in integration by parts.

For the results in this section, another proof can be found in Ref. \cite{Lavinia Heisenberg1}. In Appendix \ref{app: A}, we provide an example of using $S^{(2)}_{flat}$ to derive the effective energy-momentum tensor of gravitational waves in general relativity.

\section{The effective energy-momentum tensor of gravitational waves in general relativity}
\label{app: A}
We calculate $t_{\mu\nu}$ and $\tilde{t}_{\mu\nu}$ specifically in general relativity. Expanding $\sqrt{-g}$ to the second-order with respect to the perturbation $h_{\mu\nu}$, we have
\begin{eqnarray}
	\label{Expand sqrt -g to second order}
	\sqrt{-g}
	=
	\sqrt{-\bar{g}}
	\left(
	1
	+\frac{1}{2}h
	-\frac{1}{4}h_{\mu\nu}h^{\mu\nu}
	+\frac{1}{8}h^2
	\right)
	+\mathcal{O}\left(h^3\right).
\end{eqnarray}
Here and later in this subsection, we use the background metric %$\bar{g}^{\mu\nu}$ and $\bar{g}_{\mu\nu}$ 
to raise and lower the indices of $h_{\mu\nu}$ on the right side of the equation, and let $h \coloneqq  \bar{g}^{\mu\nu}h_{\mu\nu}$. 

For $G^{(2)}_{\mu\nu}$, we can expand the Einstein tensor $G_{\mu\nu}$ and obtain
\begin{eqnarray}
	\label{G2=}
	G^{(2)}_{\mu\nu}
	\!&=&\!
	\frac{1}{2}R^{(0)}_{\lambda\rho}h^{\lambda\rho} h_{\mu\nu}
	-\bar{g}_{\mu\nu}R^{(0)}_{\lambda\rho}h_{\sigma}^{~\rho}h^{\sigma\lambda}
	+\frac{1}{2}\bar{g}_{\mu\nu}R^{(0)}_{\lambda\rho\sigma\delta}h^{\lambda\sigma}h^{\rho\delta}
	\nonumber \\
	\!&+&\!
	\frac{1}{4}\bar{\nabla}_{\mu}h^{\lambda\rho}\bar{\nabla}_{\nu}h_{\lambda\rho}
	+\frac{1}{2}h^{\lambda\rho}\bar{\nabla}_{\nu}\bar{\nabla}_{\mu}h_{\lambda\rho}
	+\frac{1}{4}\bar{\nabla}_{\mu}h_{\nu}^{~\lambda} \bar{\nabla}_{\lambda}h
	+\frac{1}{4}\bar{\nabla}_{\nu}h_{\mu}^{~\lambda} \bar{\nabla}_{\lambda}h
	\nonumber \\
	\!&-&\!
	\frac{1}{4}\bar{\nabla}_{\lambda}h \bar{\nabla}^{\lambda} h_{\mu\nu}
	-\frac{1}{2}\bar{\nabla}_{\mu}h_{\nu}^{~\lambda} \bar{\nabla}_{\rho}h_{\lambda}^{~\rho}
	-\frac{1}{2}\bar{\nabla}_{\nu}h_{\mu}^{~\lambda} \bar{\nabla}_{\rho}h_{\lambda}^{~\rho}
	+\frac{1}{2}\bar{\nabla}^{\lambda}h_{\mu\nu}\bar{\nabla}_{\rho}h_{\lambda}^{~\rho}
	\nonumber \\
	\!&-&\!
	\frac{1}{2}h^{\lambda\rho}\bar{\nabla}_{\rho}\bar{\nabla}_{\mu}h_{\nu\lambda}
	-\frac{1}{2}h^{\lambda\rho}\bar{\nabla}_{\rho}\bar{\nabla}_{\nu}h_{\mu\lambda}
	+\frac{1}{2}h^{\lambda\rho}\bar{\nabla}_{\rho}\bar{\nabla}_{\lambda}h_{\mu\nu}
	-\frac{1}{2}h_{\mu\nu}\bar{\nabla}_{\rho}\bar{\nabla}_{\lambda}h^{\lambda\rho}
	\nonumber \\
	\!&-&\!
	\frac{1}{2}\bar{g}_{\mu\nu}h^{\lambda\rho}\bar{\nabla}_{\rho}\bar{\nabla}_{\lambda}h
	+\frac{1}{2}h_{\mu\nu}\bar{\nabla}_{\lambda}\bar{\nabla}^{\lambda}h
	-\frac{1}{2}\bar{\nabla}_{\lambda}h_{\nu\rho}\bar{\nabla}^{\rho}h_{\mu}^{~\lambda}
	+\frac{1}{2}\bar{\nabla}_{\rho}h_{\nu\lambda}\bar{\nabla}^{\rho}h_{\mu}^{~\lambda}
	\nonumber \\
	\!&+&\!
	\frac{1}{8}\bar{g}_{\mu\nu}\bar{\nabla}_{\lambda}h\bar{\nabla}^{\lambda}h
	+\frac{1}{2}\bar{g}_{\mu\nu}\bar{\nabla}_{\lambda}h^{\lambda\rho}\bar{\nabla}_{\sigma}h_{\rho}^{~\sigma}
	-\frac{1}{2}\bar{g}_{\mu\nu}\bar{\nabla}^{\rho}h\bar{\nabla}_{\lambda}h_{\rho}^{~\lambda}
	+\bar{g}_{\mu\nu}h^{\lambda\rho}\bar{\nabla}_{\sigma}\bar{\nabla}_{\rho}h_{\lambda}^{~\sigma}
	\nonumber \\
	\!&-&\!
	\frac{1}{2}\bar{g}_{\mu\nu}h^{\lambda\rho}\bar{\nabla}_{\sigma}\bar{\nabla}^{\sigma}h_{\lambda\rho}
	+\frac{1}{4}\bar{g}_{\mu\nu}\bar{\nabla}_{\rho}h_{\lambda\sigma}\bar{\nabla}^{\sigma}h^{\lambda\rho}
	-\frac{3}{8}\bar{g}_{\mu\nu}\bar{\nabla}_{\sigma}h_{\lambda\rho}\bar{\nabla}^{\sigma}h^{\lambda\rho}.
\end{eqnarray}
Here, the indices on the right side of the equation are still raised or lowered by $\bar{g}^{\mu\nu}$ and $\bar{g}_{\mu\nu}$, while $\bar{\nabla}_{\mu}$ corresponds to the covariant differentiation with respect to the background metric $\bar{g}_{\mu\nu}$. By combining the above equation with Eq. (\ref{effective energy-momentum tensor of gravitational waves in GR deltaS/delta g}), we can provide the expression for the effective energy-momentum tensor of gravitational waves $t_{\mu\nu}$ in the Isaacson picture.

Now, we employ the perturbation action method. For the action (\ref{E-H action in vacuum}), we have
\begin{eqnarray}
	\label{E-H action in vacuum S0}
	S^{(0)}\!&=&\! \frac{1}{16\pi}\int d^{4}x \sqrt{-\bar{g}} R^{(0)},
	\\
	\label{E-H action in vacuum S1}
	S^{(1)}\!&=&\! \frac{1}{16\pi}\int d^{4}x \sqrt{-\bar{g}}
	\left[
	-h^{\mu\nu}R^{(0)}_{\mu\nu}
	+\frac{1}{2}hR^{(0)}
	+\bar{\nabla}_{\nu}\bar{\nabla}_{\mu}h^{\mu\nu}
	-\bar{\nabla}_{\nu}\bar{\nabla}^{\nu}h
	\right],
	\\
	\label{E-H action in vacuum S2}
	S^{(2)}\!&=&\! \frac{1}{16\pi}\int d^{4}x \sqrt{-\bar{g}}
	\left[
	2h_{\mu}^{~\lambda}h^{\mu\nu}R^{(0)}_{\nu\lambda}
	-\frac{1}{2}hh^{\mu\nu}R^{(0)}_{\mu\nu}
	-\frac{1}{4}h_{\mu\nu}h^{\mu\nu}R^{(0)}
	\right.
	\nonumber \\
	\!&+&\!
	\frac{1}{8}h^2 R^{(0)}
	-h^{\mu\nu}h^{\lambda\rho}R^{(0)}_{\mu\lambda\nu\rho}
	+h^{\mu\nu}\bar{\nabla}_{\nu}\bar{\nabla}_{\mu}h
	-\frac{1}{4}\bar{\nabla}_{\mu}h\bar{\nabla}^{\mu}h
	\nonumber \\
	\!&-&\!
	\bar{\nabla}_{\mu}h^{\mu\nu}\bar{\nabla}_{\lambda}h_{\nu}^{~\lambda}
	+\bar{\nabla}^{\nu}h\bar{\nabla}_{\mu}h_{\nu}^{~\mu}
	-2h^{\mu\nu}\bar{\nabla}_{\lambda}\bar{\nabla}_{\nu}h_{\mu}^{~\lambda}
	+\frac{1}{2}h\bar{\nabla}_{\mu}\bar{\nabla}_{\nu}h^{\nu\mu}
	\nonumber \\
	\!&+&\!
	\left.
	h^{\mu\nu}\bar{\nabla}_{\lambda}\bar{\nabla}^{\lambda}h_{\mu\nu}
	-\frac{1}{2}h\bar{\nabla}_{\mu}\bar{\nabla}^{\mu}h
	-\frac{1}{2}\bar{\nabla}_{\nu}h_{\mu\lambda}\bar{\nabla}^{\lambda}h^{\mu\nu}
	+\frac{3}{4}\bar{\nabla}_{\lambda}h_{\mu\nu}\bar{\nabla}^{\lambda}h^{\mu\nu}
	\right].
\end{eqnarray}
Thus, using Eqs (\ref{relationship between G and perturbation action 0 order write G})-(\ref{relationship between G and perturbation action 2 order write G}), it can be calculated that
\begin{eqnarray}
	\label{G0 raise}
	G^{(0)\mu\nu}
	\!&=&\!
	R^{(0)\mu\nu}
	-\frac{1}{2}\bar{g}^{\mu\nu}R^{(0)},
	\\
	\label{G1 raise}
	G^{(1)\mu\nu}
	\!&=&\!
	\frac{1}{2}\bar{g}^{\mu\nu}h^{\lambda\rho}R^{(0)}_{\lambda\rho}
	-\frac{1}{2}h^{\nu\lambda}R^{(0)\mu}_{\lambda}
	-\frac{1}{2}h^{\mu\lambda}R^{(0)\nu}_{\lambda}
	\nonumber \\
	\!&+&\!
	\frac{1}{2}h^{\mu\nu}R^{(0)}
	-h^{\lambda\rho}R^{(0)\mu~\nu}_{~~~~\lambda~\rho}
	-\frac{1}{2}\bar{\nabla}_{\lambda}\bar{\nabla}^{\lambda}h^{\mu\nu}
	-\frac{1}{2}\bar{g}^{\mu\nu}\bar{\nabla}_{\lambda}\bar{\nabla}_{\rho}h^{\lambda\rho}
	\nonumber \\
	\!&+&\!
	\frac{1}{2}\bar{g}^{\mu\nu}\bar{\nabla}_{\lambda}\bar{\nabla}^{\lambda}h
	+\frac{1}{2}\bar{\nabla}^{\mu}\bar{\nabla}_{\lambda}h^{\nu\lambda}
	+\frac{1}{2}\bar{\nabla}^{\nu}\bar{\nabla}_{\lambda}h^{\mu\lambda}
	-\frac{1}{2}\bar{\nabla}^{\mu}\bar{\nabla}^{\nu}h,
	\\
	\label{G2 raise}
	G^{(2)\mu\nu}
	\!&=&\!
	h^{\nu\lambda}h_{\lambda}^{~\rho}R^{(0)\mu}_{~\rho}
	+h^{\mu\lambda}h_{\lambda}^{~\rho}R^{(0)\nu}_{~\rho}
	+h^{\mu\lambda}h^{\nu\rho}R^{(0)}_{\lambda\rho}
	-\frac{1}{2}h^{\mu\nu}h^{\lambda\rho}R^{(0)}_{\lambda\rho}
	\nonumber \\
	\!&-&\!
	\bar{g}^{\mu\nu}h_{\lambda}^{~\sigma}h^{\lambda\rho}R^{(0)}_{\rho\sigma}
	-\frac{1}{2}h^{\mu\lambda}h^{\nu}_{~\lambda}R^{(0)}
	+\frac{1}{2}\bar{g}^{\mu\nu}h^{\lambda\rho}h^{\sigma\delta}R^{(0)}_{\lambda\sigma\rho\delta}
	\nonumber \\
	\!&+&\!
	\frac{1}{4}\bar{\nabla}^{\mu}h^{\lambda\rho}\bar{\nabla}^{\nu}h_{\lambda\rho}
	+\frac{1}{2}h^{\lambda\rho}\bar{\nabla}^{\nu}\bar{\nabla}^{\mu}h_{\lambda\rho}
	+\frac{1}{4}\bar{\nabla}^{\mu}h^{\nu\lambda}\bar{\nabla}_{\lambda}h
	+\frac{1}{4}\bar{\nabla}^{\nu}h^{\mu\lambda}\bar{\nabla}_{\lambda}h
	\nonumber \\
	\!&+&\!
	\frac{1}{2}h^{\nu\lambda}\bar{\nabla}_{\lambda}\bar{\nabla}^{\mu}h
	+\frac{1}{2}h^{\mu\lambda}\bar{\nabla}_{\lambda}\bar{\nabla}^{\nu}h
	-\frac{1}{4}\bar{\nabla}_{\lambda}h\bar{\nabla}^{\lambda}h^{\mu\nu}
	-\frac{1}{2}\bar{\nabla}^{\mu}h^{\nu\lambda}\bar{\nabla}_{\rho}h_{\lambda}^{~\rho}
	\nonumber \\
	\!&-&\!
	\frac{1}{2}\bar{\nabla}^{\nu}h^{\mu\lambda}\bar{\nabla}_{\rho}h_{\lambda}^{~\rho}
	+\frac{1}{2}\bar{\nabla}^{\lambda}h^{\mu\nu}\bar{\nabla}_{\rho}h_{\lambda}^{~\rho}
	-\frac{1}{2}h^{\lambda\rho}\bar{\nabla}_{\rho}\bar{\nabla}^{\mu}h^{\nu}_{~\lambda}
	-\frac{1}{2}h^{\nu\lambda}\bar{\nabla}_{\rho}\bar{\nabla}^{\mu}h_{\lambda}^{~\rho}
	\nonumber \\
	\!&-&\!
	\frac{1}{2}h^{\lambda\rho}\bar{\nabla}_{\rho}\bar{\nabla}^{\nu}h^{\mu}_{~\lambda}
	-\frac{1}{2}h^{\mu\lambda}\bar{\nabla}_{\rho}\bar{\nabla}^{\nu}h_{\lambda}^{~\rho}
	+\frac{1}{2}h^{\lambda\rho}\bar{\nabla}_{\rho}\bar{\nabla}_{\lambda}h^{\mu\nu}
	-\frac{1}{2}h^{\nu\lambda}\bar{\nabla}_{\rho}\bar{\nabla}_{\lambda}h^{\mu\rho}
	\nonumber \\
	\!&-&\!
	\frac{1}{2}h^{\mu\lambda}\bar{\nabla}_{\rho}\bar{\nabla}_{\lambda}h^{\nu\rho}
	+\frac{1}{2}h^{\mu\nu}\bar{\nabla}_{\rho}\bar{\nabla}_{\lambda}h^{\lambda\rho}
	-\frac{1}{2}\bar{g}^{\mu\nu}h^{\lambda\rho}\bar{\nabla}_{\rho}\bar{\nabla}_{\lambda}h
	+\frac{1}{2}h^{\nu\lambda}\bar{\nabla}_{\rho}\bar{\nabla}^{\rho}h^{\mu}_{~\lambda}
	\nonumber \\
	\!&+&\!
	\frac{1}{2}h^{\mu\lambda}\bar{\nabla}_{\rho}\bar{\nabla}^{\rho}h^{\nu}_{~\lambda}
	-\frac{1}{2}h^{\mu\nu}\bar{\nabla}_{\rho}\bar{\nabla}^{\rho}h
	-\frac{1}{2}\bar{\nabla}_{\lambda}h^{\nu}_{~\rho}\bar{\nabla}^{\rho}h^{\mu\lambda}
	+\frac{1}{2}\bar{\nabla}_{\rho}h^{\nu}_{~\lambda}\bar{\nabla}^{\rho}h^{\mu\lambda}
	\nonumber \\
	\!&+&\!
	\frac{1}{8}\bar{g}^{\mu\nu}\bar{\nabla}_{\rho}h\bar{\nabla}^{\rho}h
	+\frac{1}{2}\bar{g}^{\mu\nu}\bar{\nabla}_{\lambda}h^{\lambda\rho}\bar{\nabla}_{\sigma}h_{\rho}^{~\sigma}
	-\frac{1}{2}\bar{g}^{\mu\nu}\bar{\nabla}^{\rho}h\bar{\nabla}_{\lambda}h_{\rho}^{~\lambda}
	+\bar{g}^{\mu\nu}h^{\lambda\rho}\bar{\nabla}_{\sigma}\bar{\nabla}_{\rho}h_{\lambda}^{~\sigma}
	\nonumber \\
	\!&-&\!
	\frac{1}{2}\bar{g}^{\mu\nu}h^{\lambda\rho}\bar{\nabla}_{\sigma}\bar{\nabla}^{\sigma}h_{\lambda\rho}
	+\frac{1}{4}\bar{g}^{\mu\nu}\bar{\nabla}_{\rho}h_{\lambda\sigma}\bar{\nabla}^{\sigma}h^{\lambda\rho}
	-\frac{3}{8}\bar{g}^{\mu\nu}\bar{\nabla}_{\sigma}h_{\lambda\rho}\bar{\nabla}^{\sigma}h^{\lambda\rho}.
\end{eqnarray}
By substituting the above equations into Eq. (\ref{G _munu with 2 order}), we can obtain the expression for $G^{(2)}_{\mu\nu}$ which is equal to Eq. (\ref{G2=}). Therefore, we can use the perturbation action method to obtain the effective energy-momentum tensor $t_{\mu\nu}$ of gravitational waves. Both methods yield consistent results.

For the effective energy-momentum tensor $\tilde{t}_{\mu\nu}$ of gravitational waves defined in Refs. \cite{Leo C. Stein,Lavinia Heisenberg1}, using Eq. (\ref{effective energy-momentum tensor of gravitational waves in some paper}) and (\ref{E-H action in vacuum S2}), we find
\begin{eqnarray}
	\label{tilde t =}
	\tilde{t}_{\mu\nu}
	\!&=&\!
	\left\langle
	\frac{h_{\lambda\rho}h^{\lambda\rho}R^{(0)}_{\mu\nu}}{32\pi}
	-\frac{h^2 R^{(0)}_{\mu\nu}}{64\pi}
	+\frac{h_{\nu}^{~\lambda}hR^{(0)}_{\mu\lambda}}{16\pi}
	-\frac{h_{\nu}^{~\lambda}h_{\lambda}^{~\rho}R^{(0)}_{\mu\rho}}{8\pi}
	\right.
	\nonumber \\
	\!&+&\!
	\frac{h_{\mu}^{~\lambda}hR^{(0)}_{\nu\lambda}}{16\pi}
	-\frac{h_{\mu}^{~\lambda}h_{\lambda}^{~\rho}R^{(0)}_{\nu\rho}}{8\pi}
	-\frac{h_{\mu}^{~\lambda}h_{\nu}^{~\rho}R^{(0)}_{\lambda\rho}}{8\pi}
	+\frac{h_{\mu\nu}h^{\lambda\rho}R^{(0)}_{\lambda\rho}}{16\pi}
	\nonumber \\
	\!&+&\!
	\frac{\bar{g}_{\mu\nu}h_{\lambda}^{~\sigma}h^{\lambda\rho}R^{(0)}_{\rho\sigma}}{8\pi}
	-\frac{\bar{g}_{\mu\nu}hh^{\rho\lambda}R^{(0)}_{\rho\lambda}}{32\pi}
	+\frac{h_{\mu}^{~\lambda}h_{\nu\lambda}R^{(0)}}{32\pi}
	-\frac{h_{\mu\nu}hR^{(0)}}{32\pi}
	\nonumber \\
	\!&-&\!
	\frac{\bar{g}_{\mu\nu}h_{\lambda\rho}h^{\lambda\rho}R^{(0)}}{64\pi}
	+\frac{\bar{g}_{\mu\nu}h^2 R^{(0)}}{128\pi}
	-\frac{\bar{g}_{\mu\nu}h^{\lambda\rho}h^{\sigma\delta}R^{(0))}_{\lambda\sigma\rho\delta}}{16\pi}
	\nonumber \\
	\!&-&\!
	\frac{\bar{\nabla}_{\mu}h^{\lambda\rho}\bar{\nabla}_{\nu}h_{\lambda\rho}}{32\pi}
	-\frac{h^{\lambda\rho}\bar{\nabla}_{\nu}\bar{\nabla}_{\mu}h_{\lambda\rho}}{16\pi}
	+\frac{h\bar{\nabla}_{\nu}\bar{\nabla}_{\mu}h}{32\pi}
	-\frac{\bar{\nabla}_{\mu}h_{\nu}^{~\lambda}\bar{\nabla}_{\lambda}h}{32\pi}
	\nonumber \\
	\!&-&\!
	\frac{\bar{\nabla}_{\nu}h_{\mu}^{~\lambda}\bar{\nabla}_{\lambda}h}{32\pi}
	-\frac{h_{\nu}^{~\lambda}\bar{\nabla}_{\lambda}\bar{\nabla}_{\mu}h}{16\pi}
	-\frac{h_{\mu}^{~\lambda}\bar{\nabla}_{\lambda}\bar{\nabla}_{\nu}h}{16\pi}
	+\frac{\bar{\nabla}_{\lambda}h\bar{\nabla}^{\lambda}h_{\mu\nu}}{32\pi}
	\nonumber \\
	\!&+&\!
	\frac{\bar{\nabla}_{\mu}h_{\nu}^{~\lambda}\bar{\nabla}_{\rho}h_{\lambda}^{~\rho}}{16\pi}
	+\frac{\bar{\nabla}_{\nu}h_{\mu}^{~\lambda}\bar{\nabla}_{\rho}h_{\lambda}^{~\rho}}{16\pi}
	-\frac{\bar{\nabla}^{\lambda}h_{\mu\nu}\bar{\nabla}_{\rho}h_{\lambda}^{~\rho}}{16\pi}
	+\frac{h^{\lambda\rho}\bar{\nabla}_{\rho}\bar{\nabla}_{\mu}h_{\nu\lambda}}{16\pi}
	\nonumber \\
	\!&-&\!
	\frac{h\bar{\nabla}_{\rho}\bar{\nabla}_{\mu}h_{\nu}^{~\rho}}{32\pi}
	+\frac{h_{\nu}^{~\lambda}\bar{\nabla}_{\rho}\bar{\nabla}_{\mu}h_{\lambda}^{~\rho}}{16\pi}
	+\frac{h^{\lambda\rho}\bar{\nabla}_{\rho}\bar{\nabla}_{\nu}h_{\mu\lambda}}{16\pi}
	-\frac{h\bar{\nabla}_{\rho}\bar{\nabla}_{\nu}h_{\mu}^{~\rho}}{32\pi}
	\nonumber \\
	\!&+&\!
	\frac{h_{\mu}^{~\lambda}\bar{\nabla}_{\rho}\bar{\nabla}_{\nu}h_{\lambda}^{~\rho}}{16\pi}
	-\frac{h^{\lambda\rho}\bar{\nabla}_{\rho}\bar{\nabla}_{\lambda}h_{\mu\nu}}{16\pi}
	+\frac{h_{\nu}^{~\lambda}\bar{\nabla}_{\rho}\bar{\nabla}_{\lambda}h_{\mu}^{~\rho}}{16\pi}
	+\frac{h_{\mu}^{~\lambda}\bar{\nabla}_{\rho}\bar{\nabla}_{\lambda}h_{\nu}^{~\rho}}{16\pi}
	\nonumber \\
	\!&-&\!
	\frac{h_{\mu\nu}\bar{\nabla}_{\rho}\bar{\nabla}_{\lambda}h^{\lambda\rho}}{16\pi}
	+\frac{\bar{g}_{\mu\nu}h^{\lambda\rho}\bar{\nabla}_{\rho}\bar{\nabla}_{\lambda}h}{16\pi}
	+\frac{h\bar{\nabla}_{\rho}\bar{\nabla}^{\rho}h_{\mu\nu}}{32\pi}
	-\frac{h_{\nu}^{~\lambda}\bar{\nabla}_{\rho}\bar{\nabla}^{\rho}h_{\mu\lambda}}{16\pi}
	\nonumber \\
	\!&-&\!
	\frac{h_{\mu}^{~\lambda}\bar{\nabla}_{\rho}\bar{\nabla}^{\rho}h_{\nu\lambda}}{16\pi}
	+\frac{h_{\mu\nu}\bar{\nabla}_{\rho}\bar{\nabla}^{\rho}h}{16\pi}
	+\frac{\bar{\nabla}_{\lambda}h_{\nu\rho}\bar{\nabla}^{\rho}h_{\mu}^{~\lambda}}{16\pi}
	-\frac{\bar{\nabla}_{\rho}h_{\nu\lambda}\bar{\nabla}^{\rho}h_{\mu}^{~\lambda}}{16\pi}
	\nonumber \\
	\!&-&\!
	\frac{\bar{g}_{\mu\nu}\bar{\nabla}_{\rho}h\bar{\nabla}^{\rho}h}{64\pi}
	-\frac{\bar{g}_{\mu\nu}\bar{\nabla}_{\lambda}h^{\lambda\rho}\bar{\nabla}_{\sigma}h_{\rho}^{~\sigma}}{16\pi}
	+\frac{\bar{g}_{\mu\nu}\bar{\nabla}^{\rho}h\bar{\nabla}_{\lambda}h_{\rho}^{~\lambda}}{16\pi}
	-\frac{\bar{g}_{\mu\nu}h^{\lambda\rho}\bar{\nabla}_{\sigma}\bar{\nabla}_{\rho}h_{\lambda}^{~\sigma}}{8\pi}
	\nonumber \\
	\!&+&\!
	\frac{\bar{g}_{\mu\nu}h\bar{\nabla}_{\sigma}\bar{\nabla}_{\rho}h^{\rho\sigma}}{32\pi}
	+\frac{\bar{g}_{\mu\nu}h^{\lambda\rho}\bar{\nabla}_{\sigma}\bar{\nabla}^{\sigma}h_{\lambda\rho}}{16\pi}
	-\frac{\bar{g}_{\mu\nu}g\bar{\nabla}_{\rho}\bar{\nabla}^{\rho}h}{32\pi}
	-\frac{\bar{g}_{\mu\nu}\bar{\nabla}_{\rho}h_{\lambda\sigma}\bar{\nabla}^{\sigma}h^{\lambda\rho}}{32\pi}
	\nonumber \\
	\!&+&\!
	\left.
	\frac{3\bar{g}_{\mu\nu}\bar{\nabla}_{\sigma}h_{\lambda\rho}\bar{\nabla}^{\sigma}h^{\lambda\rho}}{64\pi}
	\right\rangle.
\end{eqnarray}
It can be seen that $\tilde{t}_{\mu\nu}$ given by Eq. (\ref{tilde t =}) and $t_{\mu\nu}$ given by Eqs. (\ref{G2=}) and (\ref{effective energy-momentum tensor of gravitational waves in GR deltaS/delta g}) are different. 

When we consider the asymptotic Minkowski spacetime far from the source, and the gravitational waves are on-shell, general relativity requires only the transverse traceless spatial part $h^{TT}_{ij}$ of the perturbation $h_{\mu\nu}$ to be non-zero and satisfies
\begin{eqnarray}
	\label{GR asymptotic Minkowski spacetime far from the source and in the shell}
	\bar{g}_{\mu\nu}=\eta_{\mu\nu},\quad \Box h^{TT}_{ij}=0,
\end{eqnarray}
where $\Box$ is the d'Alembert operator and $h^{TT}_{ij}$ satisfies $\delta^{ij}h^{TT}_{ij}=\partial^{i}h^{TT}_{ij}=0$. At this point, we have
\begin{eqnarray}
	\label{tilde t = t = GR  asymptotic Minkowski spacetime far from the source and in the shell}
	\tilde{t}_{\mu\nu}
	=
	t_{\mu\nu}
	\!&=&\!
	\left\langle
	-\frac{\partial_{\mu}h_{TT}^{ij}\partial_{\nu}h^{TT}_{ij}}{32\pi}
	-\frac{h_{TT}^{ij}\partial_{\nu}\partial_{\mu}h^{TT}_{ij}}{16\pi}
	+\frac{h_{TT}^{ij}\partial_{j}\partial_{\mu}h^{TT}_{\nu i}}{16\pi}
	\right.
	\nonumber \\
	\!&+&\!
	\frac{h_{TT}^{ij}\partial_{j}\partial_{\nu}h^{TT}_{\mu i}}{16\pi}
	-\frac{h_{TT}^{ij}\partial_{j}\partial_{i}h^{TT}_{\mu\nu}}{16\pi}
	+\frac{\partial_{i}h^{TT}_{\nu j}\partial^{j}h_{\mu}^{TTi}}{16\pi}
	\nonumber \\
	\!&-&\!
	\left.
	\frac{\partial_{\rho}h^{TT}_{\nu i}\partial^{\rho}h_{\mu}^{TTi}}{16\pi}
	-\frac{\eta_{\mu\nu}\partial_{k}h^{TT}_{ij}\partial^{j}h_{TT}^{ik}}{32\pi}
	+\frac{3\eta_{\mu\nu}\partial_{\sigma}h^{TT}_{ij}\partial^{\sigma}h_{TT}^{ij}}{64\pi}
	\right\rangle.
\end{eqnarray}
Here, indices are raised by $\eta^{\mu\nu}$ and lowered by $\eta_{\mu\nu}$, respectively. The spatial part of $h^{TT}_{\mu\nu}$ is defined as $h^{TT}_{ij}$, while all other parts are $0$. For the sake of compactness, sometimes $TT$ is placed in the lower right corner of $h$, which does not cause confusion. As mentioned earlier, the two definitions of the effective energy-momentum tensor of gravitational waves yield the same result. 

It should also be pointed out that the properties of the averaging operation $\langle ... \rangle$ can further simplify Eq. ({\ref{tilde t = t = GR  asymptotic Minkowski spacetime far from the source and in the shell}}). The most useful properties of $\langle ... \rangle$ are \cite{Michele Maggiore,Leo C. Stein}:
\begin{itemize}
	\item [(1)] The average of terms containing an odd number of high-frequency quantities is $0$.
	\item [(2)] The average operation of the tensor divergence is $0$. (This is due to the small boundary term, which can be ignored.) For example, for any tensor $X^{\mu\nu}$, ${\left\langle \bar{\nabla}_{\mu} X^{\mu\nu} \right\rangle}=0$.
	\item [(3)] As a corollary to (2), integration by parts does not affect averaging operations. For example, for tensors $X^{\mu\nu}$ and $Y^{\mu}$, ${\left\langle Y^{\lambda} \bar{\nabla}_{\mu} X^{\mu\nu} \right\rangle}={-\left\langle X^{\mu\nu} \bar{\nabla}_{\mu} Y^{\lambda} \right\rangle}$.
\end{itemize}
Using property (3), Eq. ({\ref{tilde t = t = GR  asymptotic Minkowski spacetime far from the source and in the shell}}) can be rewritten as 
\begin{eqnarray}
	\label{tilde t = t = GR  asymptotic Minkowski spacetime far from the source and in the shell rewritten}
	\tilde{t}_{\mu\nu}
	=
	t_{\mu\nu}
	\!&=&\!
	\frac{1}{32\pi}
	\left\langle
	\partial_{\mu}h_{TT}^{ij}\partial_{\nu}h^{TT}_{ij}
	\right\rangle.
\end{eqnarray}
This is the standard expression for the effective energy-momentum tensor of gravitational waves in general relativity in textbooks \cite{Michele Maggiore}.

Now, we use general relativity as an example to demonstrate how to obtain the effective energy-momentum tensor of gravitational waves from $S^{(2)}_{flat}$. Using Eq. (\ref{E-H action in vacuum S2}), after integration by parts, $S^{(2)}_{flat}$ for general relativity is
\begin{eqnarray}
	\label{S2 flat in GR}
	S^{(2)}_{flat}\!&=&\! \frac{1}{64\pi}\int d^{4}x \sqrt{-{\eta}}
	\left[
	2h^{\mu\nu}\partial_{\nu}\partial_{\mu}h
	-h\Box h
	-2h^{\mu\nu}\partial_{\mu}\partial_{\lambda}h_{\nu}^{~\lambda}
	+h^{\mu\nu}\Box h_{\mu\nu}
	\right].
\end{eqnarray}
We write the above action as an explicit expression for $\eta^{\mu\nu}$, i.e.,
\begin{eqnarray}
	\label{S2 flat in GR write eta}
	S^{(2)}_{flat}\!&=&\! \frac{1}{64\pi}\int d^{4}x \sqrt{-{\eta}}
	\left[
	2  \eta^{\lambda\rho}\eta^{\mu\sigma}\eta^{\nu\gamma} h_{\sigma\gamma}\partial_{\mu}\partial_{\nu}h_{\lambda\rho}
	-\eta^{\mu\nu}\eta^{\lambda\rho}\eta^{\sigma\gamma}h_{\sigma\gamma}\partial_{\mu}\partial_{\nu}h_{\lambda\rho}
	\right.
	\nonumber \\
	\!&-&\!
	\left.
	2  \eta^{\lambda\rho} \eta^{\mu\sigma}\eta^{\nu\gamma}    h_{\sigma\gamma}\partial_{\mu}\partial_{\lambda}h_{\nu\rho}
	+\eta^{\lambda\rho} \eta^{\mu\sigma}\eta^{\nu\gamma} h_{\sigma\gamma}\partial_{\lambda}\partial_{\rho}h_{\mu\nu}
	\right].
\end{eqnarray}
From Eqs. (\ref{leadingorder effective energy-momentum tensor of gravitational waves in MGT}) and (\ref{M2=deltaS flat/delta g}), it can be inferred that the effective energy-momentum tensor of gravitational waves is
\begin{eqnarray}
	\label{t_munu S2flat}
	t_{\mu\nu}
	= 
	-2\left\langle\frac{\delta S_{flat}^{(2)}}{\delta \eta^{\mu\nu}}\right\rangle.
\end{eqnarray}
Thus, we have
\begin{eqnarray}
	\label{t_munu GR S2flat}
	t_{\mu\nu}
	\!&=&\! 
	-\frac{1}{32 \pi}
	\bigg\langle
	2h^{\lambda\rho}\partial_{\lambda}\partial_{\rho}h_{\mu\nu}
	+2h_{\nu\lambda}\partial_{\mu}\partial^{\lambda}h
	+2h_{\mu\lambda}\partial_{\nu}\partial^{\lambda}h
	\nonumber \\
	\!&-&\!
	h\partial_{\mu}\partial_{\nu}h
	-2h\Box h_{\mu\nu}
	-h^{\lambda\rho}\partial_{\lambda}\partial_{\mu}h_{\rho\nu}
	-h^{\lambda\rho}\partial_{\lambda}\partial_{\nu}h_{\rho\mu}
	\nonumber \\
	\!&-&\!
	h_{\nu\rho}\partial_{\mu}\partial^{\lambda}h^{\rho}_{~\lambda}
	-h_{\mu\rho}\partial_{\nu}\partial^{\lambda}h^{\rho}_{~\lambda}
	-h_{\rho\mu}\partial^{\rho}\partial^{\lambda}h_{\nu\lambda}
	-h_{\rho\nu}\partial^{\rho}\partial^{\lambda}h_{\mu\lambda}
	\nonumber \\
	\!&+&\!
	h^{\lambda\rho}\partial_{\mu}\partial_{\nu}h_{\lambda\rho}
	+h_{\nu\lambda}\Box h_{\mu}^{~\lambda}
	+h_{\mu\lambda}\Box h_{\nu}^{~\lambda}
	\nonumber \\
	\!&-&\!
	\frac{1}{2}\eta_{\mu\nu}
	\left(
	2h^{\lambda\rho}\partial_{\rho}\partial_{\lambda}h
	-h\Box h
	-2h^{\lambda\rho}\partial_{\lambda}\partial_{\sigma}h_{\rho}^{~\sigma}
	+h^{\lambda\rho}\Box h_{\lambda\rho}
	\right)
	\bigg\rangle.
\end{eqnarray}
Considering gravitational waves to be on-shell, we once again obtain 
\begin{eqnarray}
	t_{\mu\nu}
	= 
	\frac{1}{32\pi}
	\left\langle
	\partial_{\mu}h_{TT}^{ij}\partial_{\nu}h^{TT}_{ij}
	\right\rangle.
\end{eqnarray}

\section{Detailed construction of the second-order perturbation action}
\label{app: E}
When constructing the most general second-order perturbation action in Sec. \ref{sec: 4}, assumptions (1), (3), and (4) will limit the possible structures of the action. According to the viewpoint that gravity can be fully described geometrically, the second-order perturbation action in vector-tensor theory should be constructed entirely from the inherent quantities in the algebraic structures of a differential manifold and an additional vector field defined in the tangent space. In the (pseudo) Riemannian geometry considered in assumption (1), the only intrinsic quantities in its algebraic structures are: \normalsize{\textcircled{\scriptsize{1}}} the background metric $\eta_{\mu\nu}$ and the perturbation $h_{\mu\nu}$, which are derived from the inner product structure of the tangent space; \normalsize{\textcircled{\scriptsize{2}}} the four-dimensional Levi-Civita totally antisymmetric tensor $E^{\mu\nu\lambda\rho}$ $(E^{0123}=1)$, which arises from the exterior product structure of the tangent space; and \normalsize{\textcircled{\scriptsize{3}}} the partial derivative $\partial_{\mu}$, which is defined by the differential structure \cite{Y.Dong4}. Therefore, assumption (1) requires
\begin{itemize}
	\item Each term in the second-order perturbation action can be represented as a combination of $\eta_{\mu\nu}$, $\eta^{\mu\nu}$, $A^{\mu}$, $h_{\mu\nu}$, $B^{\mu}$, $E^{\mu\nu\lambda\rho}$, $\partial_{\mu}$, and the theoretical parameters.
\end{itemize}
It should be noted that due to \cite{landau}
\begin{eqnarray}
	\label{EE=delta}
	E^{\alpha\beta\gamma\sigma}E_{\mu\nu\lambda\rho}
	~=~-
	\begin{vmatrix}
		~\delta^{\alpha}_{\mu}~ & \delta^{\alpha}_{\nu}~ & \delta^{\alpha}_{\lambda}~ & \delta^{\alpha}_{\rho}~
		\\
		~\delta^{\beta}_{\mu}~ & \delta^{\beta}_{\nu}~ & \delta^{\beta}_{\lambda}~ & \delta^{\beta}_{\rho}~
		\\
		~\delta^{\gamma}_{\mu}~ & \delta^{\gamma}_{\nu}~ & \delta^{\gamma}_{\lambda}~ & \delta^{\gamma}_{\rho}~
		\\
		~\delta^{\sigma}_{\mu}~ & \delta^{\sigma}_{\nu}~ & \delta^{\sigma}_{\lambda}~ & \delta^{\sigma}_{\rho}~
	\end{vmatrix},
\end{eqnarray}
where $\delta^{\mu}_{\nu}$ is the Kronecker delta, each term in the second-order perturbation action will have at most one $E^{\mu\nu\lambda\rho}$. For assumption (3), being generally covariant requires that after performing a generalized coordinate transformation
\begin{eqnarray}
	\label{generalized coordinate transformation}
	x^\mu \rightarrow x^\mu+\xi^{\mu}(x), 
\end{eqnarray} 
where $\xi^{\mu}$ is an arbitrary function, the transformed action differs from the original action only by an integration by parts. If we require that the fields can always be written as 
\begin{eqnarray}
	\label{weak perturbation of fields}
	g_{\mu\nu}\!&=&\!\eta_{\mu\nu}+h_{\mu\nu},\quad \left|h_{\mu\nu}\right| \sim \alpha \ll 1,
	\nonumber \\
	\mathcal{A}^{\mu}\!&=&\!A^{\mu}+B^{\mu},\quad \left|\frac{B^{\mu}}{A^{\mu}}\right| \sim \alpha \ll 1,
\end{eqnarray}
before and after the transformation (\ref{generalized coordinate transformation}), then we should also have $\left|\partial_{\mu}\xi_{\nu}\right| \sim \alpha$ and 
\begin{eqnarray}
	\label{gauge transformation}
	h_{\mu\nu} \!&\rightarrow&\! h_{\mu\nu}-\partial_{\mu}\xi_{\nu}-\partial_{\nu}\xi_{\mu}
	+\mathcal{O}\left(\alpha^2\right), \nonumber\\
	B^{\mu} \!&\rightarrow&\!
	B^{\mu}+A^{\nu}\partial_{\nu}\xi^{\mu}
	+\mathcal{O}\left(\alpha^2\right).
\end{eqnarray} 
After expanding the action for the perturbations $S=S^{(0)}_{flat}+S^{(1)}_{flat}+S^{(2)}_{flat}+\mathcal{O}\left(\alpha^3\right)$, we can substitute the transformation (\ref{gauge transformation}) into the action. At this point, being generally covariant requires that for each order of $\alpha$, the action before and after the transformation can differ only by integration by parts. Since the background (\ref{vector-tensor theory background solution in Minkowski spacrtime}) is the solution to the field equations, $S^{(1)}_{flat}$ can be expressed as a tensor divergence, with $S^{(0)}_{flat}$ being constant. Therefore, considering the second-order terms of $\alpha$, we have
\begin{itemize}
	\item $S^{(2)}_{flat}$ before and after the gauge transformation $h_{\mu\nu} \rightarrow h_{\mu\nu}-\partial_{\mu}\xi_{\nu}-\partial_{\nu}\xi_{\mu}$, $B^{\mu} \rightarrow
	B^{\mu}+A^{\nu}\partial_{\nu}B^{\mu}$ can differ only by integration by parts.
\end{itemize}
Assumption (4) requires
\begin{itemize}
	\item Each term in $S^{(2)}_{flat}$ can have at most two derivative operators.
\end{itemize}
If any term in $S^{(2)}_{flat}$ includes more than two derivative operators, the linear perturbation equations obtained by varying $S^{(2)}_{flat}$ with respect to perturbations must exceed second order, contradicting assumption (4).

For now, without considering the gauge symmetry, the above requirements yield the most general second-order perturbation action (The most general meaning is that the second-order perturbation action satisfying the assumptions can always be transformed into the following form through integration by parts):
\begin{eqnarray}
	S^{(2)}_{flat}=S^{(2)}_{0}+S^{(2)}_{1}+S^{(2)}_{2}=\int d^4 x \sqrt{-\eta} \left(\mathcal{L}_{0}+\mathcal{L}_{1}+\mathcal{L}_{2}\right),
\end{eqnarray}
where
\begin{eqnarray}
	\label{L0=}
	\mathcal{L}_{0}
	\!&=&\!
	a^{(0)}_{1}A^{\mu}A^{\nu}A^{\lambda}A^{\rho}h_{\mu\nu}h_{\lambda\rho}
	+a^{(0)}_{2}A^{\lambda}A_{\rho}h_{\mu\lambda}h^{\mu\rho}
	\nonumber \\
	\!&+&\!
	a^{(0)}_{3}h_{\mu\nu}h^{\mu\nu}
	+a^{(0)}_{4}A_{\mu}A_{\nu}h^{\mu\nu}h
	+a^{(0)}_{5}h^2
	\nonumber \\
	\!&+&\!
	b^{(0)}_{1}A^{\mu}A^{\nu}A^{\lambda}h_{\mu\nu}B_{\lambda}
	+b^{(0)}_{2}A^{\mu}h_{\mu\nu}B^{\nu}
	+b^{(0)}_{3}A^{\mu}hB_{\mu}
	\nonumber \\
	\!&+&\!
	c^{(0)}_{1}A_{\mu}A_{\nu}B^{\mu}B^{\nu}
	+c^{(0)}_{2}B_{\mu}B^{\mu},
\end{eqnarray}
\begin{eqnarray}
	\label{L1=}
	\mathcal{L}_{1}
	\!&=&\!
	a^{(1)}_{1}\left(A^{\mu}A^{\nu}A^{\lambda}\partial^{\rho}h_{\mu\nu}\right)h_{\lambda\rho}
	+a^{(1)}_{2}\left(E^{\mu\lambda\sigma\gamma}A^{\nu}A^{\rho}A_{\sigma}\partial_{\gamma}h_{\mu\nu}\right)h_{\lambda\rho}
	\nonumber \\
	\!&+&\!
	a^{(1)}_{3}\left(A^{\lambda}\partial_{\rho}h_{\mu\lambda}\right)h^{\mu\rho}
	+a^{(1)}_{4}\left(E^{\lambda~\sigma\gamma}_{~\rho}A_{\sigma}\partial_{\gamma}h_{\mu\lambda}\right)h^{\mu\rho}
	\nonumber \\
	\!&+&\!
	a^{(1)}_{5}\left(\left(A \cdot \partial\right)A_{\mu}A_{\nu}h\right)h^{\mu\nu}
	+a^{(1)}_{6}\left(A_{\mu}\partial_{\nu}h\right)h^{\mu\nu}
	\nonumber \\
	\!&+&\!
	b^{(1)}_{1}\left(\left(A \cdot \partial\right)A^{\mu}A^{\nu}A^{\lambda}h_{\mu\nu}\right)B_{\lambda}
	+b^{(1)}_{2}\left(A^{\mu}A^{\nu}\partial^{\lambda}h_{\mu\nu}\right)B_{\lambda}
	+b^{(1)}_{3}\left(A^{\mu}\partial^{\nu}A^{\lambda}h_{\mu\nu}\right)B_{\lambda}
	\nonumber \\
	\!&+&\!
	b^{(1)}_{4}\left(E^{\mu\lambda\sigma\gamma}A^{\sigma}\partial_{\gamma}A^{\nu}h_{\mu\nu}\right)B_{\lambda}
	+b^{(1)}_{5}\left(\left(A \cdot \partial\right)A^{\mu}h_{\mu\lambda}\right)B^{\lambda}
	\nonumber \\
	\!&+&\!
	b^{(1)}_{6}\left(\partial^{\mu}h_{\mu\lambda}\right)B^{\lambda}
	+b^{(1)}_{7}\left(\left(A \cdot \partial\right)A^{\mu}h\right)B_{\mu}
	+b^{(1)}_{8}\left(\partial^{\mu}h\right)B_{\mu}
	\nonumber \\
	\!&+&\!
	c^{(1)}_{1}\left(A_{\mu}\partial_{\nu}B^{\mu}\right)B^{\nu}
	+c^{(1)}_{2}\left(E^{\mu\nu\lambda\rho}\partial_{\lambda}A_{\rho}B_{\mu}\right)B_{\nu},
\end{eqnarray}
\begin{eqnarray}
	\label{L2=}
	\mathcal{L}_{2}
	\!&=&\!
	a^{(2)}_{1}\left(\Box A^{\mu}A^{\nu}A^{\lambda}A^{\rho}h_{\mu\nu}\right)h_{\lambda\rho}
	+a^{(2)}_{2}\left(\left(A \cdot \partial\right)^{2} A^{\mu}A^{\nu}A^{\lambda}A^{\rho}h_{\mu\nu}\right)h_{\lambda\rho}
	\nonumber \\
	\!&+&\!
	a^{(2)}_{3}\left(\left(A \cdot \partial\right)A^{\mu}A^{\nu}A^{\lambda}\partial^{\rho}h_{\mu\nu}\right)h_{\lambda\rho}
	+a^{(2)}_{4}\left(A^{\mu}A^{\nu}\partial^{\lambda}\partial^{\rho}h_{\mu\nu}\right)h_{\lambda\rho}
	\nonumber \\
	\!&+&\!
	a^{(2)}_{5}\left(A^{\mu}A^{\lambda}\partial^{\nu}\partial^{\rho}h_{\mu\nu}\right)h_{\lambda\rho}
	+a^{(2)}_{6}\left(E^{\mu\lambda\sigma\gamma}\partial^{\nu}A^{\rho}A_{\sigma}\partial_{\gamma}h_{\mu\nu}\right)h_{\lambda\rho}
	\nonumber \\
	\!&+&\!
	a^{(2)}_{7}\left(\Box A^{\lambda}A_{\rho}h_{\mu\lambda}\right)h^{\mu\rho}
	+a^{(2)}_{8}\left(\left(A \cdot \partial\right)^{2} A^{\lambda}A_{\rho}h_{\mu\lambda}\right)h^{\mu\rho}
	+a^{(2)}_{9}\left(\left(A \cdot \partial\right)A^{\lambda}\partial_{\rho}h_{\mu\lambda}\right)h^{\mu\rho}
	\nonumber \\
	\!&+&\!
	a^{(2)}_{10}\left(\partial^{\lambda}\partial_{\rho}h_{\mu\lambda}\right)h^{\mu\rho}
	+a^{(2)}_{11}\left(\Box h_{\mu\nu}\right)h^{\mu\nu}
	+a^{(2)}_{12}\left(\left(A \cdot \partial\right)^{2} h_{\mu\nu}\right)h^{\mu\nu}
	\nonumber \\
	\!&+&\!
	a^{(2)}_{13}\left(\Box A_{\mu}A_{\nu}h\right)h^{\mu\nu}
	+a^{(2)}_{14}\left(\left(A \cdot \partial\right)^{2}A_{\mu}A_{\nu}h\right)h^{\mu\nu}
	+a^{(2)}_{15}\left(\left(A \cdot \partial\right)A_{\mu}\partial_{\nu}h\right)h^{\mu\nu}
	\nonumber \\
	\!&+&\!
	a^{(2)}_{16}\left(\partial_{\mu}\partial_{\nu}h\right)h^{\mu\nu}
	+a^{(2)}_{17}\left(\Box h\right)h
	+a^{(2)}_{18}\left(\left(A \cdot \partial\right)^{2} h\right)h
	\nonumber \\
	\!&+&\!
	b^{(2)}_{1}\left(\Box A^{\mu}A^{\nu}A^{\lambda}h_{\mu\nu}\right)B_{\lambda}
	+b^{(2)}_{2}\left(\left(A \cdot \partial\right)^{2}A^{\mu}A^{\nu}A^{\lambda}h_{\mu\nu}\right)B_{\lambda}
	\nonumber \\
	\!&+&\!
	b^{(2)}_{3}\left(\left(A \cdot \partial\right)A^{\mu}A^{\nu}\partial^{\lambda}h_{\mu\nu}\right)B_{\lambda}
	+b^{(2)}_{4}\left(\left(A \cdot \partial\right)A^{\mu}\partial^{\nu}A^{\lambda}h_{\mu\nu}\right)B_{\lambda}
	\nonumber \\
	\!&+&\!
	b^{(2)}_{5}\left(A^{\mu}\partial^{\nu}\partial^{\lambda}h_{\mu\nu}\right)B_{\lambda}
	+b^{(2)}_{6}\left(\partial^{\mu}\partial^{\nu}A^{\lambda}h_{\mu\nu}\right)B_{\lambda}
	\nonumber \\
	\!&+&\!
	b^{(2)}_{7}\left(\left(A \cdot \partial\right)E^{\mu\lambda\sigma\gamma}A_{\sigma}\partial_{\gamma}A^{\nu}h_{\mu\nu}\right)B_{\lambda}
	+b^{(2)}_{8}\left(E^{\mu\lambda\sigma\gamma}A_{\sigma}\partial_{\gamma}\partial^{\nu}h_{\mu\nu}\right)B_{\lambda}
	\nonumber \\
	\!&+&\!
	b^{(2)}_{9}\left(\Box A^{\mu}h_{\mu\lambda}\right)B^{\lambda}
	+b^{(2)}_{10}\left(\left(A \cdot \partial\right)^{2}A^{\mu}h_{\mu\lambda}\right)B^{\lambda}
	+b^{(2)}_{11}\left(\left(A \cdot \partial\right)\partial^{\mu}h_{\mu\lambda}\right)B^{\lambda}
	\nonumber \\
	\!&+&\!
	b^{(2)}_{12}\left(\Box A^{\mu} h\right)B_{\mu}
	+b^{(2)}_{13}\left(\left(A \cdot \partial\right)^{2}A^{\mu}h\right)B_{\mu}
	+b^{(2)}_{14}\left(\left(A \cdot \partial\right)\partial^{\mu} h\right)B_{\mu}
	\nonumber \\
	\!&+&\!
	c^{(2)}_{1}\left(\Box A_{\mu}A_{\nu}B^{\mu}\right)B^{\nu}
	+c^{(2)}_{2}\left(\left(A \cdot \partial\right)^{2}A_{\mu}A_{\nu}B^{\mu}\right)B^{\nu}
	+c^{(2)}_{3}\left(\left(A \cdot \partial\right)A_{\mu}\partial_{\nu}B^{\mu}\right)B^{\nu}
	\nonumber \\
	\!&+&\!
	c^{(2)}_{4}\left(\partial_{\mu}\partial_{\nu}B^{\mu}\right)B^{\nu}
	+c^{(2)}_{5}\left(\Box B_{\mu}\right)B^{\mu}
	+c^{(2)}_{6}\left(\left(A \cdot \partial\right)^{2}B_{\mu}\right)B^{\mu}.
\end{eqnarray}
Here, $S^{(2)}_{0}$ includes all terms without derivative operators, $S^{(2)}_{1}$ includes all terms with only one derivative operator, and $S^{(2)}_{2}$ includes all terms with two derivative operators. $A \cdot \partial=A^{\mu}\partial_{\mu}$, and all quantities labeled with 
$a$, $b$, and $c$, such as $a^{(0)}_{1}$, are constant parameters. In fact, when considering combinations of $\eta_{\mu\nu}$, $\eta^{\mu\nu}$, $A^{\mu}$, $h_{\mu\nu}$, $B^{\mu}$, $E^{\mu\nu\lambda\rho}$, and $\partial_{\mu}$ that satisfy the assumptions, the terms such as 
\begin{eqnarray}
	\label{term not in vector-tensor S2}
	\left(\left(A \cdot \partial\right)A^{\mu}A^{\nu}A^{\lambda}A^{\rho}h_{\mu\nu}\right)h_{\lambda\rho}	
\end{eqnarray}
can also be constructed in addition to the terms listed in Eqs. (\ref{L0=})-(\ref{L2=}). However, it is easy to see that such terms can be written in the form of tensor divergence through integration by parts, and therefore do not contribute to the action. Finally, it should be noted that, unlike pure metric theory and scalar-tensor theory \cite{Y.Dong4}, the second-order perturbation action for vector-tensor theory can include terms with only an odd number of derivative operators and terms containing $E^{\mu\nu\lambda\rho}$.

The gauge symmetry will further constrain the parameters in Eqs. (\ref{L0=})-(\ref{L2=}). After lengthy calculations, it can be found that the parameters in $\mathcal{L}_{0}$ satisfy 
\begin{eqnarray}
	\label{parameters in L0 for gauge symmetry}
	a^{(0)}_{2} \!&=&\! a^{(0)}_{3}=a^{(0)}_{4}=a^{(0)}_{5}=0,
	\nonumber \\
	b^{(0)}_{1} \!&=&\! 4 a^{(0)}_{1},~ b^{(0)}_{2}=b^{(0)}_{3}=0,
	\nonumber \\
	c^{(0)}_{1} \!&=&\! b^{(0)}_{1}=4 a^{(0)}_{1},~ c^{(0)}_{2}=0.
\end{eqnarray}
Therefore, by redefining the parameters, we can rewrite $\mathcal{L}_{0}$ as
\begin{eqnarray}
	\label{L0= gauge symmetry}
	\mathcal{L}_{0}
	\!&=&\!
	A_{(0)} A^{\mu}A^{\nu}A^{\lambda}A^{\rho}h_{\mu\nu}h_{\lambda\rho}
	+4 A_{(0)} A^{\mu}A^{\nu}A^{\lambda}h_{\mu\nu}B_{\lambda}
	+4 A_{(0)} A_{\mu}A_{\nu}B^{\mu}B^{\nu},
\end{eqnarray}
where $A_{(0)}$ is a redefined constant parameter. For $\mathcal{L}_{1}$, the constraints between parameters are
\begin{eqnarray}
	\label{parameters in L1 for gauge symmetry}
	a^{(1)}_{1} \!&=&\! a^{(1)}_{3}=a^{(1)}_{4}=a^{(1)}_{6}=0,
	\nonumber \\
	b^{(1)}_{1} \!&=&\! b^{(1)}_{3}=b^{(1)}_{5}=b^{(1)}_{6}=b^{(1)}_{8}=0,
	\nonumber \\
	b^{(1)}_{2} \!&=&\! -2 a^{(1)}_{5},~ b^{(1)}_{4}=2 a^{(1)}_{2},~ b^{(1)}_{7}=2 a^{(1)}_{5},
	\nonumber \\
	c^{(1)}_{1} \!&=&\! -4 a^{(1)}_{5},~ c^{(1)}_{2}=-a^{(1)}_{2}.
\end{eqnarray}
So, by redefining the parameters, $\mathcal{L}_{1}$ can be rewritten as
\begin{eqnarray}
	\label{L1= gauge symmetry}
	\mathcal{L}_{1}
	\!&=&\!
	A_{(1)} \left(E^{\mu\lambda\sigma\gamma}A^{\nu}A^{\rho}A_{\sigma}\partial_{\gamma}h_{\mu\nu}\right)h_{\lambda\rho}
	+B_{(1)}\left(\left(A \cdot \partial\right)A_{\mu}A_{\nu}h\right)h^{\mu\nu}
	\nonumber \\
	\!&-&\!
	2B_{(1)}\left(A^{\mu}A^{\nu}\partial^{\lambda}h_{\mu\nu}\right)B_{\lambda}
	+2A_{(1)}\left(E^{\mu\lambda\sigma\gamma}A_{\sigma}\partial_{\gamma}A^{\nu}h_{\mu\nu}\right)B_{\lambda}
	\nonumber \\
	\!&+&\!
	2B_{(1)}\left(\left(A \cdot \partial\right)A^{\mu}h\right)B_{\mu}
	\nonumber \\
	\!&-&\!
	4B_{(1)}\left(A_{\mu}\partial_{\nu}B^{\mu}\right)B^{\nu}
	-A_{(1)}\left(E^{\mu\nu\lambda\rho}\partial_{\lambda}A_{\rho}B_{\mu}\right)B_{\nu},
\end{eqnarray}
where $A_{(1)}$ and $B_{(1)}$ are redefined constant parameters. For $\mathcal{L}_{2}$, the gauge symmetry requires 
\begin{eqnarray}
	\label{parameters in L2 for gauge symmetry}
	a^{(2)}_{13} \!&=&\! -a^{(2)}_{4},~ 
	a^{(2)}_{7}=-a^{(2)}_{5},~
	a^{(2)}_{6}=0,~
	a^{(2)}_{10}=-a^{(2)}_{16}=-2a^{(2)}_{11}=2a^{(2)}_{17},~
	a^{(2)}_{15}=-a^{(2)}_{9},
	\nonumber \\
	b^{(2)}_{1} \!&=&\! 4 a^{(2)}_{1}+a^{(2)}_{3},~
	b^{(2)}_{2}=4a^{(2)}_{2},~
	b^{(2)}_{3}=a^{(2)}_{3}+2a^{(2)}_{14},~
	b^{(2)}_{4}=2a^{(2)}_{3}+2a^{(2)}_{8},
	\nonumber \\
	b^{(2)}_{5} \!&=&\! 2a^{(2)}_{5}-a^{(2)}_{9},~
	b^{(2)}_{6}=2a^{(2)}_{4}+a^{(2)}_{9},~
	b^{(2)}_{7}=b^{(2)}_{8}=0,~
	b^{(2)}_{9}=-2a^{(2)}_{5}+a^{(2)}_{9},~
	\nonumber \\
	b^{(2)}_{10} \!&=&\! 2a^{(2)}_{8},~
	b^{(2)}_{11}=a^{(2)}_{9}+4a^{(2)}_{12},~
	b^{(2)}_{12}=-2a^{(2)}_{4}-a^{(2)}_{9},~
	b^{(2)}_{13}=2a^{(2)}_{14},~
	b^{(2)}_{14}=-a^{(2)}_{9}+4a^{(2)}_{18},~
	\nonumber \\
	c^{(2)}_{1} \!&=&\! 4a^{(2)}_{1}+2a^{(2)}_{3}+a^{(2)}_{8},~
	c^{(2)}_{2}=4a^{(2)}_{2},~
	c^{(2)}_{3}=2a^{(2)}_{3}+2a^{(2)}_{8}+4a^{(2)}_{14},~
	\nonumber \\
	c^{(2)}_{4} \!&=&\! a^{(2)}_{5}+2a^{(2)}_{12}-a^{(2)}_{9}+4a^{(2)}_{18},~
	c^{(2)}_{5}=-a^{(2)}_{5}+a^{(2)}_{9}+2a^{(2)}_{12},~
	c^{(2)}_{6}=a^{(2)}_{8}.
\end{eqnarray}
Therefore, through redefining the parameters, we can express $\mathcal{L}_{2}$ as
\begin{eqnarray}
	\label{L2= gauge symmetry}
	\mathcal{L}_{2}
	\!&=&\!
	A_{(2)} \left(\Box A^{\mu}A^{\nu}A^{\lambda}A^{\rho}h_{\mu\nu}\right)h_{\lambda\rho}
	+B_{(2)} \left(\left(A \cdot \partial\right)^{2}A^{\mu}A^{\nu}A^{\lambda}A^{\rho}h_{\mu\nu}\right)h_{\lambda\rho}
	\nonumber \\
	\!&+&\!
	C_{(2)} \left(\left(A \cdot \partial\right)A^{\mu}A^{\nu}A^{\lambda}\partial^{\rho}h_{\mu\nu}\right)h_{\lambda\rho}
	+D_{(2)} \left(A^{\mu}A^{\nu}\partial^{\lambda}\partial^{\rho}h_{\mu\nu}\right)h_{\lambda\rho}
	\nonumber \\
	\!&+&\!
	E_{(2)} \left(A^{\mu}A^{\lambda}\partial^{\nu}\partial^{\rho}h_{\mu\nu}\right)h_{\lambda\rho}
	-E_{(2)} \left(\Box A^{\lambda}A_{\rho}h_{\mu\lambda}\right)h^{\mu\rho}
	\nonumber \\
	\!&+&\!
	F_{(2)} \left(\left(A \cdot \partial\right)^{2} A^{\lambda}A_{\rho}h_{\mu\lambda}\right)h^{\mu\rho}
	+G_{(2)} \left(\left(A \cdot \partial\right)A^{\lambda}\partial_{\rho}h_{\mu\lambda}\right)h^{\mu\rho}
	\nonumber \\
	\!&-&\!
	2H_{(2)} \left(\partial^{\lambda}\partial_{\rho}h_{\mu\lambda}\right)h^{\mu\rho}
	+H_{(2)} \left(\Box h_{\mu\nu}\right)h^{\mu\nu}
	+2H_{(2)} \left(\partial_{\mu}\partial_{\nu}h\right)h^{\mu\nu}
	-H_{(2)} \left(\Box h\right)h
	\nonumber \\
	\!&+&\!
	I_{(2)} \left(\left(A \cdot \partial\right)^{2}h_{\mu\nu}\right)h^{\mu\nu}
	-D_{(2)} \left(\Box A_{\mu}A_{\nu}h\right)h^{\mu\nu}
	+J_{(2)} \left(\left(A \cdot \partial\right)^{2}A_{\mu}A_{\nu}h\right)h^{\mu\nu}
	\nonumber \\
	\!&-&\!
	G_{(2)} \left(\left(A \cdot \partial\right)A_{\mu}\partial_{\nu}h\right)h^{\mu\nu}
	+K_{(2)} \left(\left(A \cdot \partial\right)^{2}h\right)h
	\nonumber \\
	\!&+&\!
	\left(4A_{(2)}+C_{(2)}\right)\left(\Box A^{\mu}A^{\nu}A^{\lambda}h_{\mu\nu}\right)B_{\lambda}
	+4B_{(2)}\left(\left(A \cdot \partial\right)^{2}A^{\mu}A^{\nu}A^{\lambda}h_{\mu\nu}\right)B_{\lambda}
	\nonumber \\
	\!&+&\!
	\left(C_{(2)}+2J_{(2)}\right)\left(\left(A \cdot \partial\right)A^{\mu}A^{\nu}\partial^{\lambda}h_{\mu\nu}\right)B_{\lambda}
	+2\left(C_{(2)}+F_{(2)}\right)\left(\left(A \cdot \partial\right)A^{\mu}\partial^{\nu}A^{\lambda}h_{\mu\nu}\right)B_{\lambda}
	\nonumber \\
	\!&+&\!
	\left(2E_{(2)}-G_{(2)}\right)\left(A^{\mu}\partial^{\nu}\partial^{\lambda}h_{\mu\nu}\right)B_{\lambda}
	+\left(2D_{(2)}+G_{(2)}\right)\left(\partial^{\mu}\partial^{\nu}A^{\lambda}h_{\mu\nu}\right)B_{\lambda}
	\nonumber \\
	\!&+&\!
	\left(-2E_{(2)}+G_{(2)}\right)\left(\Box A^{\mu}h_{\mu\lambda}\right)B^{\lambda}
	+2F_{(2)}\left(\left(A \cdot \partial\right)^{2}A^{\mu}h_{\mu\lambda}\right)B^{\lambda}
	\nonumber \\
	\!&+&\!
	\left(G_{(2)}+4I_{(2)}\right)\left(\left(A \cdot \partial\right)\partial^{\mu}h_{\mu\lambda}\right)B^{\lambda}
	+\left(-2D_{(2)}-G_{(2)}\right)\left(\Box A^{\mu} h\right)B_{\mu}
	\nonumber \\
	\!&+&\!
	2J_{(2)}\left(\left(A \cdot \partial\right)^{2}A^{\mu}h\right)B_{\mu}
	+\left(-G_{(2)}+4K_{(2)}\right)\left(\left(A \cdot \partial\right)\partial^{\mu} h\right)B_{\mu}
	\nonumber \\
	\!&+&\!
	\left(4A_{(2)}+2C_{(2)}+F_{(2)}\right)\left(\Box A_{\mu}A_{\nu}B^{\mu}\right)B^{\nu}
	+4B_{(2)}\left(\left(A \cdot \partial\right)^{2}A_{\mu}A_{\nu}B^{\mu}\right)B^{\nu}
	\nonumber \\
	\!&+&\!
	\left(2C_{(2)}+2F_{(2)}+4J_{(2)}\right)\left(\left(A \cdot \partial\right)A_{\mu}\partial_{\nu}B^{\mu}\right)B^{\nu}
	\nonumber \\
	\!&+&\!
	\left(E_{(2)}+2I_{(2)}-G_{(2)}+4K_{(2)}\right)\left(\partial_{\mu}\partial_{\nu}B^{\mu}\right)B^{\nu}
	\nonumber \\
	\!&+&\!
	\left(-E_{(2)}+G_{(2)}+2I_{(2)}\right)\left(\Box B_{\mu}\right)B^{\mu}
	+F_{(2)}\left(\left(A \cdot \partial\right)^{2}B_{\mu}\right)B^{\mu},
\end{eqnarray}
where $A_{(2)}, ..., K_{(2)}$ are redefined constant parameters.

\section{The detailed classification of vector modes}
\label{app: B}

\textbf{Case 1}: $G_{(2)}A^{2}+4H_{(2)}=0,~G_{(2)}+4I_{(2)}=0$. In such a situation, Eq. (\ref{vector mode equations 1}) is always zero. There is only Eq. (\ref{vector mode equations 2}) to constrain the values of the two vectors $\Xi_{i}$ and $\Sigma_{i}$. Now, at least one of $\Xi_{i}$ and $\Sigma_{i}$ can take any value. We believe that this is unreasonable in physics, so in this case we will not further discuss the properties of gravitational waves.

\textbf{Case 2}: $G_{(2)}A^{2}+4H_{(2)} \neq 0,~G_{(2)}+4I_{(2)}=0$. In such a situation, from Eq. (\ref{vector mode equations 1}), we know that $\Xi_{i}=0$. Therefore, there is no vector mode gravitational wave.

\textbf{Case 3}: $G_{(2)}A^{2}+4H_{(2)} = 0,~G_{(2)}+4I_{(2)} \neq 0$. In this case, from Eq. (\ref{vector mode equations 1}), we know that $\Sigma_{i}=0$ and Eq. (\ref{vector mode equations 2}) becomes
\begin{eqnarray}
	\label{vector mode equations 2 in case 3}
	2A_{(1)}E^{0ijk}A^{2}\partial_{k}\Xi_{j}
	+\mathcal{C}_{1}\partial_{0}^{2}\Xi_{i}
	-\mathcal{C}_{2}\Delta \Xi_{i}
	=0,
\end{eqnarray}
where
\begin{eqnarray}
	\label{C1}
	\mathcal{C}_{1}
	\!&=&\!
	2E_{(2)}A-2G_{(2)}A-4I_{(2)}A+2F_{(2)}A^{3},
	\\
	\label{C2}
	\mathcal{C}_{2}
	\!&=&\!
	2E_{(2)}A-G_{(2)}A.
\end{eqnarray}
Therefore, for the solution of a monochromatic plane wave
\begin{eqnarray}
	\label{Xi i=Xi i eikx}
	\Xi_{i}=\mathring{\Xi}_{i} e^{ikx},
\end{eqnarray}
the above equations can be transformed into the following matrix form:
\begin{eqnarray}
	\label{vector mode equations 2 in case 3 matrix form}
	\begin{pmatrix}
		-\mathcal{C}_{1}k_{0}^{2}+\mathcal{C}_{2}k_{3}^{2} & 
		2iA_{(1)}A^{2}k_{3} \\
		-2iA_{(1)}A^{2}k_{3}   & 
		-\mathcal{C}_{1}k_{0}^{2}+\mathcal{C}_{2}k_{3}^{2} \\
	\end{pmatrix}
	\begin{pmatrix}
		\mathring{\Xi}_{1}\\
		\mathring{\Xi}_{2}
	\end{pmatrix}
	=0.
\end{eqnarray}
Equation (\ref{vector mode equations 2 in case 3 matrix form}) can be solved using the standard methods for solving linear systems of equations. Since the true solution of the physical world is the real part of solution (\ref{Xi i=Xi i eikx}), $\Xi_{i}$ and its complex conjugate $\bar{\Xi}_{i}$ represent the same solution. Furthermore, both $\Xi_{i}$ and $\bar{\Xi}_{i}$ must either be solutions or not solutions to Eq. (\ref{vector mode equations 2 in case 3}) at the same time. This allows us to apply the condition
\begin{eqnarray}
	\label{z direction}
	k_{0} \leq 0,~k_{3} \geq 0
\end{eqnarray}
without loss of generality when considering gravitational waves propagating along the $+z$ direction.

The necessary and sufficient condition for Eq. (\ref{vector mode equations 2 in case 3}) to have a monochromatic plane wave solution is that the determinant of the coefficient matrix of Eq. (\ref{vector mode equations 2 in case 3 matrix form}) is zero, that is, 
\begin{eqnarray}
	\label{vector mode equations 2 in case 3 determinant of the coefficient matrix}
	\left(\mathcal{C}_{1}k_{0}^{2}-\mathcal{C}_{2}k_{3}^{2}\right)^{2}
	-4A_{(1)}^{2}A^{4}k_{3}^{2}
	=0.
\end{eqnarray}
In other words, we write it as
\begin{eqnarray}
	\label{vector mode equations 2 in case 3 determinant of the coefficient matrix another write}
	\mathcal{C}_{1}k_{0}^{2}
	=
	\mathcal{C}_{2}k_{3}^{2}
	\pm 2A_{(1)}A^{2}k_{3}.
\end{eqnarray}

We need to further classify and discuss Case 3.

\textbf{Case 3.1}: $\mathcal{C}_{1}=\mathcal{C}_{2}=A_{(1)}=0$. In such a situation, Eq. (\ref{vector mode equations 2 in case 3 determinant of the coefficient matrix another write}) remains constantly at zero. For any value of the wave vector, Eq. (\ref{vector mode equations 2 in case 3 matrix form}) has a plane wave solution. This is unreasonable in physics, so we rule out this case.

\textbf{Case 3.2}: $\mathcal{C}_{1}=\mathcal{C}_{2}=0,~A_{(1)} \neq 0$. In this case, $k_{3}=0$, and $k_{0}$ can take any value. This is unreasonable in physics, so we rule out this case.

\textbf{Case 3.3}: $\mathcal{C}_{1}=A_{(1)}=0,~\mathcal{C}_{2} \neq 0$. In such a situation, $k_{3}=0$, and $k_{0}$ can take any value. We rule out this case.

\textbf{Case 3.4}: $\mathcal{C}_{2}=A_{(1)}=0,~\mathcal{C}_{1} \neq 0$. In this case, $k_{0}=0$, and $k_{3}$ can take any value. We rule out this case.

\textbf{Case 3.5}: $\mathcal{C}_{1}=0,~\mathcal{C}_{2} \neq 0,~A_{(1)} \neq 0$. In such a situation, Eq. (\ref{vector mode equations 2 in case 3 determinant of the coefficient matrix another write}) becomes a quadratic equation with respect to $k_{3}$. Because the gravitational waves we are considering propagate along the $+z$ direction, we have conditation (\ref{z direction}). Therefore, from Eq. (\ref{vector mode equations 2 in case 3 determinant of the coefficient matrix another write}), $k_{3}$ has only two solutions, 
\begin{eqnarray}
	\label{k3 in case 3.5}
	k_{3}=0,~k_{3}=\left|\frac{2A_{(1)}A^{2}}{\mathcal{C}_{2}}\right|,
\end{eqnarray}
and the value of $k_{0}$ is arbitrary. This is unreasonable in physics, so we rule out this case.

\textbf{Case 3.6}: $A_{(1)}=0,~\mathcal{C}_{1} \neq 0,~ \mathcal{C}_{2}\neq 0$. For this scenario, Eq. (\ref{vector mode equations 2 in case 3 determinant of the coefficient matrix another write}) becomes $\mathcal{C}_{1}k_{0}^{2}-\mathcal{C}_{2}k_{3}^{2}=0$. Furthermore, from Eq. (\ref{vector mode equations 2 in case 3 matrix form}), $\mathring{\Xi}_{1}$ and $\mathring{\Xi}_{2}$ can take any values. Therefore, in such a situation, using Eq. (\ref{P1-P6 gauge invariant}), we can see the theory allows for two independent vector modes: the vector-$x$ mode and the vector-$y$ mode. Their wave speed satisfies 
\begin{eqnarray}
	\label{vector mode wave speed in case 3.6}
	v_{V}^{2}
	=\frac{\mathcal{C}_{2}}{\mathcal{C}_{1}}
	=\frac{2E_{(2)}-G_{(2)}}{2E_{(2)}-2G_{(2)}-4I_{(2)}+2F_{(2)}A^{2}}.
\end{eqnarray}
Here and in the following text, we require the wave speed is a positive real number. Thus, $v_{V}^{2} \textgreater 0$.

\textbf{Case 3.7}: $\mathcal{C}_{2}=0,~\mathcal{C}_{1} \neq 0,~ A_{(1)}\neq 0$. In this case, Eq. (\ref{vector mode equations 2 in case 3 determinant of the coefficient matrix another write}) becomes 
\begin{eqnarray}
	\label{k0,k3 relationship in case 3.7}
	\mathcal{C}_{1}k_{0}^{2}=\pm 2A_{(1)}A^{2}k_{3}. 
\end{eqnarray}
Therefore, in such a situation, the group velocity of the vector mode $v_{gV}$ is
\begin{eqnarray}
	\label{vector mode group velocity in case 3.7}
	v_{gV}
	=\frac{dk^{0}}{dk^{3}}
	=\left|\frac{A_{(1)}A^{2}}{\mathcal{C}_{1}k_{0}}\right|
	=\sqrt{\left|\frac{A_{(1)}A^{2}}{2\mathcal{C}_{1}k_{3}}\right|}.
\end{eqnarray}
It should be noted that $v_{gV}$ decreases with the increase of $k_{3}$. Especially, when $k_{3} \to \infty$, $v_{gV} \to 0$. And it can be seen that when 
\begin{eqnarray}
	\label{vector mode group velocity in case 3.7 k3 superluminal}
	k_{3} 
	\textless
	\left|\frac{A_{(1)}A^{2}}{2\mathcal{C}_{1}}\right|,
\end{eqnarray}
the vector mode is superluminal. Especially when $k_{3} \to 0$, $v_{gV} \to \infty$. Therefore, unless an additional mechanism prevents the spatial wave vector $k_{3}$ of the vector mode from falling within the range shown in Eq. (\ref{vector mode group velocity in case 3.7 k3 superluminal}), this will lead to superluminal phenomena, and thus this case needs to be ruled out.

The analysis of gravitational wave polarization modes requires us to further divide Case 3.7 into two cases.

\textbf{Case 3.7.1}: $A_{(1)}\mathcal{C}_{1} \textgreater 0$. Since we require $k_{0}^{2}\textgreater0,~k_{3} \textgreater 0$, only one of the two dispersion relations in Eq. (\ref{k0,k3 relationship in case 3.7}) satisfies the above condition:  $k_{0}^{2}=2A_{(1)}A^{2}k_{3}/\mathcal{C}_{1}$. Therefore, Eq. (\ref{vector mode equations 2 in case 3 matrix form}) becomes
\begin{eqnarray}
	\label{vector mode equations 2 in case 3.7.1}
	-2A_{(1)}A^{2}k_{3}\left(\mathring{\Xi}_{1}-i \mathring{\Xi}_{2}\right)
	=0.
\end{eqnarray}
It can be seen that the amplitude of the vector mode must satisfy $\mathring{\Xi}_{2}=-i\mathring{\Xi}_{1}$, so only the left-handed wave with amplitude $\mathring{\Xi}_{1}+i\mathring{\Xi}_{2} \neq 0$ exists. In this case, vector mode gravitational waves have only one degree of freedom.

\textbf{Case 3.7.2}: $A_{(1)}\mathcal{C}_{1} \textless 0$. For this scenario, Eq. (\ref{k0,k3 relationship in case 3.7}) requires $k_{0}^{2}=-2A_{(1)}A^{2}k_{3}/\mathcal{C}_{1}$ and Eq. (\ref{vector mode equations 2 in case 3 matrix form}) becomes
\begin{eqnarray}
	\label{vector mode equations 2 in case 3.7.2}
	2A_{(1)}A^{2}k_{3}\left(\mathring{\Xi}_{1}+i \mathring{\Xi}_{2}\right)
	=0.
\end{eqnarray}
In such a situation, only the right-handed wave with amplitude $\mathring{\Xi}_{1}-i\mathring{\Xi}_{2} \neq 0$ exists and vector mode gravitational waves also have only one degree of freedom.

\textbf{Case 3.8}: $\mathcal{C}_{1} \neq 0,~\mathcal{C}_{2} \neq 0,~ A_{(1)}\neq 0$. In this case, the two dispersion relations in Eq. (\ref{vector mode equations 2 in case 3 determinant of the coefficient matrix another write}) are
\begin{eqnarray}
	\label{two dispersion relations vector mode 1}
	\mathcal{C}_{1}k_{0}^{2}
	=
	\mathcal{C}_{2}k_{3}^{2}
	+ 2A_{(1)}A^{2}k_{3},
	\\
	\label{two dispersion relations vector mode 2}
	\mathcal{C}_{1}k_{0}^{2}
	=
	\mathcal{C}_{2}k_{3}^{2}
	- 2A_{(1)}A^{2}k_{3}.
\end{eqnarray}
To analyze the properties of vector mode gravitational waves corresponding to dispersion relation (\ref{two dispersion relations vector mode 1}), note that Eq. (\ref{vector mode equations 2 in case 3 matrix form}) implies 
\begin{eqnarray}
	\label{vector mode equations 2 in case 3.8 dispersion relation 1}
	-2A_{(1)}A^{2}k_{3}\left(\mathring{\Xi}_{1}-i \mathring{\Xi}_{2}\right)
	=0,
\end{eqnarray}
hence $\mathring{\Xi}_{2}=-i\mathring{\Xi}_{1}$. This indicates that the dispersion relation (\ref{two dispersion relations vector mode 1}) corresponds to the left-handed vector mode gravitational waves, whose group velocity satisfies 
\begin{eqnarray}
	\label{vector mode group velocity in case 3.8 dispersion relation 1}
	v_{gV}
	=\frac{dk^{0}}{dk^{3}}
	=\left|\frac{2\mathcal{C}_{2}k_{3}+2A_{(1)}A^{2}}{2\mathcal{C}_{1}k_{0}}\right|
	=\frac{\left|2\mathcal{C}_{2}k_{3}+2A_{(1)}A^{2}\right|}{2\sqrt{\left|\mathcal{C}_{1}\left(\mathcal{C}_{2}k_{3}^{2}+2A_{(1)}A^{2}k_{3}\right)\right|}}.
\end{eqnarray}
It can be seen that when $k_{3} \to 0$, we have $v_{gV} \to \infty$ and when $k_{3} \to \infty$, we have $v_{gV}^{2} \to \left|\mathcal{C}_{2}/\mathcal{C}_{1} \right|$. The range of group velocity not exceeding the speed of light is given by the following condition:
\begin{eqnarray}
	\label{range of group velocities less than the speed of light vector mode in case 3.8 dispersion relation 1}
	4\mathcal{C}_{2}^{2}k_{3}^{2}+8\mathcal{C}_{2}A_{(1)}A^{2}k_{3}+4A_{(1)}^{2}A^{4}
	-4\left|\mathcal{C}_{1}\left(\mathcal{C}_{2}k_{3}^{2}+2A_{(1)}A^{2}k_{3}\right)\right|
	\leq
	0.
\end{eqnarray}

For the second dispersion relation (\ref{two dispersion relations vector mode 2}), the corresponding vector mode satisfies
\begin{eqnarray}
	\label{vector mode equations 2 in case 3.8 dispersion relation 2}
	2A_{(1)}A^{2}k_{3}\left(\mathring{\Xi}_{1}+i \mathring{\Xi}_{2}\right)
	=0,
\end{eqnarray}
hence $\mathring{\Xi}_{2}=i\mathring{\Xi}_{1}$. This indicates that the dispersion relation (\ref{two dispersion relations vector mode 1}) corresponds to right-handed vector mode gravitational waves, whose group velocity satisfies 
\begin{eqnarray}
	\label{vector mode group velocity in case 3.8 dispersion relation 2}
	v_{gV}
	=\frac{dk^{0}}{dk^{3}}
	=\left|\frac{2\mathcal{C}_{2}k_{3}-2A_{(1)}A^{2}}{2\mathcal{C}_{1}k_{0}}\right|
	=\frac{\left|2\mathcal{C}_{2}k_{3}-2A_{(1)}A^{2}\right|}{2\sqrt{\left|\mathcal{C}_{1}\left(\mathcal{C}_{2}k_{3}^{2}-2A_{(1)}A^{2}k_{3}\right)\right|}}.
\end{eqnarray}
The range of group velocity not exceeding the speed of light is given by the following condition:
\begin{eqnarray}
	\label{range of group velocities less than the speed of light vector mode in case 3.8 dispersion relation 2}
	4\mathcal{C}_{2}^{2}k_{3}^{2}-8\mathcal{C}_{2}A_{(1)}A^{2}k_{3}+4A_{(1)}^{2}A^{4}
	-4\left|\mathcal{C}_{1}\left(\mathcal{C}_{2}k_{3}^{2}-2A_{(1)}A^{2}k_{3}\right)\right|
	\leq
	0.
\end{eqnarray}

We further classify Case 3.8 to discuss the existence of left-handed and right-handed waves.

\textbf{Case 3.8.1}: $\mathcal{C}_{1}\mathcal{C}_{2} \textgreater 0,~ A_{(1)}\mathcal{C}_{1}\textgreater0$. According to condition (\ref{z direction}), the wave vector must satisfy the condition  
\begin{eqnarray}
	\label{k0^2 gen 0, k3 gen 0}
	k_{3} \geq 0,~k_{0}^{2} \geq 0.
\end{eqnarray}
Using Eq. (\ref{vector mode equations 2 in case 3 determinant of the coefficient matrix another write}), we find that when $k_{3} \in \left(0, \left|2A_{(1)}A^{2}/\mathcal{C}_{2}\right|\right)$, only wave vectors that satisfy the dispersion relation (\ref{two dispersion relations vector mode 1}) meet the above condition. Therefore, within this range, the theory only supports left-handed waves.
When $k_{3} \in \left[\left|2A_{(1)}A^{2}/\mathcal{C}_{2}\right|, \infty \right)$, both dispersion relations satisfy the condition (\ref{k0^2 gen 0, k3 gen 0}), allowing for the existence of both left-handed and right-handed waves.

\textbf{Case 3.8.2}: $\mathcal{C}_{1}\mathcal{C}_{2} \textgreater 0,~ A_{(1)}\mathcal{C}_{1}\textless 0$. In this case, using Eq. (\ref{vector mode equations 2 in case 3 determinant of the coefficient matrix another write}), we find that when $k_{3} \in \left(0, \left|2A_{(1)}A^{2}/\mathcal{C}_{2}\right|\right)$, only wave vectors that satisfy the dispersion relation (\ref{two dispersion relations vector mode 2}) meet condition (\ref{k0^2 gen 0, k3 gen 0}). Therefore, within this range, the theory only supports right-handed waves. When $k_{3} \in \left[\left|2A_{(1)}A^{2}/\mathcal{C}_{2}\right|, \infty \right)$, both left-handed and right-handed waves are allowed to exist.

\textbf{Case 3.8.3}: $\mathcal{C}_{1}\mathcal{C}_{2} \textless 0,~ A_{(1)}\mathcal{C}_{1}\textgreater0$. When $k_{3} \in \left(0, \left|2A_{(1)}A^{2}/\mathcal{C}_{2}\right|\right)$, only wave vectors that satisfy the dispersion relation (\ref{two dispersion relations vector mode 1}) meet condition (\ref{k0^2 gen 0, k3 gen 0}). Therefore, within this range, the theory only supports left-handed waves. When $k_{3} \in \left[\left|2A_{(1)}A^{2}/\mathcal{C}_{2}\right|, \infty \right)$, since neither dispersion relation satisfies condition (\ref{k0^2 gen 0, k3 gen 0}), there are no vector modes present within this range.

\textbf{Case 3.8.4}: $\mathcal{C}_{1}\mathcal{C}_{2} \textless 0,~ A_{(1)}\mathcal{C}_{1}\textless0$. In this case, when $k_{3} \in \left(0, \left|2A_{(1)}A^{2}/\mathcal{C}_{2}\right|\right)$, only wave vectors that satisfy the dispersion relation (\ref{two dispersion relations vector mode 2}) meet condition (\ref{k0^2 gen 0, k3 gen 0}). Therefore, within this range, the theory only supports right-handed waves. When $k_{3} \in \left[\left|2A_{(1)}A^{2}/\mathcal{C}_{2}\right|, \infty \right)$, there are no vector modes present within this range.

\textbf{Case 4}: $G_{(2)}A^{2}+4H_{(2)} \neq 0,~G_{(2)}+4I_{(2)} \neq 0$. For this scenario, Eq. (\ref{vector mode equations 1}) requires
\begin{eqnarray}
	\label{vector mode equations 1 Xi=Sigma}
	\Xi_{i}
	=
	-\frac
	{\left(G_{(2)}+4I_{(2)}\right)A}{4H_{(2)}+G_{(2)}A^{2}}
	\Sigma_{i}.
\end{eqnarray}
Substituting Eq. (\ref{vector mode equations 1 Xi=Sigma}) into Eq. (\ref{vector mode equations 2}), we have
\begin{eqnarray}
	\label{vector mode equations 2 in case 4}
	2\mathcal{D}_{3}E^{0ijk}A^{2}\partial_{k}\Xi_{j}
	+\mathcal{D}_{1}\partial_{0}^{2}\Xi_{i}
	-\mathcal{D}_{2}\Delta \Xi_{i}
	=0,
\end{eqnarray}
where
\begin{eqnarray}
	\label{D1}
	\mathcal{D}_{1}
	\!&=&\!
	-\frac{\left(G_{(2)}+4I_{(2)}\right)A^{2}}{4H_{(2)}+G_{(2)}A^{2}}
	\left(
	2E_{(2)}-2G_{(2)}-4I_{(2)}+2F_{(2)}A^{2}
	\right)
	\nonumber \\
	\!&+&\!
	\left(
	2E_{(2)}-2G_{(2)}-4I_{(2)}+2F_{(2)}A^{2}
	\right),
	\\
	\label{D2}
	\mathcal{D}_{2}
	\!&=&\!
	\frac{\left(G_{(2)}+4I_{(2)}\right)A^{2}}{4H_{(2)}+G_{(2)}A^{2}}
	\left(G_{(2)}-2E_{(2)}\right)
	-2\left(-E_{(2)}+G_{(2)}+2I_{(2)}\right),
	\\
	\label{D3}
	\mathcal{D}_{3}
	\!&=&\!
	-\frac{A_{(1)}A\left(G_{(2)}+4I_{(2)}\right)}{4H_{(2)}+G_{(2)}A^{2}}
	+\frac{A_{(1)}}{A}.
\end{eqnarray}

To analyze the polarization modes and wave speeds of gravitational waves in Case 4, it should be noted that the forms of Eqs. (\ref{vector mode equations 2 in case 4}) and (\ref{vector mode equations 2 in case 3}) are exactly the same. Thus, the analysis is identical to that of Case 3, with the only difference being that all instances of $\mathcal{C}_{1},~\mathcal{C}_{2},~A_{(1)}$ in Case 3 are replaced by $\mathcal{D}_{1},~\mathcal{D}_{2},~\mathcal{D}_{3}$, respectively. Therefore, the specific analysis of Case 4 will not be repeated here.

\section{The detailed classification of scalar modes}
\label{app: C}
\textbf{Case 1}: $M_{1}'=M_{3}'=M_{4}'=0$. For this scenario, Eq. (\ref{scalar mode equation 3 new}) remains zero. Therefore, Eqs. (\ref{scalar mode equation 1 new}) and (\ref{scalar mode equation 2 new}) need to constrain three variables $\phi$, $\Omega$, and $\Psi$. At least one variable can take any value, so we rule out this case.

\textbf{Case 2}: $M_{1}' \neq 0,~M_{3}'=M_{4}'=0$. In this case, Eq. (\ref{scalar mode equation 3 new}) requires $\phi=0$, and Eqs. (\ref{scalar mode equation 1 new}) and (\ref{scalar mode equation 2 new}) respectively become
\begin{eqnarray}
	\label{scalar mode equation 1 new in case 2}
	4B_{(1)}A\Omega
	+\Lambda_{3}'\partial_{0}\Omega
	+\Lambda_{4}'\partial_{0}^{2}\Psi
	+\Lambda_{5}\Delta\Psi
	=0,
	\\
	-4A^{3}A_{(0)}\Omega
	-\frac{3B_{(1)}A^{3}}{2H_{(2)}}
	\left(
	+N_{3}\partial_{0}\Omega
	+N_{4}\partial_{0}^{2}\Psi
	\right)
	+2B_{(1)}A^{2}\Delta\Psi
	\nonumber \\
	\label{scalar mode equation 2 new in case2}
	+K_{5}'\partial_{0}^{2}\Omega
	+K_{6}'\Delta\Omega
	+K_{7}'\partial_{0}\Delta\Psi
	-\frac{K_{3}N_{4}}{2H_{(2)}}\partial_{0}^{3}\Psi
	=0.
\end{eqnarray}

For the solution of a monochromatic plane wave
\begin{eqnarray}
	\label{scalar=scalar eikx}
	\phi=\mathring{\phi}e^{ikx},~
	\Omega=\mathring{\Omega}e^{ikx},~
	\Psi=\mathring{\Psi}e^{ikx},
\end{eqnarray}
the above equations can be transformed into the following matrix form:
\begin{eqnarray}
	\label{scalar mode equations 2 in case 2 matrix form}
	\begin{pmatrix}
		4B_{(1)}A+i\Lambda_{3}'k_{0} & ~
		-\Lambda_{4}'k_{0}^{2}-\Lambda_{5}k_{3}^{2} \\
		\mathcal{A}_{21} & ~
		\mathcal{A}_{22}\\
	\end{pmatrix}
	\begin{pmatrix}
		\mathring{\Omega}\\
		\mathring{\Psi}
	\end{pmatrix}
	=0,
\end{eqnarray}
where
\begin{eqnarray}
	\label{A21,A22 in case 2 matrix form}
	\mathcal{A}_{21}
	\!&=&\!
	-4A_{(0)}A^{3}-3i\frac{B_{(1)}A^{3}N_{3}}{2H_{(2)}}k_{0}-K_{5}'k_{0}^{2}-K_{6}'k_{3}^{2} ,
	\nonumber \\
	\mathcal{A}_{22}
	\!&=&\!
	3\frac{B_{(1)}A^{3}N_{4}}{2H_{(2)}}k_{0}^{2}-2B_{(1)}A^{2}k_{3}^{2}-iK_{7}'k_{0}k_{3}^{2}+i\frac{k_{3}N_{4}}{2H_{(2)}}k_{0}^{3}.
\end{eqnarray}
Equation (\ref{scalar mode equations 2 in case 2 matrix form}) with non-zero solutions equivalently requires that the determinant of the coefficient matrix be zero, that is,
\begin{eqnarray}
	\label{determinant of the coefficient matrix is 0 in case 2}
	\!&&\!
	\left(
	\frac{6A^{4}B_{(1)}^{2}N_{4}}{H_{(2)}}
	+K_{7}'\Lambda_{3}'k_{3}^{2}
	-4A^{3}A_{(0)}\Lambda_{4}'
	-K_{6}'\Lambda_{4}'k_{3}^{2}
	-K_{5}'\Lambda_{5}
	\right)k_{0}^{2}
	\nonumber \\
	\!&-&\!
	\left(
	\frac{N_{4}K_{3}\Lambda_{3}'}{2H_{(2)}}
	+K_{5}'\Lambda_{4}'
	\right)k_{0}^{4}
	+\left(
	-8A^{3}B_{(1)}^{2}k_{3}^{2}
	-4A^{3}A_{(0)}\Lambda_{5}k_{3}^{2}
	-K_{6}'\Lambda_{5}k_{3}^{4}
	\right)
	\nonumber \\
	\!&+&\!i
	\left[
	\left(
	\frac{2AB_{(1)}N_{4}k_{3}}{H_{(2)}}
	+\frac{3A^{3}B_{(1)}N_{4}\Lambda_{3}'}{2H_{(2)}}
	-\frac{3A^{3}B_{(1)}N_{3}\Lambda_{4}'}{2H_{(2)}}
	\right)k_{0}^{3}
	\right.
	\nonumber \\
	\!&+&\!
	\left.
	\left(
	-4AB_{(1)}K_{7}'k_{3}^{2}
	-2A^{2}B_{(1)}\Lambda_{3}'k_{3}^{2}
	-\frac{3A^{3}B_{(1)}N_{3}\Lambda_{5}k_{3}^{2}}{2H_{(2)}}
	\right)k_{0}
	\right]
	=0.
\end{eqnarray}
Considering that both the real and imaginary parts of the above equation should be zero simultaneously, and since the theoretical parameters and $k_{0}$, $k_{3}$ are real numbers, this equivalently requires the following two equations to hold simultaneously:
\begin{eqnarray}
	\label{real part determinant of the coefficient matrix is 0 in case 2}
	\!&&\!
	\left(
	\frac{6A^{4}B_{(1)}^{2}N_{4}}{H_{(2)}}
	+K_{7}'\Lambda_{3}'k_{3}^{2}
	-4A^{3}A_{(0)}\Lambda_{4}'
	-K_{6}'\Lambda_{4}'k_{3}^{2}
	-K_{5}'\Lambda_{5}
	\right)k_{0}^{2}
	\nonumber \\
	\!&-&\!
	\left(
	\frac{N_{4}K_{3}\Lambda_{3}'}{2H_{(2)}}
	+K_{5}'\Lambda_{4}'
	\right)k_{0}^{4}
	+\left(
	-8A^{3}B_{(1)}^{2}k_{3}^{2}
	-4A^{3}A_{(0)}\Lambda_{5}k_{3}^{2}
	-K_{6}'\Lambda_{5}k_{3}^{4}
	\right)
	=0,
	\\
	\label{imaginary part determinant of the coefficient matrix is 0 in case 2}
	\!&&\!
	\left(
	\frac{2AB_{(1)}N_{4}k_{3}}{H_{(2)}}
	+\frac{3A^{3}B_{(1)}N_{4}\Lambda_{3}'}{2H_{(2)}}
	-\frac{3A^{3}B_{(1)}N_{3}\Lambda_{4}'}{2H_{(2)}}
	\right)k_{0}^{3}
	\nonumber \\
	\!&+&\!
	\left(
	-4AB_{(1)}K_{7}'k_{3}^{2}
	-2A^{2}B_{(1)}\Lambda_{3}'k_{3}^{2}
	-\frac{3A^{3}B_{(1)}N_{3}\Lambda_{5}k_{3}^{2}}{2H_{(2)}}
	\right)k_{0}
	=0.
\end{eqnarray}

The above equation can be used to solve for the wave speed of the scalar mode. However, it is necessary to discuss and classify the parameter space.

\textbf{Case 2.1}: $\frac{2AB_{(1)}N_{4}k_{3}}{H_{(2)}}
+\frac{3A^{3}B_{(1)}N_{4}\Lambda_{3}'}{2H_{(2)}}
-\frac{3A^{3}B_{(1)}N_{3}\Lambda_{4}'}{2H_{(2)}} \neq 0$.
From Eq. (\ref{imaginary part determinant of the coefficient matrix is 0 in case 2}), we can conclude that $k_{0}$ needs to satisfy one of the following two conditions:
\begin{eqnarray}
	\label{imaginary part is 0 in case 2.1 two solution of k0 1}
	k_{0}\!&=&\!0,
	\\
	\label{imaginary part is 0 in case 2.1 two solution of k0 2}
	k_{0}^{2}
	\!&=&\!
	\frac
	{8H_{(2)}AB_{(1)}K_{7}'+4H_{(2)}A^{2}B_{(1)}\Lambda_{3}'+3A^{3}B_{(1)}N_{3}\Lambda_{5}}
	{4AB_{(1)}N_{4}K_{3}+3A^{3}B_{(1)}N_{4}\Lambda_{3}'-3A^{3}B_{(1)}N_{3}\Lambda_{4}'}k_{3}^{2}.
\end{eqnarray}
For condition (\ref{imaginary part is 0 in case 2.1 two solution of k0 1}), Eq. (\ref{real part determinant of the coefficient matrix is 0 in case 2}) becomes 
\begin{eqnarray}
	\label{real part determinant of the coefficient matrix is 0 in case 2.1 when k0=0}
	-8A^{3}B_{(1)}^{2}k_{3}^{2}
	-4A^{3}A_{(0)}\Lambda_{5}k_{3}^{2}
	-K_{6}'\Lambda_{5}k_{3}^{4}
	=0.
\end{eqnarray}
This is an algebraic equation with respect to $k_{3}$, providing the allowed solutions for $k_{3}$. Therefore, solutions with $k_{0}=0$ and $k_{3}$ satisfying Eq. (\ref{real part determinant of the coefficient matrix is 0 in case 2.1 when k0=0}) are possible wave vectors. However, since these do not actually propagate, we do not consider them.

For condition (\ref{imaginary part is 0 in case 2.1 two solution of k0 2}), there are two possible cases for substituting Eq. (\ref{imaginary part is 0 in case 2.1 two solution of k0 2}) into Eq. (\ref{real part determinant of the coefficient matrix is 0 in case 2}). The first case is when Eq. (\ref{real part determinant of the coefficient matrix is 0 in case 2}) is not always zero. In such a situation, it becomes an algebraic equation with respect to $k_{3}$. This results in $k_{3}$ having at most a finite number of discrete solutions. Therefore, there are only a finite number of possible wave vectors, and they are discretely distributed. Whether this case is physically reasonable still requires further theoretical and experimental consideration. The second case is that Eq. (\ref{real part determinant of the coefficient matrix is 0 in case 2}) always holds. In this case, the wave vector is continuously distributed. The dispersion relation of the scalar mode is given by Eq. (\ref{imaginary part is 0 in case 2.1 two solution of k0 2}), and the speed of scalar mode gravitational waves satisfies
\begin{eqnarray}
	\label{vs in case 2}
	v_{S}^{2}=
	\frac
	{8H_{(2)}AB_{(1)}K_{7}'+4H_{(2)}A^{2}B_{(1)}\Lambda_{3}'+3A^{3}B_{(1)}N_{3}\Lambda_{5}}
	{4AB_{(1)}N_{4}K_{3}+3A^{3}B_{(1)}N_{4}\Lambda_{3}'-3A^{3}B_{(1)}N_{3}\Lambda_{4}'}.
\end{eqnarray}

\textbf{Case 2.2}: $\frac{2AB_{(1)}N_{4}k_{3}}{H_{(2)}}
+\frac{3A^{3}B_{(1)}N_{4}\Lambda_{3}'}{2H_{(2)}}
-\frac{3A^{3}B_{(1)}N_{3}\Lambda_{4}'}{2H_{(2)}}=0,~-4AB_{(1)}K_{7}'
-2A^{2}B_{(1)}\Lambda_{3}'
-\frac{3A^{3}B_{(1)}N_{3}\Lambda_{5}}{2H_{(2)}} \neq0$.
From Eq. (\ref{imaginary part determinant of the coefficient matrix is 0 in case 2}), at least one of the following two conditions needs to be met:
\begin{eqnarray}
	\label{imaginary part is 0 in case 2.2 two solution of k0 1}
	k_{0}\!&=&\!0,
	\\
	\label{imaginary part is 0 in case 2.2 two solution of k0 2}
	k_{3}
	\!&=&\!
	0.
\end{eqnarray}
For condition (\ref{imaginary part is 0 in case 2.2 two solution of k0 1}), there are no propagating waves, so we do not consider this situation. For condition (\ref{imaginary part is 0 in case 2.2 two solution of k0 2}), substitute $k_{3}=0$ into Eq. (\ref{real part determinant of the coefficient matrix is 0 in case 2}). If the resulting equation only has a solution of $k_{0}=0$, then there are no propagating scalar mode gravitational waves. If the obtained equation has non-zero $k_{0}$ solutions, superluminal phenomena occur, which should be ruled out.

\textbf{Case 2.3}: $\frac{2AB_{(1)}N_{4}k_{3}}{H_{(2)}}
+\frac{3A^{3}B_{(1)}N_{4}\Lambda_{3}'}{2H_{(2)}}
-\frac{3A^{3}B_{(1)}N_{3}\Lambda_{4}'}{2H_{(2)}}=0,~-4AB_{(1)}K_{7}'
-2A^{2}B_{(1)}\Lambda_{3}'
-\frac{3A^{3}B_{(1)}N_{3}\Lambda_{5}}{2H_{(2)}}=0$. Equation (\ref{imaginary part determinant of the coefficient matrix is 0 in case 2}) always holds, and in this case, the dispersion relation only needs to be solved using Eq. (\ref{real part determinant of the coefficient matrix is 0 in case 2}). Since Eq. (\ref{real part determinant of the coefficient matrix is 0 in case 2}) is an algebraic equation with respect to $k_{0}^{2}$, we can formalize it as
\begin{eqnarray}
	\label{quadratic equation with respect to k02}
	\alpha_{1}k_{0}^{4}+\alpha_{2}k_{0}^{2}+\alpha_{3}=0,
\end{eqnarray}
where $k_{3}$ appears in the coefficients of the equation in the form of $k_{3}^{2}$ or $k_{3}^{4}$. Specifically, $\alpha_{1}$ is a constant, $\alpha_{2}$ is a linear polynomial of $k_{3}^{2}$, and $\alpha_{3}$ is a quadratic polynomial of $k_{3}^{2}$. The specific values of $\alpha_{1}$, $\alpha_{2}$, and $\alpha_{3}$ can be determined from Eq. (\ref{real part determinant of the coefficient matrix is 0 in case 2}). 

Depending on whether $\alpha_{1}$, $\alpha_{2}$, and $\alpha_{3}$ are zero, parameters can classify several cases between $k_{0}^{2}$ and $k_{3}^{2}$: (1) The equation is quadratic and has two dispersion relations; (2) The equation is linear and has one dispersion relation; (3) The equation becomes $\alpha_{3}\left[k_{3}^{2}\right]=0$, thereby constraining $k_{3}$, while $k_{0}$ can take any value; (4) The equation is always zero and $k_{0}$, $k_{3}$ can take any values; and (5) the equation does not have a solution that satisfies (\ref{z direction}). In the last case, there are no scalar mode gravitational waves. The second-to-last and third-to-last cases are physically unreasonable and need to be ruled out. The dispersion relation of the first two cases can be directly derived from the general solution of either a quadratic or a linear equation. It should be noted that $k_{0}^{2}$ is generally not proportional to $k_{3}^{2}$.

Finding the relation between scalar mode amplitudes requires solving Eq. (\ref{scalar mode equations 2 in case 2 matrix form}). For a certain dispersion relation that we determined in the previous discussion, Eq. (\ref{scalar mode equations 2 in case 2 matrix form}) also presents two possible cases. The first case is that the dispersion relation makes the coefficient matrix of Eq. (\ref{scalar mode equations 2 in case 2 matrix form}) zero, allowing $\mathring{\Omega}$ and $\mathring{\Psi}$ to take any value. For the second case, the rank of the coefficient matrix is one. In such a situation, the following relationship will be satisfied between $\mathring{\Omega}$ and $\mathring{\Psi}$:
\begin{eqnarray}
	\label{relationship between Omega and Psi in case 2}
	\left(4B_{(1)}A+i\Lambda_{3}'k_{0}\right)\Omega
	=
	\left(\Lambda_{4}'k_{0}^{2}+\Lambda_{5}k_{3}^{2}\right)\Psi.
\end{eqnarray}
In this case, $\mathring{\Omega}$ and $\mathring{\Psi}$ are not independent of each other; their ratio is a complex number. 

However, regardless of which sub case mentioned above, in Case 2, $\phi=0$. Therefore, according to Eq. (\ref{P1-P6 gauge invariant}), the scalar polarization mode is determined only by $\Theta$, which is determined by Eq. (\ref{scalar mode Theta}), and generally there is only one scalar polarization mode. (In fact, the term ``generally" here implies the need for further discussion of the parameters in Eqs. (\ref{relationship between Omega and Psi in case 2}) and (\ref{scalar mode Theta}). For instance, when $N_{3}=N_{4}=0$ in Eq. (\ref{scalar mode Theta}), $\Theta=0$, there are no scalar mode gravitational waves. However, such situations are extremely rare compared to the cases where $\Theta \neq 0 $, and they are straightforward to analyze. Due to space constraints, we will not delve into similar situations here and in the following text, but will focus on discussing the vast majority of cases. It should be noted that in this and subsequent analyses, a comprehensive analysis needs to consider these special cases.) It is a mixed mode of breathing mode and longitudinal mode, with equal amplitudes, i.e., $P_{1}=P_{6}$. 

\textbf{Case 3}: $M_{3}' \neq 0,~M_{1}'=M_{4}'=0$. In this case, Eq. (\ref{scalar mode equation 3 new}) requires $\Omega=0$, and Eqs. (\ref{scalar mode equation 1 new}) and (\ref{scalar mode equation 2 new}) respectively become
\begin{eqnarray}
	\label{scalar mode equation 1 new in case 3}
	4B_{(1)}A^{2}\phi
	+\Lambda_{1}'\partial_{0}\phi
	+\Lambda_{4}'\partial_{0}^{2}\Psi
	+\Lambda_{5}\Delta\Psi
	=0,
	\\
	-4A^{4}A_{(0)}\phi
	-\frac{3B_{(1)}A^{3}}{2H_{(2)}}
	\left(
	N_{1}\partial_{0}\phi
	+N_{4}\partial_{0}^{2}\Psi
	\right)
	+2B_{(1)}A^{2}\Delta\Psi
	\nonumber \\
	\label{scalar mode equation 2 new in case 3}
	+K_{1}'\partial_{0}^{2}\phi
	+K_{2}'\Delta\phi
	+K_{7}'\partial_{0}\Delta\Psi
	-\frac{K_{3}N_{4}}{2H_{(2)}}\partial_{0}^{3}\Psi
	=0.
\end{eqnarray}
Therefore, for the monochromatic plane wave solution (\ref{scalar=scalar eikx}), the above equations can be expressed in the following matrix form:
\begin{eqnarray}
	\label{scalar mode equations 2 in case 3 matrix form}
	\begin{pmatrix}
		4B_{(1)}A^{2}+i\Lambda_{1}'k_{0} & ~
		-\Lambda_{4}'k_{0}^{2}-\Lambda_{5}k_{3}^{2} \\
		\mathcal{A}_{21} & ~
		\mathcal{A}_{22}\\
	\end{pmatrix}
	\begin{pmatrix}
		\mathring{\phi}\\
		\mathring{\Psi}
	\end{pmatrix}
	=0,
\end{eqnarray}
where
\begin{eqnarray}
	\label{A21,A22 in case 3 matrix form}
	\mathcal{A}_{21}
	\!&=&\!
	-4A_{(0)}A^{4}-3i\frac{B_{(1)}A^{3}N_{1}}{2H_{(2)}}k_{0}-K_{1}'k_{0}^{2}-K_{2}'k_{3}^{2} ,
	\nonumber \\
	\mathcal{A}_{22}
	\!&=&\!
	3\frac{B_{(1)}A^{3}N_{4}}{2H_{(2)}}k_{0}^{2}-2B_{(1)}A^{2}k_{3}^{2}-iK_{7}'k_{0}k_{3}^{2}+i\frac{k_{3}N_{4}}{2H_{(2)}}k_{0}^{3}.
\end{eqnarray}
Similar to Case 2, if the equation has non-zero solutions, it is equivalent to requiring the determinant of the coefficient matrix to be zero. Thus, the wave vector needs to satisfy both the real and imaginary parts of the determinant being zero:
\begin{eqnarray}
	\label{real part determinant of the coefficient matrix is 0 in case 3}
	\!&&\!
	\left(
	\frac{6A^{6}B_{(1)}^{2}N_{4}}{H_{(2)}}
	+K_{7}'\Lambda_{1}'k_{3}^{2}
	-4A^{4}A_{(0)}\Lambda_{4}'
	-K_{2}'\Lambda_{4}'k_{3}^{2}
	-K_{1}'\Lambda_{5}
	\right)k_{0}^{2}
	\nonumber \\
	\!&-&\!
	\left(
	\frac{N_{4}K_{3}\Lambda_{1}'}{2H_{(2)}}
	+K_{1}'\Lambda_{4}'
	\right)k_{0}^{4}
	+\left(
	-8A^{5}B_{(1)}^{2}k_{3}^{2}
	-4A^{4}A_{(0)}\Lambda_{5}k_{3}^{2}
	-K_{2}'\Lambda_{5}k_{3}^{4}
	\right)
	=0,
	\\
	\label{imaginary part determinant of the coefficient matrix is 0 in case 3}
	\!&&\!
	\left(
	\frac{2A^{2}B_{(1)}N_{4}k_{3}}{H_{(2)}}
	+\frac{3A^{4}B_{(1)}N_{4}\Lambda_{1}'}{2H_{(2)}}
	-\frac{3A^{4}B_{(1)}N_{1}\Lambda_{4}'}{2H_{(2)}}
	\right)k_{0}^{3}
	\nonumber \\
	\!&+&\!
	\left(
	-4A^{2}B_{(1)}K_{7}'k_{3}^{2}
	-2A^{3}B_{(1)}\Lambda_{1}'k_{3}^{2}
	-\frac{3A^{4}B_{(1)}N_{1}\Lambda_{5}k_{3}^{2}}{2H_{(2)}}
	\right)k_{0}
	=0.
\end{eqnarray}
For the same considerations as in Case 2, when $\frac{2A^{2}B_{(1)}N_{4}k_{3}}{H_{(2)}}
+\frac{3A^{4}B_{(1)}N_{4}\Lambda_{1}'}{2H_{(2)}}
-\frac{3A^{4}B_{(1)}N_{1}\Lambda_{4}'}{2H_{(2)}} \neq 0$, we ignore the solution with $k_{0}=0$ and only consider the case where 
\begin{eqnarray}
	\label{imaginary part is 0 in case 3 two solution of k0 2}
	k_{0}^{2}
	\!&=&\!
	\frac
	{8H_{(2)}A^{2}B_{(1)}K_{7}'+4H_{(2)}A^{3}B_{(1)}\Lambda_{1}'+3A^{4}B_{(1)}N_{1}\Lambda_{5}}
	{4A^{2}B_{(1)}N_{4}K_{3}+3A^{4}B_{(1)}N_{4}\Lambda_{1}'-3A^{4}B_{(1)}N_{1}\Lambda_{4}'}k_{3}^{2}.
\end{eqnarray}
By substituting Eq. (\ref{imaginary part is 0 in case 3 two solution of k0 2}) into Eq. (\ref{real part determinant of the coefficient matrix is 0 in case 3}), we can still discuss it in two cases. In the first case, Eq. (\ref{real part determinant of the coefficient matrix is 0 in case 3}) still transforms into an algebraic equation with respect to $k_{3}$, and for the second case, the wave vector is continuously distributed. The speed of scalar mode gravitational waves satisfies
\begin{eqnarray}
	\label{vs in case 3}
	v_{S}^{2}=
	\frac
	{8H_{(2)}A^{2}B_{(1)}K_{7}'+4H_{(2)}A^{3}B_{(1)}\Lambda_{1}'+3A^{4}B_{(1)}N_{1}\Lambda_{5}}
	{4A^{2}B_{(1)}N_{4}K_{3}+3A^{4}B_{(1)}N_{4}\Lambda_{1}'-3A^{4}B_{(1)}N_{1}\Lambda_{4}'}.
\end{eqnarray}

For other cases, the analysis of the dispersion relation is identical to that in Case 2.2 and Case 2.3, as long as 
$\alpha_{1}$, $\alpha_{2}$ and, $\alpha_{3}$ in Eq. (\ref{quadratic equation with respect to k02}) are considered as the parameters corresponding to Eq. (\ref{real part determinant of the coefficient matrix is 0 in case 3}). 

For the analysis of the amplitude of scalar mode gravitational waves, by substituting the considered dispersion relation into Eq. (\ref{scalar mode equations 2 in case 3 matrix form}), we can conclude that there are also two possible cases for discussion. 

In the case where the coefficient matrix is zero, $\mathring{\phi}$ and $\mathring{\Psi}$ can take any values. According to Eq. (\ref{scalar mode Theta}), this indicates that $\phi$ and $\Theta$ can generally take any values. Furthermore, according to Eq. (\ref{P1-P6 gauge invariant}), scalar gravitational waves allow for two independent polarization modes: the breathing mode and the longitudinal mode.

For the case where the rank of the coefficient matrix is one, the following relationship will be satisfied between $\phi$ and $\Psi$:
\begin{eqnarray}
	\label{relationship between Omega and Psi in case 3}
	\left(4B_{(1)}A^{2}+i\Lambda_{1}'k_{0}\right)\phi
	=
	\left(\Lambda_{4}'k_{0}^{2}+\Lambda_{5}k_{3}^{2}\right)\Psi.
\end{eqnarray}
For this scenario, $\phi$ and $\Theta$ are not independent, and the amplitude ratio between them forms a complex number. Scalar mode gravitational waves have only one polarization mode, which is a mixture of two modes: (1) a pure longitudinal mode (is determined by $\phi$), and (2) a mixed mode of the breathing mode and longitudinal mode, with equal amplitude for both (is determined by $\Theta$). There exists a phase difference between these two mixed modes. The specific values of the amplitude ratio and phase difference of these two mixed modes can be easily obtained from equations (\ref{relationship between Omega and Psi in case 3}), (\ref{scalar mode Theta}), and (\ref{P1-P6 gauge invariant}). Therefore, we will not list them in this paper.

\textbf{Case 4}: $M_{4}' \neq 0,~M_{1}'=M_{3}'=0$. In such a situation, Eq. (\ref{scalar mode equation 3 new}) requires $\Psi=0$, and Eqs. (\ref{scalar mode equation 1 new}) and (\ref{scalar mode equation 2 new}) respectively become
\begin{eqnarray}
	\label{scalar mode equation 1 new in case 4}
	4B_{(1)}A^{2}\phi
	+4B_{(1)}A\Omega
	+\Lambda_{1}'\partial_{0}\phi
	+\Lambda_{3}'\partial_{0}\Omega
	=0,
	\\
	-4A^{3}A_{(0)}\left(\Omega+A\phi\right)
	-\frac{3B_{(1)}A^{3}}{2H_{(2)}}
	\left(
	N_{1}\partial_{0}\phi
	+N_{3}\partial_{0}\Omega
	\right)
	\nonumber \\
	\label{scalar mode equation 2 new in case 4}
	+K_{1}'\partial_{0}^{2}\phi
	+K_{2}'\Delta\phi
	+K_{5}'\partial_{0}^{2}\Omega
	+K_{6}'\Delta\Omega
	=0.
\end{eqnarray}
When considering monochromatic plane wave solutions, the above equation can be expressed in the following matrix form:
\begin{eqnarray}
	\label{scalar mode equations 2 in case 4 matrix form}
	\begin{pmatrix}
		4B_{(1)}A^{2}+i\Lambda_{1}'k_{0} & ~
		4B_{(1)}A+i\Lambda_{3}'k_{0} \\
		\mathcal{A}_{21} & ~
		\mathcal{A}_{22}\\
	\end{pmatrix}
	\begin{pmatrix}
		\mathring{\phi}\\
		\mathring{\Omega}
	\end{pmatrix}
	=0,
\end{eqnarray}
where
\begin{eqnarray}
	\label{A21,A22 in case 4 matrix form}
	\mathcal{A}_{21}
	\!&=&\!
	-4A_{(0)}A^{4}-3i\frac{B_{(1)}A^{3}N_{1}}{2H_{(2)}}k_{0}-K_{1}'k_{0}^{2}-K_{2}'k_{3}^{2} ,
	\nonumber \\
	\mathcal{A}_{22}
	\!&=&\!
	-4A^{3}A_{(0)}-3i\frac{B_{(1)}A^{3}N_{3}k_{0}}{2H_{(2)}}-K_{5}'k_{0}^{2}-K_{6}'k_{3}^{2}.
\end{eqnarray}

Similarly, requiring the determinant of the coefficient matrix to be zero is equivalent to requiring the wave vector to satisfy the following equations simultaneously:
\begin{eqnarray}
	\label{real part determinant of the coefficient matrix is 0 in case 4}
	\!&&\!
	\left(
	-4B_{(1)}A^{2}K_{5}'
	+\frac{3\Lambda_{1}'B_{(1)}A^{3}N_{3}}{2H_{(2)}}
	+4K_{1}'B_{(1)}A
	-3\frac{\Lambda_{3}'B_{(1)}A^{3}N_{1}}{2H_{(2)}}
	\right)k_{0}^{2}
	\nonumber \\
	\!&+&\!
	\left(
	-4B_{(1)}K_{6}'A^{2}+4B_{(1)}K_{2}'A
	\right)k_{3}^{2}
	=0,
	\\
	\label{imaginary part determinant of the coefficient matrix is 0 in case 4}
	\!&&\!
	\left(
	-K_{5}'\Lambda_{1}'+K_{1}'\Lambda_{3}'
	\right)k_{0}^{3}
	+\left(
	-\frac{6B_{(1)}^{2}A^{5}N_{3}}{H_{(2)}}
	-4A_{(0)}A^{3}\Lambda_{1}'
	-K_{6}'\Lambda_{1}'k_{3}^{2}
	\right.
	\nonumber \\
	\!&+&\!
	\left.
	4A_{(0)}\Lambda_{3}'A^{4}
	+\Lambda_{3}'K_{2}'k_{3}^{2}
	+6\frac{B_{(1)}^{2}A^{4}N_{1}}{H_{(2)}}
	\right)k_{0}
	=0.
\end{eqnarray}
We need to classify and discuss the parameters as in Case 2.

\textbf{Case 4.1}: $-4B_{(1)}A^{2}K_{5}'
+\frac{3\Lambda_{1}'B_{(1)}A^{3}N_{3}}{2H_{(2)}}
+4K_{1}'B_{(1)}A
-3\frac{\Lambda_{3}'B_{(1)}A^{3}N_{1}}{2H_{(2)}} \neq 0$. From Eq. (\ref{real part determinant of the coefficient matrix is 0 in case 4}), the wave vector satisfies
\begin{eqnarray}
	\label{imaginary part is 0 in case 4 two solution of k0 2}
	k_{0}^{2}
	\!&=&\!
	\frac
	{8H_{(2)}A^{2}B_{(1)}K_{6}'-8H_{(2)}B_{(1)}K_{2}'A}
	{8H_{(2)}K_{1}'B_{(1)}A-8H_{(2)}B_{(1)}K_{5}'A^{2}+3\Lambda_{1}'B_{(1)}N_{3}A^{3}-3\Lambda_{3}'B_{(1)}N_{1}A^{3}}k_{3}^{2}.
\end{eqnarray}
Substituting Eq. (\ref{imaginary part is 0 in case 4 two solution of k0 2}) into Eq. (\ref{imaginary part determinant of the coefficient matrix is 0 in case 4}), the discussion can be divided into two cases based on whether the resulting equation is always zero. This discussion is completely similar to Case 2.1, as before, which will not be repeated here. Therefore, the speed of scalar mode gravitational waves satisfies
\begin{eqnarray}
	\label{vs in case 4}
	v_{S}^{2}=
	\frac
	{8H_{(2)}A^{2}B_{(1)}K_{6}'-8H_{(2)}B_{(1)}K_{2}'A}
	{8H_{(2)}K_{1}'B_{(1)}A-8H_{(2)}B_{(1)}K_{5}'A^{2}+3\Lambda_{1}'B_{(1)}N_{3}A^{3}-3\Lambda_{3}'B_{(1)}N_{1}A^{3}}.
\end{eqnarray}

\textbf{Case 4.2}: $-4B_{(1)}A^{2}K_{5}'
+\frac{3\Lambda_{1}'B_{(1)}A^{3}N_{3}}{2H_{(2)}}
+4K_{1}'B_{(1)}A
-3\frac{\Lambda_{3}'B_{(1)}A^{3}N_{1}}{2H_{(2)}} = 0$, $-4B_{(1)}K_{6}'A^{2}+4B_{(1)}K_{2}'A \neq 0$. From Eq. (\ref{real part determinant of the coefficient matrix is 0 in case 4}), we have $k_{3}=0$. Similar to the discussion in Case 2.2, substituting $k_{3}=0$ into Eq. (\ref{imaginary part is 0 in case 4 two solution of k0 2}) yields an algebraic equation with respect to $k_{0}$. If the obtained equation only allows $k_{0}=0$, then there are no propagating scalar mode gravitational waves. If a solution with $k_{0} \neq 0$ exists, then superluminal phenomena occur, and such parameters need to be ruled out.

\textbf{Case 4.3}: $-4B_{(1)}A^{2}K_{5}'
+\frac{3\Lambda_{1}'B_{(1)}A^{3}N_{3}}{2H_{(2)}}
+4K_{1}'B_{(1)}A
-3\frac{\Lambda_{3}'B_{(1)}A^{3}N_{1}}{2H_{(2)}} = 0$, $-4B_{(1)}K_{6}'A^{2}+4B_{(1)}K_{2}'A = 0$. . For this case, Eq. (\ref{real part determinant of the coefficient matrix is 0 in case 4}) is always zero, so only Eq. (\ref{imaginary part determinant of the coefficient matrix is 0 in case 4}) needs to be solved. It can be seen that $k_{0}=0$ is a solution to the equation. However, such a solution cannot represent propagating gravitational waves, so we only consider the case where $k_{0} \neq 0$. For this scenario, dividing Eq. (\ref{imaginary part determinant of the coefficient matrix is 0 in case 4}) by $k_{0}$ yields an algebraic equation that satisfies the following form:
\begin{eqnarray}
	\label{linear equation with respect to k02 in case 4.3}
	\beta_{1}k_{0}^{2}+\beta_{2}=0,
\end{eqnarray}
where $k_{3}$ appears in the coefficients of the equation in the form of $k_{3}^{2}$. Specifically, $\beta_{1}$ is a constant and $\beta_{2}$ is a linear polynomial of $k_{3}^{2}$. The specific values of $\beta_{1}$ and $\beta_{2}$ can be determined from Eq. (\ref{imaginary part determinant of the coefficient matrix is 0 in case 4}). 

We can further classify the parameters of Eq. (\ref{linear equation with respect to k02 in case 4.3}) by considering whether $\beta_{1}$ and $\beta_{2}$ are zero, resulting in the following cases:
(1) the equation is linear and has one dispersion relation: $k_{0}^{2}=-\beta_{2}/\beta_{1}$; (It should be noted that $k_{0}^{2}$ is generally not proportional to $k_{3}^{2}$.) (2) the equation becomes $\beta_{2}\left[k_{3}^{2}\right]=0$, thereby constraining $k_{3}$, while $k_{0}$ can take any value; (3) the equation is always zero and $k_{0}$, $k_{3}$ can take any values; and (4) the equation does not have a solution that satisfies (\ref{z direction}). In the last case, there are no scalar mode gravitational waves. The second-to-last and third-to-last cases are physically unreasonable and need to be ruled out.

For the analysis of the amplitude of scalar mode gravitational waves, we substitute the considered dispersion relation into Eq. (\ref{scalar mode equations 2 in case 4 matrix form}). If the coefficient matrix is zero, then $\phi$ and $\Omega$ can take any value. Furthermore, from Eq. (\ref{scalar mode Theta}), $\phi$ and $\Theta$ can take any values. Therefore, scalar gravitational waves have two independent polarization modes: the breathing mode and the longitudinal mode. If the rank of the coefficient matrix is one, then there exists a relationship:
\begin{eqnarray}
	\label{relationship between Omega and Psi in case 4}
	\left(4B_{(1)}A^{2}+i\Lambda_{1}'k_{0}\right)\phi
	=
	\left(4B_{(1)}A+i\Lambda_{3}'k_{0}\right)\Omega.
\end{eqnarray}
At this point, $\phi$ and $\Theta$ are not independent, and the amplitude ratio between them forms a complex number. Scalar mode gravitational waves have only one polarization mode, which is a mixture of two modes: (1) a pure longitudinal mode (determined by $\phi$), and (2) a mixed mode of breathing mode and longitudinal mode, with equal amplitude for both (determined by $\Theta$).

\textbf{Case 5}: $M_{1}'=0,~M_{3}' \neq 0,~M_{4}' \neq 0$. In this case, Eq. (\ref{scalar mode equation 3 new}) requires 
\begin{eqnarray}
	\label{Omega in case 5}
	\Omega=-\frac{M_{4}'}{M_{3}'}\partial_{0}\Psi,
\end{eqnarray}
and Eqs. (\ref{scalar mode equation 1 new}) and (\ref{scalar mode equation 2 new}) respectively become
\begin{eqnarray}
	\label{scalar mode equation 1 new in case 5}
	4B_{(1)}A^{2}\phi
	-\frac{4B_{(1)}AM_{4}'}{M_{3}'}\partial_{0}\Psi
	+\Lambda_{1}'\partial_{0}\phi
	+\left(
	-\frac{\Lambda_{3}'M_{4}'}{M_{3}'}
	+\Lambda_{4}'
	\right)\partial_{0}^{2}\Psi
	+\Lambda_{5}\Delta\Psi
	=0,
	\\
	\label{scalar mode equation 2 new in case 5}
	-4A_{(0)}A^{4}\phi
	+\frac{4A_{(0)}A^{3}M_{4}'}{M_{3}'}\partial_{0}\Psi
	-\frac{3B_{(1)}A^{3}N_{1}}{2H_{(2)}}\partial_{0}\phi
	\nonumber \\
	+\left(
	\frac{3B_{(1)}A^{3}N_{3}M_{4}'}{2H_{(2)}M_{3}'}
	-\frac{3B_{(1)}A^{3}N_{4}}{2H_{(2)}}
	\right)\partial_{0}^{2}\Psi
	+2B_{(1)}A^{2}\Delta\Psi
	+K_{1}'\partial_{0}^{2}\phi
	+K_{2}'\Delta\phi
	\nonumber \\
	+\left(
	-\frac{K_{5}'M_{4}'}{M_{3}'}
	-\frac{K_{3}N_{4}}{2H_{(2)}}
	\right)\partial_{0}^{3}\Psi
	+\left(
	-\frac{M_{4}'K_{6}'}{M_{3}'}
	+K_{7}'
	\right)\partial_{0}\Delta\Psi
	=0.
\end{eqnarray}
When considering monochromatic plane wave solutions, the equations above can be expressed in matrix form as:
\begin{eqnarray}
	\label{scalar mode equations 2 in case 5 matrix form}
	\begin{pmatrix}
		4B_{(1)}A^{2}+i\Lambda_{1}'k_{0} & ~
		\mathcal{A}_{12} \\
		-4A_{(0)}A^{4}-3i\frac{B_{(1)}A^{3}N_{1}}{2H_{(2)}}k_{0}-K_{1}'k_{0}^{2}-K_{2}'k_{3}^{2} & ~
		\mathcal{A}_{22}\\
	\end{pmatrix}
	\begin{pmatrix}
		\mathring{\phi}\\
		\mathring{\Psi}
	\end{pmatrix}
	=0,
\end{eqnarray}
where
\begin{eqnarray}
	\label{A12,A22 in case 5 matrix form}
	\mathcal{A}_{12}
	\!&=&\!
	-i\frac{4B_{(1)}AM_{4}'}{M_{3}'}k_{0}
	+\frac{\Lambda_{3}'M_{4}'}{M_{3}'}k_{0}^{2}
	-\Lambda_{4}'k_{0}^{2}
	-\Lambda_{5}k_{3}^{2},
	\nonumber \\
	\mathcal{A}_{22}
	\!&=&\!
	i\frac{4A_{(0)}A^{3}M_{4}'}{M_{3}'}k_{0}
	-\frac{3B_{(1)}A^{3}N_{3}M_{4}'}{2H_{(2)}M_{3}'}k_{0}^{2}
	+\frac{3B_{(1)}A^{3}N_{4}}{2H_{(2)}}k_{0}^{2}
	-2B_{(1)}A^{2}k_{3}^{2}
	\nonumber \\
	\!&+&\!
	i\frac{K_{5}'M_{4}'}{M_{3}'}k_{0}^{3}
	+i\frac{K_{3}N_{4}}{2H_{(2)}}k_{0}^{3}
	+i\left(
	\frac{M_{4}'K_{6}'}{M_{3}'}
	-K_{7}'
	\right)k_{0}k_{3}^{2}.
\end{eqnarray}

We require the determinant of the coefficient matrix of the above equation to be zero, that is, 
\begin{eqnarray}
	\label{real part determinant of the coefficient matrix is 0 in case 5}
	\!&&\!
	\left(
	-\frac{K_{3}N_{4}\Lambda_{1}'}{2H_{(2)}}
	-\frac{K_{5}'M_{4}'\Lambda_{1}'}{M_{3}'}
	+\frac{K_{1}'M_{4}'\Lambda_{3}'}{M_{3}'}
	-K_{1}'\Lambda_{4}'
	\right)k_{0}^{4}
	+\left(
	\frac{6A^{5}B_{(1)}^{2}N_{4}}{H_{(2)}}
	-\Lambda_{5}K_{1}'k_{3}^{2}
	\right.
	\nonumber \\
	\!&+&\!
	\frac{6B_{(1)}^{2}A^{4}N_{1}M_{4}'}{H_{(2)}M_{3}'}
	-\frac{6B_{(1)}^{2}A^{5}N_{3}M_{4}'}{H_{(2)}M_{3}'}
	+K_{7}'\Lambda_{1}'k_{3}^{2}
	-\frac{4A_{(0)}A^{3}M_{4}'\Lambda_{1}'}{M_{3}'}
	-\frac{K_{6}'M_{4}'\Lambda_{1}'k_{3}^{2}}{M_{3}'}
	\nonumber \\
	\!&+&\!
	\left.
	\frac{4A_{(0)}A^{4}M_{4}'\Lambda_{3}'}{M_{3}'}
	+\frac{K_{2}'M_{4}'\Lambda_{3}'k_{3}^{2}}{M_{3}'}
	-4A_{(0)}A^{4}\Lambda_{4}'
	-K_{2}'\Lambda_{4}'k_{3}^{2}
	\right)k_{0}^{2}
	\nonumber \\
	\!&+&\!
	\left(
	-8B_{(1)}^{2}A^{4}k_{3}^{2}
	-4A_{(0)}A^{4}\Lambda_{5}k_{3}^{2}
	-K_{2}'\Lambda_{5}k_{3}^{4}
	\right)
	=0,
	\\
	\label{imaginary part determinant of the coefficient matrix is 0 in case 5}
	\!&&\!
	\left(
	\frac{2B_{(1)}A^{2}N_{4}K_{3}}{H_{(2)}}
	-\frac{4B_{(1)}AK_{1}'M_{4}'}{M_{3}'}
	+\frac{4B_{(1)}A^{2}K_{5}'M_{4}'}{M_{3}'}
	+\frac{3B_{(1)}A^{3}N_{4}\Lambda_{1}'}{2H_{(2)}}
	\right.
	\nonumber \\
	\!&-&\!
	\left.
	\frac{3B_{(1)}A^{3}N_{3}M_{4}'\Lambda_{1}'}{2H_{(2)}M_{3}'}
	+\frac{3B_{(1)}A^{3}N_{1}M_{4}'\Lambda_{3}'}{2H_{(2)}M_{3}'}
	-\frac{3B_{(1)}A^{3}N_{1}\Lambda_{4}'}{2H_{(2)}}
	\right)k_{0}^{3}
	\nonumber \\
	\!&+&\!
	\left(
	-\frac{3B_{(1)}A^{3}N_{1}\Lambda_{5}k_{3}^{2}}{2H_{(2)}}
	-4B_{(1)}A^{2}K_{7}'k_{3}^{2}
	-\frac{4B_{(1)}AK_{2}'M_{4}'k_{3}^{2}}{M_{3}'}
	\right.
	\nonumber \\
	\!&+&\!
	\left.
	\frac{4B_{(1)}A^{2}K_{6}'M_{4}'k_{3}^{2}}{M_{3}'}
	-2B_{(1)}A^{2}\Lambda_{1}'k_{3}^{2}
	\right)k_{0}
	=0.
\end{eqnarray}
The analysis of the dispersion relation in Case 5 is entirely similar to that in Case 2. In Case 2, we classified the parameter space of Eq. (\ref{imaginary part determinant of the coefficient matrix is 0 in case 2}), while in Case 5, we classified the parameter space of Eq. (\ref{imaginary part determinant of the coefficient matrix is 0 in case 5}). Equation (\ref{imaginary part is 0 in case 2.1 two solution of k0 2}) in Case 2 corresponds to
\begin{eqnarray}
	\label{imaginary part is 0 in case 5.1 two solution of k0 2}
	k_{0}^{2}=v_{S}^{2}k_{3}^{2}
\end{eqnarray}
in Case 5, where
\begin{eqnarray}
	\label{vS in case 5.1}
	v_{S}^{2}=\frac{\mathcal{D}_{1}}{\mathcal{D}_{2}}.
\end{eqnarray}
Here,
\begin{eqnarray}
	\label{D1,D2 in case 5.1}
	\mathcal{D}_{1}
	\!&=&\!
	\frac{3B_{(1)}A^{3}N_{1}\Lambda_{5}}{2H_{(2)}}
	+4B_{(1)}A^{2}K_{7}'
	+\frac{4B_{(1)}AK_{2}'M_{4}'}{M_{3}'}
	-\frac{4B_{(1)}A^{2}K_{6}'M_{4}'}{M_{3}'}
	+2B_{(1)}A^{2}\Lambda_{1}',
	\nonumber \\
	\mathcal{D}_{2}
	\!&=&\!
	\frac{2B_{(1)}A^{2}N_{4}K_{3}}{H_{(2)}}
	-\frac{4B_{(1)}AK_{1}'M_{4}'}{M_{3}'}
	+\frac{4B_{(1)}A^{2}K_{5}'M_{4}'}{M_{3}'}
	+\frac{3B_{(1)}A^{3}N_{4}\Lambda_{1}'}{2H_{(2)}}
	\nonumber \\
	\!&-&\!
	\frac{3B_{(1)}A^{3}N_{3}M_{4}'\Lambda_{1}'}{2H_{(2)}M_{3}'}
	+\frac{3B_{(1)}A^{3}N_{1}M_{4}'\Lambda_{3}'}{2H_{(2)}M_{3}'}
	-\frac{3B_{(1)}A^{3}N_{1}\Lambda_{4}'}{2H_{(2)}}.
\end{eqnarray}
And in this case, the values of $\alpha_{1}$, $\alpha_{2}$, and $\alpha_{3}$ in Eq. (\ref{quadratic equation with respect to k02}) are taken according to Eq. (\ref{real part determinant of the coefficient matrix is 0 in case 5}). The specific analysis will not be repeated.

For the analysis of the amplitude of scalar mode gravitational waves, considering the dispersion relation we are examining, if the coefficient matrix is zero in Eq. (\ref{scalar mode equations 2 in case 5 matrix form}), then $\phi$ and $\Psi$ are independent of each other. Scalar gravitational waves generally have two independent polarization modes. If the rank of the coefficient matrix is one, there is a relationship:
\begin{eqnarray}
	\label{relationship between phi and Psi in case 5}
	\left(4B_{(1)}A^{2}+i\Lambda_{1}'k_{0}\right)\phi
	+
	\left(
	-i\frac{4B_{(1)}AM_{4}'k_{0}}{M_{3}'}
	+\frac{\Lambda_{3}'M_{4}'k_{0}^{2}}{M_{3}'}
	-\Lambda_{4}'k_{0}^{2}
	-\Lambda_{5}k_{3}^{2}
	\right)\Psi
	=0.
\end{eqnarray}
Here, $\phi$ and $\Theta$ are not independent, and the amplitude ratio between them forms a complex number. Scalar mode gravitational waves have only one polarization mode, which is a mixture of two modes: (1) a pure longitudinal mode (determined by $\phi$), and (2) a mixed mode of the breathing mode and longitudinal mode, with equal amplitude for both (determined by $\Theta$).

\textbf{Case 6}: $M_{3}'=0,~M_{1}' \neq 0,~M_{4}' \neq 0$. In this case, Eq. (\ref{scalar mode equation 3 new}) requires 
\begin{eqnarray}
	\label{phi in case 6}
	\phi=-\frac{M_{4}'}{M_{1}'}\partial_{0}\Psi,
\end{eqnarray}
and Eqs. (\ref{scalar mode equation 1 new}) and (\ref{scalar mode equation 2 new}) respectively become
\begin{eqnarray}
	\label{scalar mode equation 1 new in case 6}
	-\frac{4B_{(1)}A^{2}M_{4}'}{M_{1}'}\partial_{0}\Psi
	+4B_{(1)}A\Omega
	+\left(
	-\frac{\Lambda_{1}'M_{4}'}{M_{1}'}
	+\Lambda_{4}'
	\right)\partial_{0}^{2}\Psi
	+\Lambda_{5}\Delta\Psi
	+\Lambda_{3}'\partial_{0}\Omega
	=0,
	\\
	\label{scalar mode equation 2 new in case 6}
	-4A_{(0)}A^{3}\Omega
	+\frac{4A_{(0)}A^{4}M_{4}'}{M_{1}'}\partial_{0}\Psi
	-\frac{3B_{(1)}A^{3}N_{3}}{2H_{(2)}}\partial_{0}\Omega
	\nonumber \\
	+\left(
	\frac{3B_{(1)}A^{3}N_{1}M_{4}'}{2H_{(2)}M_{1}'}
	-\frac{3B_{(1)}A^{3}N_{4}}{2H_{(2)}}
	\right)\partial_{0}^{2}\Psi
	+2B_{(1)}A^{2}\Delta\Psi
	+K_{5}'\partial_{0}^{2}\Omega
	+K_{6}'\Delta\Omega
	\nonumber \\
	+\left(
	-\frac{K_{1}'M_{4}'}{M_{1}'}
	-\frac{K_{3}N_{4}}{2H_{(2)}}
	\right)\partial_{0}^{3}\Psi
	+\left(
	-\frac{K_{2}'M_{4}'}{M_{1}'}
	+K_{7}'
	\right)\partial_{0}\Delta\Psi
	=0.
\end{eqnarray}
Considering monochromatic plane wave solutions, the equations above can be expressed in matrix form as:
\begin{eqnarray}
	\label{scalar mode equations 2 in case 6 matrix form}
	\begin{pmatrix}
		4B_{(1)}A+i\Lambda_{3}'k_{0} & ~
		\mathcal{A}_{12} \\
		-4A_{(0)}A^{3}-3i\frac{B_{(1)}A^{2}N_{3}}{2H_{(2)}}k_{0}-K_{5}'k_{0}^{2}-K_{6}'k_{3}^{2} & ~
		\mathcal{A}_{22}\\
	\end{pmatrix}
	\begin{pmatrix}
		\mathring{\Omega}\\
		\mathring{\Psi}
	\end{pmatrix}
	=0,
\end{eqnarray}
where
\begin{eqnarray}
	\label{A12,A22 in case 6 matrix form}
	\mathcal{A}_{12}
	\!&=&\!
	-i\frac{4B_{(1)}A^{2}M_{4}'}{M_{1}'}k_{0}
	+\frac{\Lambda_{1}'M_{4}'}{M_{1}'}k_{0}^{2}
	-\Lambda_{4}'k_{0}^{2}
	-\Lambda_{5}k_{3}^{2},
	\nonumber \\
	\mathcal{A}_{22}
	\!&=&\!
	i\frac{4A_{(0)}A^{4}M_{4}'}{M_{1}'}k_{0}
	-\frac{3B_{(1)}A^{3}N_{1}M_{4}'}{2H_{(2)}M_{1}'}k_{0}^{2}
	+\frac{3B_{(1)}A^{3}N_{4}}{2H_{(2)}}k_{0}^{2}
	-2B_{(1)}A^{2}k_{3}^{2}
	\nonumber \\
	\!&+&\!
	i\frac{K_{1}'M_{4}'}{M_{1}'}k_{0}^{3}
	+i\frac{K_{3}N_{4}}{2H_{(2)}}k_{0}^{3}
	+i\left(
	\frac{M_{4}'K_{2}'}{M_{1}'}
	-K_{7}'
	\right)k_{0}k_{3}^{2}.
\end{eqnarray}

The determinant of the coefficient matrix of Eq. (\ref{scalar mode equations 2 in case 6 matrix form}) is zero, which is equivalent to the following equations:
\begin{eqnarray}
	\label{real part determinant of the coefficient matrix is 0 in case 6}
	\!&&\!
	\left(
	-\frac{K_{3}N_{4}\Lambda_{3}'}{2H_{(2)}}
	-\frac{K_{1}'M_{4}'\Lambda_{3}'}{M_{1}'}
	+\frac{K_{5}'M_{4}'\Lambda_{1}'}{M_{1}'}
	-K_{5}'\Lambda_{4}'
	\right)k_{0}^{4}
	+\left(
	\frac{6A^{4}B_{(1)}^{2}N_{4}}{H_{(2)}}
	-\Lambda_{5}K_{1}'k_{3}^{2}
	\right.
	\nonumber \\
	\!&+&\!
	\frac{6B_{(1)}^{2}A^{5}N_{3}M_{4}'}{H_{(2)}M_{1}'}
	-\frac{6B_{(1)}^{2}A^{4}N_{1}M_{4}'}{H_{(2)}M_{1}'}
	+K_{7}'\Lambda_{3}'k_{3}^{2}
	-\frac{4A_{(0)}A^{4}M_{4}'\Lambda_{3}'}{M_{1}'}
	-\frac{K_{2}'M_{4}'\Lambda_{3}'k_{3}^{2}}{M_{1}'}
	\nonumber \\
	\!&+&\!
	\left.
	\frac{4A_{(0)}A^{3}M_{4}'\Lambda_{1}'}{M_{1}'}
	+\frac{K_{6}'M_{4}'\Lambda_{1}'k_{3}^{2}}{M_{1}'}
	-4A_{(0)}A^{3}\Lambda_{4}'
	-K_{6}'\Lambda_{4}'k_{3}^{2}
	\right)k_{0}^{2}
	\nonumber \\
	\!&+&\!
	\left(
	-8B_{(1)}^{2}A^{3}k_{3}^{2}
	-4A_{(0)}A^{3}\Lambda_{5}k_{3}^{2}
	-K_{6}'\Lambda_{5}k_{3}^{4}
	\right)
	=0,
	\\
	\label{imaginary part determinant of the coefficient matrix is 0 in case 6}
	\!&&\!
	\left(
	\frac{2B_{(1)}AN_{4}K_{3}}{H_{(2)}}
	-\frac{4B_{(1)}A^{2}K_{5}'M_{4}'}{M_{1}'}
	+\frac{4B_{(1)}AK_{1}'M_{4}'}{M_{1}'}
	+\frac{3B_{(1)}A^{3}N_{4}\Lambda_{3}'}{2H_{(2)}}
	\right.
	\nonumber \\
	\!&-&\!
	\left.
	\frac{3B_{(1)}A^{3}N_{1}M_{4}'\Lambda_{3}'}{2H_{(2)}M_{1}'}
	+\frac{3B_{(1)}A^{3}N_{3}M_{4}'\Lambda_{1}'}{2H_{(2)}M_{1}'}
	-\frac{3B_{(1)}A^{3}N_{3}\Lambda_{4}'}{2H_{(2)}}
	\right)k_{0}^{3}
	\nonumber \\
	\!&+&\!
	\left(
	-\frac{3B_{(1)}A^{3}N_{3}\Lambda_{5}k_{3}^{2}}{2H_{(2)}}
	-4B_{(1)}AK_{7}'k_{3}^{2}
	-\frac{4B_{(1)}A^{2}K_{6}'M_{4}'k_{3}^{2}}{M_{1}'}
	\right.
	\nonumber \\
	\!&+&\!
	\left.
	\frac{4B_{(1)}AK_{2}'M_{4}'k_{3}^{2}}{M_{1}'}
	-2B_{(1)}A^{2}\Lambda_{3}'k_{3}^{2}
	\right)k_{0}
	=0.
\end{eqnarray}
The analysis of the dispersion relation is also entirely parallel to Case 2. Here, we classify the parameter space of Eq. (\ref{imaginary part determinant of the coefficient matrix is 0 in case 6}). In this case, Eq. (\ref{imaginary part is 0 in case 2.1 two solution of k0 2}) corresponds to
\begin{eqnarray}
	\label{imaginary part is 0 in case 6.1 two solution of k0 2}
	k_{0}^{2}=v_{S}^{2}k_{3}^{2},
\end{eqnarray}
where
\begin{eqnarray}
	\label{vS in case 6.1}
	v_{S}^{2}=\frac{\mathcal{D}_{1}}{\mathcal{D}_{2}},
\end{eqnarray}
and
\begin{eqnarray}
	\label{D1,D2 in case 6.1}
	\mathcal{D}_{1}
	\!&=&\!
	\frac{3B_{(1)}A^{3}N_{3}\Lambda_{5}}{2H_{(2)}}
	+4B_{(1)}AK_{7}'
	+\frac{4B_{(1)}A^{2}K_{6}'M_{4}'}{M_{1}'}
	-\frac{4B_{(1)}AK_{2}'M_{4}'}{M_{1}'}
	+2B_{(1)}A^{2}\Lambda_{3}',
	\nonumber \\
	\mathcal{D}_{2}
	\!&=&\!
	\frac{2B_{(1)}AN_{4}K_{3}}{H_{(2)}}
	-\frac{4B_{(1)}A^{2}K_{5}'M_{4}'}{M_{1}'}
	+\frac{4B_{(1)}AK_{1}'M_{4}'}{M_{1}'}
	+\frac{3B_{(1)}A^{3}N_{4}\Lambda_{3}'}{2H_{(2)}}
	\nonumber \\
	\!&-&\!
	\frac{3B_{(1)}A^{3}N_{1}M_{4}'\Lambda_{3}'}{2H_{(2)}M_{1}'}
	+\frac{3B_{(1)}A^{3}N_{3}M_{4}'\Lambda_{1}'}{2H_{(2)}M_{1}'}
	-\frac{3B_{(1)}A^{3}N_{3}\Lambda_{4}'}{2H_{(2)}}.
\end{eqnarray}
The values of $\alpha_{1}$, $\alpha_{2}$, and $\alpha_{3}$ in Eq. (\ref{quadratic equation with respect to k02}) are taken according to Eq. (\ref{real part determinant of the coefficient matrix is 0 in case 6}). The specific analysis will not be repeated.

For the analysis of the amplitude of scalar mode gravitational waves, considering the dispersion relation we are examining, if the coefficient matrix is zero in Eq. (\ref{scalar mode equations 2 in case 6 matrix form}), then $\Omega$ and $\Psi$ are independent of each other. Scalar gravitational waves generally have two independent polarization modes: the breathing mode and the longitudinal mode. If the rank of the coefficient matrix is one, there is a relationship
\begin{eqnarray}
	\label{relationship between phi and Psi in case 6}
	\left(4B_{(1)}A+i\Lambda_{3}'k_{0}\right)\Omega
	+
	\left(
	-i\frac{4B_{(1)}A^{2}M_{4}'k_{0}}{M_{1}'}
	+\frac{\Lambda_{1}'M_{4}'k_{0}^{2}}{M_{1}'}
	-\Lambda_{4}'k_{0}^{2}
	-\Lambda_{5}k_{3}^{2}
	\right)\Psi
	=0.
\end{eqnarray}
Here, $\phi$ and $\Theta$ are not independent, and the amplitude ratio between them forms a complex number. Scalar mode gravitational waves have only one polarization mode, which is a mixture of two modes: (1) a pure longitudinal mode (determined by $\phi$), and (2) a mixed mode of the breathing mode and longitudinal mode, with equal amplitude for both (determined by $\Theta$).

\textbf{Case 7}: $M_{4}'=0,~M_{1}' \neq 0,~M_{3}' \neq 0$. In this case, Eq. (\ref{scalar mode equation 3 new}) requires 
\begin{eqnarray}
	\label{Omega in case 7}
	\Omega=-\frac{M_{1}'}{M_{3}'}\phi,
\end{eqnarray}
and Eqs. (\ref{scalar mode equation 1 new}) and (\ref{scalar mode equation 2 new}) respectively become
\begin{eqnarray}
	\label{scalar mode equation 1 new in case 7}
	\left(
	4B_{(1)}A^{2}
	-\frac{4B_{(1)}AM_{1}'}{M_{3}'}
	\right)\phi
	+\left(
	\Lambda_{1}'
	-\frac{\Lambda_{3}'M_{1}'}{M_{3}'}
	\right)\partial_{0}\phi
	+\Lambda_{4}'\partial_{0}^{2}\Psi
	+\Lambda_{5}\Delta\Psi
	=0,
	\\
	\label{scalar mode equation 2 new in case 7}
	\left(
	-4A_{(0)}A^{4}
	+\frac{4A_{(0)}A^{3}M_{1}'}{M_{3}'}
	\right)\phi
	+\left(
	-\frac{3B_{(1)}A^{3}N_{1}}{2H_{(2)}}
	+\frac{3B_{(1)}A^{3}N_{3}M_{1}'}{2H_{(2)}M_{3}'}
	\right)\partial_{0}\phi
	\nonumber \\
	-\frac{3B_{(1)}A^{3}N_{4}}{2H_{(2)}}\partial_{0}^{2}\Psi
	+2B_{(1)}A^{2}\Delta\Psi
	+\left(
	K_{1}'
	-\frac{K_{5}'M_{1}'}{M_{3}'}
	\right)\partial_{0}^{2}\phi
	\nonumber \\
	+\left(
	K_{2}'-\frac{K_{6}'M_{1}'}{M_{3}'}
	\right)\Delta\phi
	+K_{7}'\partial_{0}\Delta\Psi
	-\frac{K_{3}N_{4}}{2H_{(2)}}\partial_{0}^{3}\Psi
	=0.
\end{eqnarray}
Considering monochromatic plane wave solutions, the equation above can be expressed in matrix form as:
\begin{eqnarray}
	\label{scalar mode equations 2 in case 7 matrix form}
	\begin{pmatrix}
		4B_{(1)}A^{2}-\frac{4B_{(1)}AM_{1}'}{M_{3}'}+i\left(\Lambda_{1}'-\frac{\Lambda_{3}'M_{1}'}{M_{3}'}\right)k_{0} & ~
		-\Lambda_{4}'k_{0}^{2}-\Lambda_{5}k_{3}^{2}\\
		\mathcal{A}_{21} & ~
		\mathcal{A}_{22}\\
	\end{pmatrix}
	\begin{pmatrix}
		\mathring{\phi}\\
		\mathring{\Psi}
	\end{pmatrix}
	=0,
\end{eqnarray}
where
\begin{eqnarray}
	\label{A21,A22 in case 7 matrix form}
	\mathcal{A}_{21}
	\!&=&\!
	-4A_{(0)}A^{4}
	+\frac{4A_{(0)}A^{3}M_{1}'}{M_{3}'}
	+i\left(
	-\frac{3B_{(1)}A^{3}N_{1}}{2H_{(2)}}
	+\frac{3B_{(1)}A^{3}N_{3}M_{1}'}{2H_{(2)}M_{3}'}
	\right)k_{0}
	\nonumber \\
	\!&-&\!
	\left(
	K_{1}'-\frac{K_{5}'M_{1}'}{M_{3}'}
	\right)k_{0}^{2}
	-\left(
	K_{2}'-\frac{K_{6}'M_{1}'}{M_{3}'}
	\right)k_{3}^{2},
	\nonumber \\
	\mathcal{A}_{22}
	\!&=&\!
	\frac{3B_{(1)}A^{3}N_{4}}{2H_{(2)}}k_{0}^{2}
	-2B_{(1)}A^{2}k_{3}^{2}
	-iK_{7}'k_{0}k_{3}^{2}
	+i\frac{K_{3}N_{4}}{2H_{(2)}}k_{0}^{3}.
\end{eqnarray}

The determinant of the coefficient matrix of Eq. (\ref{scalar mode equations 2 in case 7 matrix form}) is zero, which is equivalent to the following equations:
\begin{eqnarray}
	\label{real part determinant of the coefficient matrix is 0 in case 7}
	\!&&\!
	\left(
	-\frac{N_{4}K_{3}\Lambda_{1}'}{2H_{(2)}}
	+\frac{N_{4}K_{3}M_{1}'\Lambda_{3}'}{2H_{(2)}M_{3}'}
	+\frac{K_{5}M_{1}'\Lambda_{4}'}{M_{3}'}
	-K_{1}'\Lambda_{4}'
	\right)k_{0}^{4}
	+\left(
	\frac{6B_{(1)}A^{5}N_{4}}{H_{(2)}}
	\right.
	\nonumber \\
	\!&-&\!
	\frac{6B_{(1)}^{2}A^{4}N_{4}M_{1}'}{H_{(2)}M_{3}'}
	+\frac{K_{5}\Lambda_{5}M_{1}'k_{3}^{2}}{M_{3}'}
	-\Lambda_{5}K_{1}'k_{3}^{2}
	+K_{7}'\Lambda_{1}'k_{3}^{2}
	-\frac{M_{1}'K_{7}'\Lambda_{3}'k_{3}^{2}}{M_{3}'}
	\nonumber \\
	\!&-&\!
	\left.
	4A_{(0)}A^{4}\Lambda_{4}'
	+\frac{4A_{(0)}A^{3}M_{1}'\Lambda_{4}'}{M_{3}'}
	-K_{2}'\Lambda_{4}'k_{3}^{2}
	+\frac{M_{1}'K_{6}'\Lambda_{4}'k_{3}^{2}}{M_{3}'}
	\right)k_{0}^{2}
	\nonumber \\
	\!&+&\!
	\left(
	-8B_{(1)}^{2}A^{4}k_{3}^{2}
	-4A_{(0)}A^{4}\Lambda_{5}k_{3}^{2}
	+\frac{8B_{(1)}^{2}A^{3}M_{1}'k_{3}^{2}}{M_{3}'}
	+\frac{4A_{(0)}A^{3}\Lambda_{5}M_{1}'k_{3}^{2}}{M_{3}'}
	\right.
	\nonumber  \\
	\!&-&\!
	\left.
	\Lambda_{5}K_{2}'k_{3}^{4}
	+\frac{\Lambda_{5}M_{1}'K_{6}'k_{3}^{4}}{M_{3}'}
	\right)
	=0,
	\\
	\label{imaginary part determinant of the coefficient matrix is 0 in case 7}
	\!&&\!
	\left(
	\frac{2B_{(1)}A^{2}N_{4}K_{3}}{H_{(2)}}
	-\frac{2B_{(1)}AN_{4}K_{3}M_{1}'}{H_{(2)}M_{3}'}
	+\frac{3B_{(1)}A^{3}N_{4}\Lambda_{1}'}{2H_{(2)}}
	-\frac{3B_{(1)}A^{3}N_{4}\Lambda_{3}'M_{1}'}{2H_{(2)}M_{3}'}
	\right.
	\nonumber \\
	\!&-&\!
	\left.
	\frac{3B_{(1)}A^{3}N_{1}\Lambda_{4}'}{2H_{(2)}}
	+\frac{3B_{(1)}A^{3}N_{3}M_{1}'\Lambda_{4}'}{2H_{(2)}M_{3}'}
	\right)k_{0}^{3}
	+\left(
	-\frac{3B_{(1)}A^{3}N_{1}\Lambda_{5}}{2H_{(2)}}
	+\frac{3B_{(1)}A^{3}N_{3}\Lambda_{5}M_{1}'}{2H_{(2)}M_{3}'}
	\right.
	\nonumber \\
	\!&-&\!
	\left.
	4B_{(1)}A^{2}K_{7}'
	+\frac{4B_{(1)}AM_{1}'K_{7}'}{M_{3}'}
	-2B_{(1)}A^{2}\Lambda_{1}'
	+\frac{2B_{(1)}A^{2}M_{1}'\Lambda_{3}'}{M_{3}'}
	\right)k_{3}^{2}k_{0}=0.
\end{eqnarray}
The analysis of the dispersion relation is also entirely parallel to Case 2. Here, we classify  the parameter space of Eq. (\ref{imaginary part determinant of the coefficient matrix is 0 in case 7}). Now, Eq. (\ref{imaginary part is 0 in case 2.1 two solution of k0 2}) corresponds to
\begin{eqnarray}
	\label{imaginary part is 0 in case 7.1 two solution of k0 2}
	k_{0}^{2}=v_{S}^{2}k_{3}^{2},
\end{eqnarray}
where
\begin{eqnarray}
	\label{vS in case 7.1}
	v_{S}^{2}=\frac{\mathcal{D}_{1}}{\mathcal{D}_{2}},
\end{eqnarray}
and
\begin{eqnarray}
	\label{D1,D2 in case 7.1}
	\mathcal{D}_{1}
	\!&=&\!
	\frac{3B_{(1)}A^{3}N_{1}\Lambda_{5}}{2H_{(2)}}
	-\frac{3B_{(1)}A^{3}N_{3}\Lambda_{5}M_{1}'}{2H_{(2)}M_{3}'}
	+4B_{(1)}A^{2}K_{7}'
	\nonumber \\
	\!&-&\!
	\frac{4B_{(1)}AM_{1}'K_{7}'}{M_{3}'}
	+2B_{(1)}A^{2}\Lambda_{1}'
	-\frac{2B_{(1)}A^{2}M_{1}'\Lambda_{3}'}{M_{3}'},
	\nonumber \\
	\mathcal{D}_{2}
	\!&=&\!
	\frac{2B_{(1)}A^{2}N_{4}K_{3}}{H_{(2)}}
	-\frac{2B_{(1)}AN_{4}K_{3}M_{1}'}{H_{(2)}M_{3}'}
	+\frac{3B_{(1)}A^{3}N_{4}\Lambda_{1}'}{2H_{(2)}}
	\nonumber \\
	\!&-&\!
	\frac{3B_{(1)}A^{3}N_{4}\Lambda_{3}'M_{1}'}{2H_{(2)}M_{3}'}
	-\frac{3B_{(1)}A^{3}N_{1}\Lambda_{4}'}{2H_{(2)}}
	+\frac{3B_{(1)}A^{3}N_{3}M_{1}'\Lambda_{4}'}{2H_{(2)}M_{3}'}.
\end{eqnarray}
The values of $\alpha_{1}$, $\alpha_{2}$, and $\alpha_{3}$ in Eq. (\ref{quadratic equation with respect to k02}) are taken according to Eq. (\ref{real part determinant of the coefficient matrix is 0 in case 7}). The specific analysis will not be repeated.

For the analysis of the amplitude of scalar mode gravitational waves, considering the dispersion relation we are examining, if the coefficient matrix is zero in Eq. (\ref{scalar mode equations 2 in case 7 matrix form}), then $\phi$ and $\Psi$ are independent of each other. Scalar gravitational waves generally have two independent polarization modes. If the rank of the coefficient matrix is one, there is a relationship
\begin{eqnarray}
	\label{relationship between phi and Psi in case 7}
	\left[
	4B_{(1)}A^{2}
	-\frac{4B_{(1)}AM_{1}'}{M_{3}'}
	+i\left(
	\Lambda_{1}'
	-\frac{\Lambda_{3}'M_{1}'}{M_{3}'}
	\right)k_{0}
	\right]\phi
	-\left(
	\Lambda_{4}'k_{0}^{2}
	+\Lambda_{5}k_{3}^{2}
	\right)\Psi
	=0.
\end{eqnarray}
Here, $\phi$ and $\Theta$ are not independent, and the amplitude ratio between them forms a complex number. Scalar mode gravitational waves have only one polarization mode, which is a mixture of two modes: (1) a pure longitudinal mode (determined by $\phi$), and (2) a mixed mode of the breathing mode and longitudinal mode, with equal amplitude for both (determined by $\Theta$).

\textbf{Case 8}: $M_{1}' \neq 0,~M_{3}' \neq 0,~M_{4}' \neq 0$. In this case, Eq. (\ref{scalar mode equation 3 new}) requires 
\begin{eqnarray}
	\label{Omega in case 8}
	\Omega=-\frac{M_{1}'}{M_{3}'}\phi-\frac{M_{4}'}{M_{3}'}\partial_{0}\Psi,
\end{eqnarray}
and Eqs. (\ref{scalar mode equation 1 new}) and (\ref{scalar mode equation 2 new}) respectively become
\begin{eqnarray}
	\label{scalar mode equation 1 new in case 8}
	\left(
	4B_{(1)}A^{2}
	-\frac{4B_{(1)}AM_{1}'}{M_{3}'}
	\right)\phi
	-\frac{4B_{(1)}AM_{4}'}{M_{3}'}\partial_{0}\Psi
	+\left(
	\Lambda_{1}'
	-\frac{\Lambda_{3}'M_{1}'}{M_{3}'}
	\right)\partial_{0}\phi
	\nonumber \\
	+\left(
	\Lambda_{4}'
	-\frac{\Lambda_{3}'M_{4}'}{M_{3}'}
	\right)\partial_{0}^{2}\Psi
	+\Lambda_{5}\Delta\Psi
	=0,
	\\
	\label{scalar mode equation 2 new in case 8}
	\left(
	-4A_{(0)}A^{4}
	+\frac{4A_{(0)}A^{3}M_{1}'}{M_{3}'}
	\right)\phi
	+\frac{4A_{(0)}A^{3}A_{4}'}{M_{3}'}\partial_{0}\Psi
	+2B_{(1)}A^{2}\Delta\Psi
	\nonumber \\
	+\left(
	-\frac{3B_{(1)}A^{3}N_{1}}{2H_{(2)}}
	+\frac{3B_{(1)}A^{3}N_{3}M_{1}'}{2H_{(2)}M_{3}'}
	\right)\partial_{0}\phi
	+\left(
	-\frac{3B_{(1)}A^{3}N_{4}}{2H_{(2)}}
	+\frac{3B_{(1)}A^{3}N_{3}M_{4}'}{2H_{(2)}M_{3}'}
	\right)\partial_{0}^{2}\Psi
	\nonumber \\
	+\left(
	K_{1}'-\frac{K_{5}'M_{1}'}{M_{3}'}
	\right)\partial_{0}^{2}\phi
	+\left(
	K_{2}'-\frac{K_{6}'M_{1}'}{M_{3}'}
	\right)\Delta\phi
	+\left(
	K_{7}'-\frac{K_{6}'M_{4}'}{M_{3}'}
	\right)\Delta \partial_{0}\Psi
	\nonumber \\
	+\left(
	-\frac{K_{3}N_{4}}{2H_{(2)}}
	-\frac{K_{5}'M_{4}'}{M_{3}'}
	\right)\partial_{0}^{3}\Psi
	=0.
\end{eqnarray}
Considering monochromatic plane wave solutions, the equations above can be expressed as:
\begin{eqnarray}
	\label{scalar mode equations 2 in case 8 matrix form}
	\begin{pmatrix}
		\mathcal{A}_{11} & ~
		\mathcal{A}_{12}\\
		\mathcal{A}_{21} & ~
		\mathcal{A}_{22}\\
	\end{pmatrix}
	\begin{pmatrix}
		\mathring{\phi}\\
		\mathring{\Psi}
	\end{pmatrix}
	=0,
\end{eqnarray}
where
\begin{eqnarray}
	\label{A11,A12,A21,A22 in case 8 matrix form}
	\mathcal{A}_{11}
	\!&=&\!
	4B_{(1)}A^{2}-\frac{4B_{(1)}AM_{1}'}{M_{3}'}+i\left(\Lambda_{1}'-\frac{\Lambda_{3}'M_{1}'}{M_{3}'}\right)k_{0},
	\\
	\mathcal{A}_{12}
	\!&=&\!
	-i\frac{4B_{(1)}AM_{4}'}{M_{3}'}k_{0}
	-\left(
	\Lambda_{4}'
	-\frac{\Lambda_{3}'M_{4}'}{M_{3}'}
	\right)k_{0}^{2}
	-\Lambda_{5}k_{3}^{2},
	\\
	\mathcal{A}_{21}
	\!&=&\!
	-4A_{(0)}A^{4}
	+\frac{4A_{(0)}A^{3}M_{1}'}{M_{3}'}
	+i\left(
	-\frac{3B_{(1)}A^{3}N_{1}}{2H_{(2)}}
	+\frac{3B_{(1)}A^{3}N_{3}M_{1}'}{2H_{(2)}M_{3}'}
	\right)k_{0}
	\nonumber \\
	\!&-&\!
	\left(
	K_{1}'-\frac{K_{5}'M_{1}'}{M_{3}'}
	\right)k_{0}^{2}
	-\left(
	K_{2}'-\frac{K_{6}'M_{1}'}{M_{3}'}
	\right)k_{3}^{2},
	\nonumber \\
	\mathcal{A}_{22}
	\!&=&\!
	i\frac{4A_{(0)}A^{3}M_{4}'}{M_{3}'}k_{0}
	-\left(
	-\frac{3B_{(1)}A^{3}N_{4}}{2H_{(2)}}
	+\frac{3B_{(1)}A^{3}N_{3}M_{4}'}{2H_{(2)}M_{3}'}
	\right)k_{0}^{2}
	\nonumber \\
	\!&-&\!
	2B_{(1)}A^{2}k_{3}^{2}
	-i\left(
	K_{7}'-\frac{K_{6}'M_{4}'}{M_{3}'}
	\right)k_{0}k_{3}^{2}
	+i\left(
	\frac{K_{3}N_{4}}{2H_{(2)}}
	+\frac{K_{5}'M_{4}'}{M_{3}'}
	\right)k_{0}^{3}.
\end{eqnarray}

The determinant of the coefficient matrix of Eq. (\ref{scalar mode equations 2 in case 8 matrix form}) is zero, which is equivalent to the following equations:
\begin{eqnarray}
	\label{real part determinant of the coefficient matrix is 0 in case 8}
	\!&&\!
	\left(
	-\frac{N_{4}K_{3}\Lambda_{1}'}{2H_{(2)}}
	-\frac{M_{4}'K_{5}'\Lambda_{1}'}{M_{3}'}
	+\frac{N_{4}K_{3}M_{1}'\Lambda_{3}'}{2H_{(2)}M_{3}'}
	+\frac{M_{4}'K_{1}'\Lambda_{3}'}{M_{3}'}
	-K_{1}'\Lambda_{4}'
	+\frac{M_{1}'K_{5}'\Lambda_{4}'}{M_{3}'}
	\right)k_{0}^{4}
	\nonumber \\
	\!&+&\!
	\left(
	\frac{6B_{(1)}^{2}A^{5}N_{4}}{H_{(2)}}
	-\frac{6B_{(1)}^{2}A^{4}N_{4}M_{1}'}{H_{(2)}M_{3}'}
	+\frac{6B_{(1)}^{2}A^{4}N_{1}M_{4}'}{H_{(2)}M_{3}'}
	-\frac{6B_{(1)}^{2}A^{5}N_{3}M_{4}'}{H_{(2)}M_{3}'}
	-\Lambda_{5}K_{1}'k_{3}^{2}
	\right.
	\nonumber \\
	\!&+&\!
	\frac{\Lambda_{5}M_{1}'K_{5}'}{M_{3}'}k_{3}^{2}
	-\frac{4A_{(0)}A^{3}M_{4}'\Lambda_{1}'}{M_{3}'}
	-\frac{M_{4}'K_{6}'\Lambda_{1}'k_{3}^{2}}{M_{3}'}
	+K_{7}'\Lambda_{1}'k_{3}^{2}
	+\frac{4A_{(0)}A^{4}M_{4}'\Lambda_{3}'}{M_{3}'}
	\nonumber \\
	\!&+&\!
	\frac{M_{4}'K_{2}'\Lambda_{3}'k_{3}^{2}}{M_{3}'}
	-\frac{M_{1}'K_{7}'\Lambda_{3}'k_{3}^{2}}{M_{3}'}
	-4A_{(0)}A^{4}\Lambda_{4}'
	+\frac{4A_{(0)}A^{3}M_{1}'\Lambda_{4}'}{M_{3}'}
	-K_{2}'\Lambda_{4}'k_{3}^{2}
	\nonumber \\
	\!&+&\!
	\left.
	\frac{M_{1}'K_{6}'\Lambda_{4}'k_{3}^{2}}{M_{3}'}
	\right)k_{0}^{2}
	+\left(
	-8B_{(1)}^{2}A^{4}k_{3}^{2}
	-4A_{(0)}A^{4}\Lambda_{5}k_{3}^{2}
	+\frac{8B_{(1)}^{2}A^{3}M_{1}'k_{3}^{2}}{M_{3}'}
	\right.
	\nonumber \\
	\!&+&\!
	\left.
	\frac{4A_{(0)}A^{3}\Lambda_{5}M_{1}'k_{3}^{2}}{M_{3}'}
	-\Lambda_{5}K_{2}'k_{3}^{4}
	+\frac{\Lambda_{5}M_{1}'K_{6}'k_{3}^{4}}{M_{3}'}
	\right)
	=0,
	\\
	\label{imaginary part determinant of the coefficient matrix is 0 in case 8}
	\!&&\!
	\left(
	\frac{2B_{(1)}A^{2}N_{4}K_{3}}{H_{(2)}}
	-\frac{2B_{(1)}AN_{4}K_{3}M_{1}'}{H_{(2)}M_{3}'}
	-\frac{4B_{(1)}AM_{4}'K_{1}'}{M_{3}'}
	+\frac{4B_{(1)}A^{2}M_{4}'K_{5}'}{M_{3}'}
	\right.
	\nonumber \\
	\!&+&\!
	\frac{3B_{(1)}A^{3}N_{4}\Lambda_{1}'}{2H_{(2)}}
	-\frac{3B_{(1)}A^{3}N_{3}M_{4}'\Lambda_{1}'}{2H_{(2)}M_{3}'}
	-\frac{3B_{(1)}A^{3}N_{4}M_{1}'\Lambda_{3}'}{2H_{(2)}M_{3}'}
	+\frac{3B_{(1)}A^{3}N_{1}M_{4}'\Lambda_{3}'}{2H_{(2)}M_{3}'}
	\nonumber \\
	\!&-&\!
	\left.
	\frac{3B_{(1)}A^{3}N_{1}\Lambda_{4}'}{2H_{(2)}}
	+3\frac{B_{(1)}A^{3}N_{3}M_{1}'\Lambda_{4}'}{2H_{(2)}M_{3}'}
	\right)k_{0}^{3}
	+\left(
	-\frac{3B_{(1)}A^{3}N_{1}\Lambda_{5}k_{3}^{2}}{2H_{(2)}}
	+\frac{3B_{(1)}A^{3}N_{3}\Lambda_{5}M_{1}'k_{3}^{2}}{2H_{(2)}M_{3}'}
	\right.
	\nonumber \\
	\!&-&\!
	\frac{4B_{(1)}AM_{4}'K_{2}'k_{3}^{2}}{M_{3}'}
	+\frac{4B_{(1)}A^{2}M_{4}'K_{6}'k_{3}^{2}}{M_{3}'}
	-4B_{(1)}A^{2}K_{7}'k_{3}^{2}
	+\frac{4B_{(1)}AM_{1}'K_{7}'k_{3}^{2}}{M_{3}'}
	\nonumber \\
	\!&-&\!
	\left.
	2B_{(1)}A^{2}\Lambda_{1}'k_{3}^{2}
	+\frac{2B_{(1)}A^{2}M_{1}'\Lambda_{3}'k_{3}^{2}}{M_{3}'}
	\right)k_{0}
	=0.
\end{eqnarray}
The analysis of the dispersion relation is entirely parallel to Case 2. Here, we classify the parameter space of Eq. (\ref{imaginary part determinant of the coefficient matrix is 0 in case 8}). In this case, Eq. (\ref{imaginary part is 0 in case 2.1 two solution of k0 2}) still yields a constant wave speed similar to Eq. (\ref{imaginary part is 0 in case 5.1 two solution of k0 2}). The only difference is that the values of $\mathcal{D}_{1}$ and $\mathcal{D}_{2}$ in Eq. (\ref{D1,D2 in case 5.1}) are replaced by
\begin{eqnarray}
	\label{D1,D2 in case 8.1}
	\mathcal{D}_{1}
	\!&=&\!
	\frac{3B_{(1)}A^{3}N_{1}\Lambda_{5}}{2H_{(2)}}
	-\frac{3B_{(1)}A^{3}N_{3}\Lambda_{5}M_{1}'}{2H_{(2)}M_{3}'}
	+\frac{4B_{(1)}AM_{4}'K_{2}'}{M_{3}'}
	-\frac{4B_{(1)}A^{2}M_{4}'K_{6}'}{M_{3}'}
	\nonumber \\
	\!&+&\!
	4B_{(1)}A^{2}K_{7}'
	-\frac{4B_{(1)}AM_{1}'K_{7}'}{M_{3}'}
	+2B_{(1)}A^{2}\Lambda_{1}'
	-\frac{2B_{(1)}A^{2}M_{1}'\Lambda_{3}'}{M_{3}'},
	\nonumber \\
	\mathcal{D}_{2}
	\!&=&\!
	\frac{2B_{(1)}A^{2}N_{4}K_{3}}{H_{(2)}}
	-\frac{2B_{(1)}AN_{4}K_{3}M_{1}'}{H_{(2)}M_{3}'}
	-\frac{4B_{(1)}AM_{4}'K_{1}'}{M_{3}'}
	+\frac{4B_{(1)}A^{2}M_{4}'K_{5}'}{M_{3}'}
	\nonumber \\
	\!&+&\!
	\frac{3B_{(1)}A^{3}N_{4}\Lambda_{1}'}{2H_{(2)}}
	-\frac{3B_{(1)}A^{3}N_{3}M_{4}'\Lambda_{1}'}{2H_{(2)}M_{3}'}
	-\frac{3B_{(1)}A^{3}N_{4}M_{1}'\Lambda_{3}'}{2H_{(2)}M_{3}'}
	+\frac{3B_{(1)}A^{3}N_{1}M_{4}'\Lambda_{3}'}{2H_{(2)}M_{3}'}
	\nonumber \\
	\!&-&\!
	\frac{3B_{(1)}A^{3}N_{1}\Lambda_{4}'}{2H_{(2)}}
	+3\frac{B_{(1)}A^{3}N_{3}M_{1}'\Lambda_{4}'}{2H_{(2)}M_{3}'}.
\end{eqnarray}
And the values of $\alpha_{1}$, $\alpha_{2}$, and $\alpha_{3}$ in Eq. (\ref{quadratic equation with respect to k02}) are taken according to Eq. (\ref{real part determinant of the coefficient matrix is 0 in case 8}). The specific analysis will not be repeated.

For the analysis of the amplitude of scalar mode gravitational waves, considering the dispersion relation we are examining, if the coefficient matrix is zero in Eq. (\ref{scalar mode equations 2 in case 8 matrix form}), then scalar gravitational waves generally have two independent polarization modes. If the rank of the coefficient matrix is one, there is a relationship
\begin{eqnarray}
	\label{relationship between phi and Psi in case 8}
	\left[
	4B_{(1)}A^{2}
	-\frac{4B_{(1)}AM_{1}'}{M_{3}'}
	+i\left(
	\Lambda_{1}'
	-\frac{\Lambda_{3}'M_{1}'}{M_{3}'}
	\right)k_{0}
	\right]\phi
	\nonumber \\
	-\left[
	i\frac{4B_{(1)}AM_{4}'}{M_{3}'}k_{0}
	+\left(
	\Lambda_{4}'-\frac{\Lambda_{3}'M_{4}'}{M_{3}'}
	\right)k_{0}^{2}
	+\Lambda_{5}k_{3}^{2}
	\right]
	\Psi
	=0.
\end{eqnarray}
Here, $\phi$ and $\Theta$ are interdependent, and their amplitude ratio constitutes a complex number. Similar to Case 3, scalar mode gravitational waves have only one polarization mode.

\section{Gravitational wave polarizations in generalized Proca theory}
\label{app: D}

Now, we use generalized Proca theory as an example to demonstrate how to directly derive the gravitational wave polarization properties of a specific theory from the generalized analysis provided in Sec. \ref{sec: 6}.

Generalized Proca theory is a relatively general second-order vector-tensor theory, for which the action is given by \cite{L. Heisenberg3}
\begin{eqnarray}
	\label{GP action}
	S\left[g_{\mu\nu},\mathcal{A}^{\mu}\right] = \int d^4x \sqrt{-g}~
	\Bigl(\mathcal{L}_{F}+\mathcal{L}_{2}+\mathcal{L}_{3}+\mathcal{L}_{4}+\mathcal{L}_{5}\Bigl),
\end{eqnarray}
where
\begin{eqnarray}
	\label{LF}
	\mathcal{L}_{F}&=&-\frac{1}{4}F_{\mu\nu}F^{\mu\nu},
	\\
	\label{L2}
	\mathcal{L}_{2}&=&G_{2}[X],
	\\
	\label{L3}
	\mathcal{L}_{3}&=&G_{3}[X]\nabla_{\mu}\mathcal{A}^{\mu},
	\\
	\label{L4}
	\mathcal{L}_{4}&=&G_{4}[X]{R}
	+G_{4,X}[X]\left[
	\left(\nabla_{\mu}\mathcal{A}^{\mu}\right)^{2}
	+c_{2}\nabla_{\mu}\mathcal{A}_{\nu}\nabla^{\mu}\mathcal{A}^{\nu}
	-\left(1+c_{2}\right)\nabla_{\mu}\mathcal{A}_{\nu}\nabla^{\nu}\mathcal{A}^{\mu}
	\right],
	\\
	\label{L5}
	\nonumber
	\mathcal{L}_{5}&=&G_{5}[X]\left({R}_{\mu\nu}-\frac{1}{2}{g}_{\mu\nu}{R}\right)\nabla^{\mu}\mathcal{A}^{\nu}
	\\ \nonumber
	&-&\frac{1}{6}G_{5,X}[X]
	\left[
	\left(\nabla_{\mu}\mathcal{A}^{\mu}\right)^{3}
	-3d_{2}\nabla_{\mu}\mathcal{A}^{\mu}\nabla_{\nu}\mathcal{A}_{\lambda}\nabla^{\nu}\mathcal{A}^{\lambda}
	-3\left(1-d_{2}\right)\nabla_{\mu}\mathcal{A}^{\mu}\nabla_{\nu}\mathcal{A}_{\lambda}\nabla^{\lambda}\mathcal{A}^{\nu}
	\right.\nonumber
	\\
	&+&
	\left.
	\left(2-3d_{2}\right)\nabla_{\mu}\mathcal{A}_{\nu}\nabla^{\lambda}\mathcal{A}^{\mu}\nabla^{\nu}\mathcal{A}_{\lambda}
	+3d_{2}\nabla_{\mu}\mathcal{A}_{\nu}\nabla^{\lambda}\mathcal{A}^{\mu}\nabla_{\lambda}\mathcal{A}^{\nu}
	\right].
\end{eqnarray}
Here, $F_{\mu\nu}=\nabla_{\mu}\mathcal{A}_{\nu}-\nabla_{\nu}\mathcal{A}_{\mu}$, $X=-\frac{1}{2}g_{\mu\nu}\mathcal{A}^{\mu}\mathcal{A}^{\nu}$, $c_{2}$ and $d_{2}$ are constants. In addition, $G_{n,X}={dG_{n}}/{dX}$, $G_{n,XX}={dG_{n,X}}/{dX}$ ($n=2,3,4,5$).

In order to use our generalized analysis to determine the gravitational wave polarization modes of  generalized Proca theory, the action (\ref{GP action}) should be expanded to second-order with respect to perturbations (\ref{weak perturbation of fields}). Subsequently, by directly comparing the second-order perturbation action with Eqs. (\ref{L0 gauge}), (\ref{L1 gauge}), and (\ref{L2 gauge}), or by comparing the equations obtained through the perturbation action method with Eqs. (\ref{delta S2flat/delta h}) and (\ref{delta S2flat/delta B}), we find that the parameters corresponding to generalized Proca theory take the following form: 
\begin{eqnarray}
	\label{parameters in GP action}
	A_{(0)} \!&=&\! \frac{1}{4}\mathring{G}_{2,XX},
	\nonumber \\
	A_{(1)} \!&=&\!0,~ B_{(1)}=-\frac{1}{2}\mathring{G}_{3,X},
	\nonumber \\
	A_{(2)} \!&=&\! B_{(2)}=C_{(2)}=F_{(2)}=G_{(2)}=J_{(2)}=0,
	 \\
	D_{(2)} \!&=&\! -\mathring{G}_{4,X},~E_{(2)}=\mathring{G}_{4,X}+2c_{2}\mathring{G}_{4,X}-1,
	\nonumber \\
	H_{(2)} \!&=&\! \frac{1}{2}\mathring{G}_{4},
	~I_{(2)}=\frac{1}{2}\mathring{G}_{4,X},
	~K_{(2)}=-\frac{1}{2}\mathring{G}_{4,X}, \nonumber
\end{eqnarray}
where the notation ``$\circ$" above the letter means that the corresponding function takes the background value. It should also be noted that all quantities here may differ by the same multiplicative factor as those obtained by directly comparing the action. This does not affect any conclusions, as multiplying the entire action by a constant does not alter the field equations.

As can be seen from Eq. (\ref{vT=}), the necessary and sufficient condition for tensor mode gravitational waves in generalized Proca theory not to propagate at the speed of light is $\mathring{G}_{4,X} \neq 0$. For vector mode gravitational waves, since $G_{(2)}=0$ in this case, it can be inferred that if the tensor mode propagates at the speed of light, there is no vector mode in generalized Proca theory. Comparing the analysis in Sec. \ref{vector modes}, we find that the case in generalized Proca theory generally corresponds to Case 4. As mentioned earlier, the analysis of Case 4 is entirely parallel to Case 3. In fact, it generally  corresponds to a case similar to Case 3.6, where the speed of vector mode gravitational waves is a constant and independent of the wave vector.

For scalar mode gravitational waves, the corresponding parameters are
\begin{eqnarray}
	\label{parameters in scalar mode GP action}
	\Lambda'_{1} \!&=&\! 8c_{2}\mathring{G}_{4,X}A-4A+\frac{8\mathring{G}_{4,X}A\left(\mathring{G}_{4,X}A^{2}+\mathring{G}_{4}\right)}{\mathring{G}_{4}},
	\nonumber \\
	\Lambda_{2} \!&=&\! -4\mathring{G}_{4,X}A,
	\nonumber \\
	\Lambda'_{3} \!&=&\!\Lambda'_{4} =4c_{2}\mathring{G}_{4,X}-2+\frac{8\mathring{G}_{4,X}^{2}A^{2}}{\mathring{G}_{4}},
	\nonumber \\
	K'_{1} \!&=&\! K_{3} = K'_{5}=0,
	\nonumber \\
	K'_{2} \!&=&\! -8c_{2}\mathring{G}_{4,X}A^{2}+4A^{2}-\frac{4\left(\mathring{G}_{4,X}A^{2}+\mathring{G}_{4}\right)^{2}}{\mathring{G}_{4}},
	\nonumber \\
	K'_{6} \!&=&\! K'_{7} =-4c_{2}\mathring{G}_{4,X}A+2A-\frac{4\mathring{G}_{4,X}A\left(\mathring{G}_{4,X}A^{2}+\mathring{G}_{4}\right)}{\mathring{G}_{4}},
	 \\
	M'_{1} \!&=&\!-4c_{2}\mathring{G}_{4,X}A^{2}+2A^{2}-4\mathring{G}_{4,X}A^{2}-4\mathring{G}_{4},
	\nonumber \\
	M_{2} \!&=&\! 2\mathring{G}_{4},
	\nonumber \\
	M'_{3} \!&=&\! M'_{4}=-2c_{2}\mathring{G}_{4,X}A+A-4\mathring{G}_{4,X}A,
	\nonumber \\
	N_{1} \!&=&\! 2\mathring{G}_{4,X}A^{2}+2\mathring{G}_{4},
	\nonumber \\
	N_{3} \!&=&\! N_{4}=2\mathring{G}_{4,X}A. \nonumber
\end{eqnarray}
It can be seen that it belongs to Case 8 in Sec. \ref{scalar modes}. For this scenario, Eq. (\ref{imaginary part determinant of the coefficient matrix is 0 in case 8}) always holds, while Eq. (\ref{real part determinant of the coefficient matrix is 0 in case 8}) provides a wave speed solution that is independent of the wave vector. Therefore, generalized Proca theory generally allows for the existence of scalar mode gravitational waves with a wave speed independent of frequency, and the specific expression for this speed is given by Eq. (\ref{real part determinant of the coefficient matrix is 0 in case 8}). 

Due to $M'_{3} = M'_{4}$, Eq. (\ref{Omega in case 8}) becomes
\begin{eqnarray}
	\label{Omega in case 8 in GP theory}
	\Omega=-\frac{M_{1}'}{M_{3}'}\phi-\partial_{0}\Psi.
\end{eqnarray}
Furthermore, noting that $N_{3}=N_{4}$, it can be inferred that Eq. (\ref{scalar mode Theta}) becomes
\begin{eqnarray}
	\label{scalar mode Theta in GP theory}
	\Theta
	=
	-\frac{N_{1}-N_{3}M'_{1}/M'_{3}}{2H_{(2)}}\phi.
\end{eqnarray}
It can be seen that there is no phase difference between $\phi$ and $\Theta$ that contribute to the scalar mode, resulting in only one independent scalar mode in generalized Proca theory, which is a mixture of the breathing mode and the longitudinal mode. The amplitude ratio of the two mixed modes is also given by Eq. (\ref{scalar mode Theta in GP theory}).

The above conclusion is consistent with our previous analysis of the gravitational wave polarization modes in generalized Proca theory as presented in Ref. \cite{Y.Dong3}. From this example, it can be seen that under the broad assumptions of this paper, even without considering parity-breaking terms containing $E^{\mu\nu\lambda\rho}$, generalized Proca theory is only a very special case within the most general second-order vector-tensor theory.

\end{CJK}
\end{document}